\documentclass[12pt,preprint]{aastex}
\usepackage{lineno}
\usepackage{graphicx}
\usepackage{epstopdf}

\renewcommand {\deg}   {\mbox{$^\circ$}}

\newcommand   {\kms}   {\mbox{km\,s$^{-1}$}}
\renewcommand {\ga}    {\mbox{\rlap{\hbox{\lower5pt\hbox{$\sim$}}}\hbox{$>$}}}
\renewcommand {\la}    {\mbox{\rlap{\hbox{\lower5pt\hbox{$\sim$}}}\hbox{$<$}}}

\begin{document}


\def\kms {\hbox{km{\hskip0.1em}s$^{-1}$}} 
\def\msol{\hbox{$\hbox{M}_\odot$}}
\def\lsol{\hbox{$\hbox{L}_\odot$}}
\def\kms{km s$^{-1}$}
\def\Blos{B$_{\rm los}$}
\def\etal   {{\it et al. }}                     
\def\psec           {$.\negthinspace^{s}$}
\def\pasec          {$.\negthinspace^{\prime\prime}$}
\def\pdeg           {$.\kern-.25em ^{^\circ}$}
\def\degree{\ifmmode{^\circ} \else{$^\circ$}\fi}
\def\ee #1 {\times 10^{#1}}          
\def\ut #1 #2 { \, \textrm{#1}^{#2}} 
\def\u #1 { \, \textrm{#1}}          
\def\nH {n_\mathrm{H}}

\def\ddeg   {\hbox{$.\!\!^\circ$}}              
\def\deg    {$^{\circ}$}                        
\def\le     {$\leq$}                            
\def\sec    {$^{\rm s}$}                        
\def\msol   {\hbox{$M_\odot$}}                  
\def\i      {\hbox{\it I}}                      
\def\v      {\hbox{\it V}}                      
\def\dasec  {\hbox{$.\!\!^{\prime\prime}$}}     
\def\asec   {$^{\prime\prime}$}                 
\def\dasec  {\hbox{$.\!\!^{\prime\prime}$}}     
\def\dsec   {\hbox{$.\!\!^{\rm s}$}}            
\def\min    {$^{\rm m}$}                        
\def\hour   {$^{\rm h}$}                        
\def\amin   {$^{\prime}$}                       
\def\lsol{\, \hbox{$\hbox{L}_\odot$}}
\def\sec    {$^{\rm s}$}                        
\def\etal   {{\it et al. }}                     

\def\xbar   {\hbox{$\overline{\rm x}$}}         

\shorttitle{structure function}
\shortauthors{zadeh}

\title{Interacting Cosmic Rays  with Molecular Clouds: 
A Bremsstrahlung Origin of Diffuse  High Energy Emission\\ 
from  the Inner 2\deg$\times$1\deg\ of the 
Galactic Center}

\author{F. Yusef-Zadeh$^1$, J.W. Hewitt$^2$, M. Wardle$^3$, V. Tatischeff$^4$, D. A. Roberts$^1$, W. Cotton$^5$,
H. Uchiyama$^6$,   M.  Nobukawa$^6$, T. G.  Tsuru$^6$,
C. Heinke$^7$ \& M. Royster$^1$}
\affil{$^1$Department of Physics and Astronomy, Northwestern University, Evanston, IL 60208}
\affil{$^2$Code 662, NASA Goddard Space Flight Center, Greenbelt, MD 20771}
\affil{$^3$Department of Physics \& Astronomy, and Research Center for Astronomy, Astrophysics \& Astrophotonics, 
Macquarie University, Sydney NSW 2109, Australia}
\affil{$^4$Centre de Spectrom\'etrie Nucl\'eaire et de Spectrom\'etrie de Masse, IN2P3/CNRS and Univ Paris-Sud, 91405 
Orsay Campus, France}
\affil{$^5$NRAO, 520 Edgemont Road, Charlottesville, VA 22903, USA}
\affil{$^6$Cosmic Ray Group, Physics, Kyoto Univ. Kitashirakawa-Oiwake-Cho, Sakyo, Kyoto 606-8502, Japan}
\affil{$^7$Department of Physics, University of Alberta, Room 238 CEB, Edmonton, AB, T6G 2G7, Canada}

\begin{abstract} 
The high energy activity in the inner few degrees of the Galactic center is traced by diffuse 
radio, X-ray and $\gamma$-ray emission. The  physical relationship between 
different components of  diffuse gas 
emitting at multiple wavelengths is a focus of this work. We first present radio continuum 
observations using Green Bank Telescope and model the nonthermal spectrum in terms of a broken 
power-law distribution of $\sim$GeV  electrons emitting synchrotron radiation.  
We show that the emission detected by Fermi is primarily due to 
nonthermal bremsstrahlung produced by  the  population of synchrotron  emitting 
electrons  in the GeV energy range 
interacting with neutral gas. 
The extrapolation of  the electron population 
measured from radio data to low and high energies can also 
explain the origin of FeI 6.4 
keV line and diffuse TeV emission, as observed with Suzaku, XMM-Newton, Chandra and the H.E.S.S. observatories. 
The inferred physical quantities from modeling multiwavelength emission in the context of 
bremsstrahlung emission from the inner $\sim300\times120$ parsecs  of the Galactic center are 
constrained to have the 
cosmic ray ionization rate $\sim1-10\times10^{-15}$ s$^{-1}$, molecular gas heating rate 
elevating the gas temperature to 75-200K, fractional ionization of molecular gas $10^{-6}$ to 
$10^{-5}$, large scale magnetic field $10-20\mu$G, the density of diffuse and dense 
molecular gas $\sim100$ and $\sim10^3$ cm$^{-3}$ over  300pc and 50pc  pathlengths, and 
the variability of FeI K$\alpha$ 6.4 keV line emission on yearly time scales. Important 
implications of our study are that GeV electrons emitting in radio can  explain the 
GeV $\gamma$-rays detected by Fermi and that the cosmic ray irradiation model, like the model of the X-ray 
irradiation triggered by past activity of Sgr A*, can also explain the origin of the variable 
6.4 keV emission from Galactic center molecular clouds. 
\end{abstract}





\section{Introduction}

The Galactic center hosts several sources of energetic activity: X-ray flare activity from 
Sgr~A*, nonthermal linear filaments, supernova remnants interacting with molecular clouds, 
colliding winds of massive stars, pulsars, transient radio and X-ray sources and a population of hard 
X-ray sources (Muno et al. 2006, 2009;  Koyama et al. 1996; Tsuboi, Ukita \& Handa  1997; Baganoff et 
al. 2001; Murakami et al. 2001; 
 Deneva,   Cordes, and  Lazio  2009).
 This region  also hosts  massive molecular clouds containing 
pockets of current and past massive star formation (see Jones et al. 2011 and references 
therein). The most prominent clouds are associated with Sgr B2 and Sgr C, 
the 50, 40, 20 and 
--30 \kms\,   complexes as well as the cloud G0.11-0.11 adjacent to the radio continuum Arc near 
$l \sim$ 0.2$^\circ$. 
Molecular clouds are traditionally studied by molecular line observations at 
millimeter wavelengths. However, 
diffuse high energy emission has also been detected 
from Galactic center molecular clouds.  
These unique Galactic center molecular clouds that emit the  6.4\, keV X-ray line, 
GeV and TeV radiation as well as  rotationally excited millimeter lines 
help to bridge a gap in understanding the radiation 
processes that operate at low and high energies.


The Galactic center region hosts  warm molecular gas as well as  a number of 
synchrotron emitting radio sources. 
A high cosmic ray ionization rate is  estimated from H$_3^+$ measurements of this region 
(Oka  et al. 2005).  
It is then natural to consider 
the interaction of  cosmic ray electrons  that produce radio emission 
with ambient gas in
the context of  nonthermal bremsstrahlung. 
We study this  interaction in detail and 
investigate the origin of the high 
cosmic ray ionization rate and  high molecular gas temperature
(Oka  et al. 2005; H\"uttemeister et al.  1993).  
We show the distribution of
GeV $\gamma$-ray emission observed by $Fermi$ and 
model  the emission by studying 
the population of nonthermal electrons using radio  data. Furthermore, the extrapolation of the radio spectrum  of the
 GeV population to 
10 keV as well as  a young population of electrons extrapolated to  TeV energies 
can also explain the observed emission at X-ray  and TeV energy range, respectively. 
In particular, 
the fluorescent FeI K$\alpha$ line emission at 6.4 keV and diffuse TeV 
emission are  recognized to trace the molecular clouds of the Galactic center. 
It has been   suggested that the fluorescent  6.4 keV emission 
results 
from X-ray irradiation (Sunyaev, Markovitch \& Pavlinsky 1993). 
The source of the emission is considered to be a 
 hypothetical transient source associated with the massive black hole at the Galactic
center, Sgr A*, and that we are now seeing its echo in the 6.4 keV line emission 
(Koyama et al.  1996; Murakami et al. 2001; Ponti et al. 2010). This event requires a hard X-ray luminosity of
$\sim$10$^{39}$ erg s$^{-1}$ from Sgr A*. The year-to-year time variability of 6.4 keV line emission
has also been used as a strong support for the  irradiation scenario. 
In this picture, the yearly variability is due to X-ray 
fronts  from multiple 
outbursts from Sgr A* which occurred in the last few hundred years. 
The origin of the 6.4
keV line emission from neutral iron due to low-energy cosmic ray electrons and
protons of 
neutral gas has also been considered (Yusef-Zadeh, Wardle \& Roy 2007a; Chernyshov et al. 2011). 
More recently, the origin of the 6.4 keV line emission from the Arches cluster 
has also been explained  in terms of cosmic-ray ion bombardment
of molecular gas surrounding the cluster 
(Tatischeff, Decourchelle \& Maurin 2012). Here, we reinvestigate the cosmic ray irradiation 
picture in the context of  nonthermal bremsstrahlung. 
It is important to determine the 6.4 keV emission contributed by  each  of these  
two models as they provide evidence  for the past activity of Sgr A* or 
for a large
population of low energy cosmic rays (LECR) in the Galactic center region. 


We begin by describing radio observations using Green Bank Telescope (GBT), $\gamma$-ray observations with the
$Fermi$ Large Area Telescope (LAT), and X-ray line observations in $\S2$.  
In $\S$2.1, the spectrum of radio emission between 325 MHz and 8.5 GHz
is modeled in order to separate thermal and nonthermal radio components. 
In $\S$3 we estimate the cosmic ray ionization rate and compare it with that
measured from H$_3^+$ absorption lines  (Oka et al. 2005; Goto et al. 2011). 
We also  account for the  warm molecular gas as observed throughout the Galactic
center and the origin of 6.4 keV emission from Galactic center molecular clouds. 
Sections 2 and 3 discuss the interaction of low energy cosmic ray particles with 
molecular gas,  whereas 
$\S4$ discusses   the high energy tail of cosmic rays interacting with molecular gas
to produce $\gamma$-ray emission detected by $Fermi$ and the H.E.S.S. telescopes.

\section {Nonthermal Radiation from  Diffuse Gas}

Nonthermal radio continuum emission is used to probe the population of cosmic ray
electrons.  
These cosmic ray electrons 
may interact with the reservoir of molecular gas distributed in the
Galactic center. 
An accurate measure of the cosmic ray flux traced at radio wavelengths 
is critical to investigate the
origin of $\gamma$-ray  and X-ray emission  in the context of bremsstrahlung emission. We 
first discuss radio measurements of the Galactic center to estimate the total
nonthermal radio flux, followed by the analysis of $\gamma$-ray data from $Fermi$.

\subsection{The Separation of Thermal and Nonthermal Radio Emission}

The distribution of relativistic electrons is  traced by synchrotron continuum emission at low radio
frequencies. However, the large scale   study of  radio continuum emission from the 
inner two hundred parsecs
of the Galaxy  shows that the diffuse component  is due to  a mixture of thermal
and nonthermal emission (e.g., Law et al. 2008). Thus, it is difficult  to separate these two components
spatially as their emission overlaps at radio wavelengths. 
Even more
challenging is that some of the diffuse and extended sources have a spectral index,
$\alpha$, where the flux density F$_{\nu} \propto \nu^{-\alpha}$, 
that is flatter or harder 
than $\alpha$=0.5 (p=$2\alpha$+1=2 corresponding
to energy spectrum E$^{-p}$). Apart from the large-scale diffuse nonthermal
emission  on
a scale of several  degrees (LaRosa et al. 2005; Crocker et al. 2010), there are several  discrete sources of 
nonthermal emission. One is the population of nonthermal 
filamentary structures found
throughout this region. These synchrotron filaments  can be as long as  $>15'$ (or 36 parsecs at the 
8 kpc distance to the Galactic center), and narrow
($\approx5-10''$ corresponding to 0.2-0.4 pc).  Polarization studies of the filaments
trace an organized magnetic field which runs perpendicular to the Galactic plane (Yusef-Zadeh,
Morris and Chance 1984; Lang, Goss \& Morris  2002; Nord et al. 2004). Nonthermal
emission also arises from supernova remnants in the Galactic center, some of which are
interacting with molecular clouds, such as  Sgr A East (SNR G0.0-0.0, e.g. Tsuboi et
al. 2011).  Another source of nonthermal emission is the population of 
pulsars that could contribute to nonthermal emission from this region (Johnston et al. 2006; 
Deneva, Cordes and  Lazio 2009; Wharton et al. 2011).  
Lastly,  populations of compact stellar  sources 
could produce nonthermal
radiation from colliding winds in  massive binary systems; two such examples have been 
detected in the Arches cluster and Sgr B2 (Yusef-Zadeh et al. 2003;
Yusef-Zadeh, Wardle \& Roy 2007a). 

A quantitative estimate of the relative amount of 
thermal and nonthermal emission from the Galactic center was 
made by Law et al. (2008) 
based on radio continuum data at 5 and 8.5 GHz taken with the GBT
over the region between 357.5\deg\ $ < l <  
$1.2\deg\, and --0.6\deg $< b < $0.4\deg\,. 
These authors assumed that thermal and  nonthermal sources
are separated from each other and identified them  from the 
spectral index $\alpha$ values measured between  5 and 8 GHz. It was concluded that 85\%
and 76\% of continuum radio flux from individual sources is due to nonthermal processes 
at 5 and 8 GHz, respectively. Earlier studies  claimed  that 
$\sim$50\% of the continuum emission at 5 GHz is due to nonthermal emission 
(Schmidt et al. 1980; Mezger and Pauls 1979). The discrepancy  in the  ratio of 
nonthermal 
to thermal emission could be due to the flat  spectrum of some of
the nonthermal sources, thus complicating the identification of thermal and nonthermal 
sources. 
LaRosa et al.  (2005) studied diffuse
radio continuum emission from the inner 6\degree$\times$2\degree of the Galactic center at
75 and 327 MHz. They found a strong  diffuse nonthermal structure with integrated flux
density of 7000 Jy at 330 MHz. The spectral index value between 330 and 74 MHz gave
$\alpha > 0.7$ which is a lower limit due to thermal absorption at 74 MHz.
In another study, 
Crocker et al. (2010) investigated the spectrum of nonthermal 
emission from the inner 3\degree$\times$2\degree and found a spectral  break of 0.6
at 1.7 GHz.

To measure   the distribution of radio flux  from the inner 2\deg$\times1$\deg\ of the
Galactic center region, we  integrated the total continuum flux at 0.325, 1.40, 8.5  and 5
GHz based on GBT observations described by Law et al. (2008) who focused only 
on 8.5 and 5 GHz data. A  region 
away from the Galactic plane
was selected having minimum contamination by Galactic center sources. 
To construct a background
subtracted image, a noise  map was first constructed from  the region that has been
mapped by GBT. The noise map and a DC offset were then subtracted from
the entire image. 
Figure 1a 
shows  a continuum subtracted image at 1.415 GHz  from
the inner $\sim5$\deg$\times$5\deg\ of the Galactic center. 
Prominent radio continuum sources along the Galactic equator 
such as Sgr A near l$\sim0$\deg\,, radio continuum Arc near 
l$\sim0.2$\deg\,,
Sgr B near l$\sim0.7$\deg\,, Sgr C near l$\sim-0.6$\deg\,,  Sgr D near l$\sim1.2$\deg\,
and the bright nonthermal source the Tornado nebula 
near l $\sim -2.5$\deg\, show  peaks in  contours of 1.415 GHz 
emission (Yusef-Zadeh, Hewitt and Cotton 2004). 
The grayscale image in Figure 1b 
shows weak extended structures distributed away from the Galactic ridge.  
Extended  features distributed at  positive-latitudes  
are known as Galactic center radio lobes (e.g., Law et al. 2008). 
There are also large-scale features at negative latitudes  near l $=-0.9$\deg\, extending 
to b $\sim-1$\deg\, associated with two supernova remnants  G359.1-0.5 and G359.0-0.9
(Reich 1982; Reich and Reich 1986). 
A new feature, G359.02+0.27, is  a long vertical 
structure running perpendicular to the plane extending
toward more negative latitudes  near b$\sim-1.6$\deg\, between  l $\sim-0.075$\deg\,  and l$\sim$0.27\deg.   


We measured the  flux from the brightest region of the maps at four different frequencies 
all based   on GBT observations and presented the flux in Table 1. 
The first two columns of this table  show
the frequency and integrated flux from 
the inner 2\deg$\times$0.85\deg\ of the Galactic
center. The DC offset and the RMS noise per beam,  measured from blank regions of individual
images of the survey, are  listed in the last two columns, respectively.
To illustrate the distribution of flux as a function of radius   from the Galactic center, 
we made azimuthally averaged 
radial profiles of radio emission at all four frequencies, as shown in 
Figure 2a-d. 
We used MIRIAD task {\tt ellint} to integrate elliptical annuli with an aspect ratio of two
centered on Sgr A*.  The width of each annulus is one pixel corresponding to 
20$''$, 30$''$, 30$''$ and 600$''$ at 8.5, 4.85, 1.415  and 0.325 GHz, respectively 
The RMS in Jy/beam  was calculated and then scaled by the square root of the number of beams 
in each  annulus. 
These plots  show considerable flux  variations
as a function of frequency, suggesting that  thermal 
and nonthermal features  dominate the total observed flux at high  and low frequencies, respectively.


To estimate the contribution of  thermal and nonthermal emission, 
we use the 
integrated flux  to derive
spectral index values $\alpha^{325\rm MHz}_{1.4 \rm GHz}=0.17\pm0.01$,
$\alpha^{1.4\rm GHz}_{4.5\rm GHz}=0.58\pm0.01$, $\alpha^{4.5\rm GHz}_{8.5\rm GHz}=1.14\pm0.01$. The
spectral index distribution is relatively flat at low frequencies whereas it becomes
steeper at high frequencies. The variation  of the spectral index is consistent with  thermal
emission from HII regions ($F_{\nu}\propto\nu^{-0.1}$) 
which becomes optically thick (F$_{\nu}\propto\nu^{2}$) 
at low frequencies. The flattening of the spectral index between 325 MHz and 1.415 GHz 
could result from the decrease  of thermal flux  due to 
free-free absorption of  thermal gas that becomes opaque at low frequencies. 

The true percentage of  thermal and nonthermal emission 
from the complex region of the
inner  2\deg$\times$0.85\deg\, is very difficult to measure directly. 
In order to separate the intrinsic flux of thermal and nonthermal emission, we assumed 
that the two components are spatially mixed with or separate from each other following
Gregory and Seaquist (1974).  
In this case, the observed flux in this model is given by 
$$ S_\nu = \Omega \left(S_{NT} \exp(-\tau_\nu)  + B_\nu(T)\right)\times (1-\exp(-\tau_\nu)) $$ 
where $S_{NT}$ $\propto 
\nu^{-\alpha}$ is  the nonthermal flux in the absence of free-free absorption, 
$\tau_\nu$ is 
the free-free optical depth at frequency   $\nu$,  $B_{\nu}$(T)   is Planck's function at 
the temperature T, and 
$\Omega$ is the solid angle subtended by the source.
These calculations account for the spectral turnover at low frequencies 
due to opacity  of diffuse thermal emission. Figure 3
shows  the flux from
the inner 2\deg$\times$0.85\deg\ as a function of frequency. 
The  solid black 
curve  represents  the  $\chi^2$-fit to total flux which is itself 
 decomposed into  thermal and nonthermal components, shown 
as long and short dashed lines, respectively. 
We  fix the thermal contribution at 4.85 GHz to be 25\% (Law et al. 2008) and assume  a kinetic 
temperature of  5000K.  We adopt a broken power-law for the  unabsorbed nonthermal
emission and assume that this 
component lies behind the thermal emission.  
With this model we find $\nu^{-0.25}$  below 3.3 
GHz and  $\nu^{-1.7}$  above 3.3 GHz, with an unabsorbed nonthermal flux of 
2450  Jy at 325 MHz. 
Because the optical depth is only  significant below 200 MHz, there is no
difference between assuming that the thermal and nonthermal components are mixed in versus
having the thermal emission be a foreground screen. 
Using a 25\% contribution from 
thermal emission at 8.5 GHz (Law et al.  2008) corresponds to an
emission measure $E\sim$10$^{4}$ cm$^{-6}$ pc. This gives average electron density
$n_e\sim$6 cm$^{-3}$ assuming that it  is uniformly distributed over $L\sim$288 pc.

The study of the cosmic rays in the Galactic disk has recently suggested a need for a low-energy 
break in the spectrum of cosmic ray electrons (Strong, Orlando and Jaffe 2011), This is not 
dissimilar to the spectral break that we infer from the cosmic ray electrons in the Galactic center.  
An unusual aspect of the spectrum of radio emission from the Galactic center is the large change in 
particle distribution index $\sim$1.7.  The energy spectrum of electrons corresponding to a broken 
power-law is hard corresponding to $p$=1.5 at low energies whereas the spectrum is steep at high 
energies with $p$=3.2. A non-standard possibility that can account for such a large change in the 
spectral index value is the contribution of electrons and positrons produced as the byproduct of dark 
matter annihilation. In this picture, the electrons and positrons created through the annihilation of 
a relatively light ($\sim$5-10 GeV) dark matter particle can provide a new population of electrons at 
energies less than the annihilation energy of WIMPS. Although this picture is by no means unique in 
explaining the large change in the spectral index, spectral distribution of electrons and positrons 
which emit synchrotron radiation are consistent with the observed spectral shape of electrons for 
individual nonthermal radio filaments at high frequencies (Linden, Hooper and Yusef-Zadeh 
2011).


\subsection {Gamma-ray Emission from the Galactic Center}

Extended $\gamma$-ray emission within the inner 2\degree\ has been detected at TeV energies by the 
H.E.S.S. atmospheric Cherenkov telescope. The Galactic center ridge of $\gamma$-ray emission (hereafter, 
Galactic ridge) appears well correlated with the molecular gas distribution in the inner region 
(Aharonian et al. 2006). 

The Galactic center is also a prominent sources of GeV $\gamma$-rays. The {\it Compton Gamma Ray Observatory} identified a 
source coincident with the Galactic center, 2EG J1746--2852, at energies of 0.2-10 GeV (Thompson et al. 1995). With greatly 
improved sensitivity and spatial resolution, the $Fermi$ LAT resolves multiple GeV sources in the inner 2\degree\ of the 
Galaxy. A bright point source coincident with the position of Sgr A is reported in the LAT First and Second Source Catalogs 
(Abdo et al. 2010a; Nolan et al. 2012, hereafter 1FGL and 2FGL). Emission from the central source, 2FGL J1745.6--2858, shows a 
peak at a few GeV.

Several explanations have been proposed for the GeV emission in the Galactic center detected by $Fermi$. Chernyakova et al. 
(2011) propose that this central gamma-ray source is produced by the diffusion of cosmic ray protons into the surrounding 
dense molecular gas in the inner 10 pc. A separate analysis by Hooper \& Goodenough (2011) claims the existence of a diffuse 
$\gamma$-ray source in   the inner 
degree of the Galactic center, on top the the Galactic diffuse background. This emission peaks at 2-4 GeV, 
which they interpret as the possible annihilation of dark matter. Alternatively this feature could arise from a population of 
millisecond pulsars in the region (Abazajian 2011) or may be due to improperly accounting for known point sources in the 
region (Boyarsky, Malyshev \& Ruchaysky  2011).

Interestingly, the hard spectrum of the filaments of the Arc which emit radio synchrotron radiation could be a strong source 
of cosmic rays responsible for the excess $\gamma$-ray emission within the inner 30\arcmin\ of the Galactic center (Linden, 
Hooper \& Yusef-Zadeh 2011). The Galactic center nonthermal filaments of the radio Arc are unique in the Galaxy and have a 
harder spectrum compared to typical nonthermal radio sources. We will argue in \S4 that the interaction of relativistic 
electrons with molecular gas in the Galactic center produces significant bremsstrahlung radiation, and that the distribution 
of diffuse $\gamma$-ray emission correlates well with the distribution of both nonthermal radio continuum emission and the 6.4 
keV K$\alpha$ line emission. Here, we analyze the $\gamma$-ray emission from the Galactic center using three years of $Fermi$ 
LAT data, in order to characterize emission from both the central $gamma$-ray source, as well as nonthermal emission from the 
Galactic ridge.

\subsubsection {$Fermi$ LAT Observations}

$Fermi$ LAT detects $\gamma$-rays between $\sim$20 MeV to $>$300 GeV in an all-sky scanning mode, observing the entire sky 
every $\sim$3 hours (Atwood et al. 2009). Events are detected by the LAT tracker in both the "front" and "back" sections, 
which are combined in this analysis. Events were selected within a radius of interest of 30 degrees from the Galactic center, 
and for times between 2008 August 4 and 2011 August 4, and at energies between 1 GeV and 100 GeV. The angular resolution 
(68\%\ containment angle for events at incident angle) is $\sim$0\ddeg9 at 1 GeV, increasing to 0\ddeg2 at the highest 
energies. The point spread function is detailed on the Fermi Science Support Center (FSSC) 
webpage\footnote{http://fermi.gsfc.nasa.gov/ssc/data/analysis/documentation/Cicerone/Cicerone\_LAT\_IRFs/IRF\_PSF.html}. The 
energy resolution of the LAT is 8--10\%\ between 100 MeV and 100 GeV. The systematic uncertainties in the IRF are energy 
dependent: 8\% at 100 MeV, 5\% at 560 MeV, 10\% at 10 
GeV\footnote{http://fermi.gsfc.nasa.gov/ssc/data/analysis/LAT\_caveats.html}.

Data is analyzed using the $Fermi$ Science Tools (v9r15p2) with the "P7SOURCE\_V6" instrument response functions. Only source 
class events with Earth zenith angles less than 100$\degree$ have been used to reduce contamination from the Earth limb. We 
use the standard maximum likelihood fitting, with photons binned in 0.05 degree pixels within a 10\deg$\times$10\deg\ region 
centered on Sgr A. Data is also binned spectrally with 4 log-normal bins per decade in energy between 1 and 100 GeV. We 
restrict our analysis only to these high energies at which the PSF of the LAT is sufficient to spatially resolve sources 
within the central 2$\times$1\degree\ region.

For source modeling we include all sources in the 2FGL source catalog (Nolan et al. 2012). Additionally, we include the 
standard isotropic model which accounts for the extragalactic diffuse background and residual instrumental background 
('iso\_p7v6source.txt') and Galactic diffuse model which accounts for interactions between cosmic rays and the Galactic 
interstellar medium and photon field ('gal\_2yearp7v6\_v0.fits'). The Galactic diffuse model\footnote{Details are available at 
the FSSC: http://fermi.gsfc.nasa.gov/ssc/data/access/lat/Model\_details/Pass7\_galactic.html} is derived from a fit to 2 years 
of LAT data using Galacto-centric rings derived from tracers of the interstellar gas distribution (HI and CO) and a model of 
inverse Compton emission calculated using GALPROP (Strong, Moskalenko \& Reimer 2004).


To estimate the systematic uncertainty from the diffuse model, we change the best-fit normalization 
of the diffuse components by $\pm$6\%, following the method used for analysis of unresolved or 
small-scale sources, such as Galactic supernova remnants (Abdo et al. 2010b). This value was 
determined by using different versions of the Galactic diffuse emission generated by GALPROP (Strong 
et al. 2004) to compare the gamma-ray intensity of nearby source-free regions of the Galactic plane 
with that expected from the models (Abdo et al. 2010b). We note that use of the Galactic diffuse 
model is only appropriate for analysis of small diameter sources, and has been employed for other 
extended GeV sources as large as a few degrees in extension.

An additional source of uncertainty arises as we must assume a morphology for GeV emission from the inner $\sim$2\degree\ of 
the Galaxy in our likelihood fit. Figure 4a shows a smoothed counts map at $\ge$1 GeV, after subtraction of the isotropic and 
Galactic diffuse templates. Prominent emission is seen from the vicinity of Sgr A, with fainter emission extending along the 
Galactic plane. Additionally, there are point sources which lie off the plane of the Galaxy, and a faint complex of emission 
coincident with TeV source H.E.S.S. J1745-303. For comparison with the distribution of nonthermal emission from the same 
region, Figure 4b shows the positions of 2FGL sources superimposed on 20cm radio continuum emission observed by the GBT. 
Ellipses indicate the 68\%\ error in the source localization. These sources are listed in Table 2, and are described in the 
following section. The 2FGL catalog decomposes emission above the Galactic diffuse model as individual point sources, however 
we also test the hypothesis that the emission arises from an extended component along the Galactic ridge.

To obtain the best possible model of the emission, we relocalize the positions of all 2FGL point sources in the inner 2 
degrees and refit their spectra using only $>$1 GeV data. We search for un-modeled point sources by creating test statistic 
maps of the residual emission. The test statistic is a measure of the significance of adding a source to a model, defined as 
TS = 2 log($\mathcal{L}_1$/$\mathcal{L}_0$) where $\mathcal{L}$ is the Poisson likelihood, and the subscripts 0 and 1 refer to 
the original model and a model with an additional source, respectively. Within 2\deg\ of the Galactic center, we find two 
significant sources with TS $>$ 25 that are not listed in 2FGL. A source is found near the location of Sgr C, 
$\alpha$,$\delta$(J2000) = 266.044,-29.323, previously identified as 1FGL J1744.0-2931c. Another source is found 0.5 degrees 
away from the Galactic ridge at positive latitudes, $\alpha$,$\delta$(J2000) = 265.005,-28.533, and has a very soft spectrum, 
$\Gamma$=2.9$\pm$0.2, above 1 GeV. Adding these two additional sources improved the model of GeV emission, as shown in Table 
2. We refer to this model as the "2FGL refit" model, hereafter.
 
Sources are initially assumed to have a power-law spectrum. For highly significant sources we attempted to replace a simple 
power-law spectral model with a broken power-law model of the form:

\begin{equation} 
\frac{dN}{dE} = N_0 \times \left \{ \begin{array}{l l} (E/E_b)^{\Gamma_1} & if E < E_b \\ (E/E_b)^{\Gamma_2} & if E > E_b\end{array} \right \}
\end{equation}

For Sgr A* we find a broken power-law improves the fit. The best fit spectral parameters are $\Gamma_1$=1.9, $\Gamma_2$=3.0, and E$_b$=3 GeV.

\subsubsection{Point Sources in the Inner Galaxy}

We briefly summarize point sources detected in the inner 2\degree . Source 2FGL J1745.6-2858 corresponds to the position of 
Sgr A*. This source has been studied in detail by Chernyakova et al. (2011) and Linden, Lovegrove \& Profumo (2012). The 
$Fermi$ LAT spectrum connects to that of the detected H.E.S.S. TeV source at $\sim$100 GeV, with a softening of the spectrum 
between 1-100 GeV, and a hardening of the spectrum at $>$100 GeV to multi-TeV energies. In addition to emission from Sgr A*, 
there are two point sources corresponding to the locations of the Arc and Sgr B (sources 2FGL J1746.6-2851c and J1747.3-2825c, 
respectively). Faint emission was reported in the vicinity of Sgr C in 1FGL, but the source was not present in 2FGL. We note 
these three point sources closest to the Galactic center (corresponding with the Arc, Sgr B and Sgr C) appear coincident with 
the diffuse emission detected at TeV energies by H.E.S.S. We explore this further in the following section.

Other sources are also present within the inner 2\deg\ that are not thought to be associated with the Galactic center region. 
The Mouse pulsar corresponds to source 2FGL J1747.1-3000, with detected $\gamma$-ray pulsations. However, this pulsar is known 
to lie at a distance of only 5 kpc (Camilo, et al. 2002). Two sources (2FGL J1743.9-3039c and J1745.5-3028c) appear to be 
counterparts to the extended TeV source H.E.S.S. J1745-303. These sources may be related to the SNR G359.1-0.5, known to be 
interacting with molecular clouds, or may be candidate pulsar wind nebulae (Aharonian et al. 2008). In either case these 
sources lie outside the Galactic ridge. There are also three sources detected more than half a degree off the Galactic plane, 
with no readily apparent counterparts (2FGL J1738.9-2908, J1748.6-2913, J1754.1-2930). Similarly, the newly detected source 
``bkgA" also lies above the Galactic plane, and has no apparent multiwavelength counterpart. The off-plane GeV sources show no 
correlation with the diffuse background model, or large scale structures seen towards the Galactic center, and are therefore 
unlikely to have any relation to the Galactic ridge.

\subsubsection{Emission from the Galactic Ridge}

To probe whether GeV emission is present on extended spatial scales we replace the three point sources 
associated with the Arc, Sgr B and Sgr C, with an extended spatial template. We then maximize the likelihood
using the extended spatial template plus a point source which accounts for emission from Sgr A. 
We separately apply four template models: 20cm radio continuum, X-ray FeI K$\alpha$ line emission, 
H.E.S.S. diffuse TeV emission, and CS 1--0 integrated line intensity representing 
the distribution of dense gas in the region. All templates span roughly the inner 2\deg$\times$1\deg .
In the 20cm and H.E.S.S. templates, the central emission from Sgr A* is removed, as it is 
clearly detected as a point source with a unique spectrum. 
Fitting the spatial templates from other wavelengths gives a means of comparing them with the morphology of GeV
emission from the Galactic ridge above the modeled Galactic diffuse.

First, we add an extended spatial template using the CS 1--0 map (Tsuboi, Handa \& Ukita  1999). As CS is an 
optically thin tracer of dense gas, this model probes the distribution of dense clouds in the 
Galactic center region. However, we find replacing the three point sources with the CS template 
results in a significantly lower global likelihood than the point source model. This is likely due to 
the fact that CS emission is detected at Galactic longitudes$>$1\degree\ while TeV and GeV emission do not appear to 
extend this far from the Galactic center.

We do find an improved fit when the point sources in the Galactic ridge are replaced by other spatial 
templates representing the distribution of X-ray FeI K$\alpha$ line, H.E.S.S. TeV and diffuse 20cm 
radio emission. The X-ray data that we used are based on Chandra observations (Yusef-Zadeh et al. 2007b). 
Table 3 presents the 6 models which we fit to the LAT data. The 2FGL point source 
models have 12 degrees of freedom (Sgr A with a broken power-law, three point sources with power-law 
spectra, and the isotropic and Galactic diffuse normalizations). The extended models require 4 fewer 
degrees of freedom since the three point sources are replaced by one source. We note that the TeV and 
radio templates provide a better fit to the data than the X-ray line template, though this may be due 
to non-uniform sensitivity of the X-ray observations.

We conclude that the GeV emission from the Galactic ridge is well-correlated with the extended morphologies observed at radio, 
TeV and X-ray wavelengths. We note that the Galactic ridge emission has sufficient statistics to fit with a broken power-law 
spectral model. Using the TeV template, we find the best fit spectral parameters are $\Gamma_1$=1.8, $\Gamma_2$=3.0, and 
E$_b$=2.5 GeV. However, we caution that despite the improvement in the likelihood, the spatial template and point source 
models are not nested, so a significance of the improvement of the extended templates over the point source model cannot be 
stated. We also note that while simply increasing or decreasing the normalization of the diffuse Galactic model cannot fit the 
emission observed by the LAT in the Galactic ridge, we have not performed an in-depth study of the diffuse emission for the 
Galaxy. However, that the GeV emission is consistent with the morphology of extended nonthermal emission observed at other 
wavelengths is suggestive of a common origin. We discuss in detail a plausible model for nonthermal emission from the Galactic 
ridge in \S4.


\begin{deluxetable}{lrrr}
\tablewidth{0pt}
\tablecaption{Integrated Radio Flux from the Inner 2\degree$\times$0.85\degree}
\tablehead{
\colhead{Frequency (Hz)}&
\colhead{Flux$\pm\sigma$ (Jy)} &
\colhead{DC Offset (Jy)} &
\colhead{RMS Noise (Jy/beam)}
}
\startdata
3.25$\times10^8$  &  3.57$\times10^3\pm10.1$ & 127 & 5.8\\
1.41$\times10^9$ & 2.77$\times10^3\pm1.1$ & 8.1 & 1.3$\times10^{-1}$ \\
4.85$\times10^9$ & 1.35$\times10^3\pm0.6$ & 1.44$\times10^{-2}$ & 2$\times10^{-2}$  \\
8.5$\times10^9$ & 7.11$\times10^2\pm0.8$ & 1.2$\times10^{-2}$ & 1.5$\times10^{-2}$   
\enddata
\label{tab:fitparams}
\end{deluxetable}

\begin{deluxetable}{lllrrrl}
\tablecaption{Detected $\gamma$-ray Sources in the Inner 2\degree\ of the Galaxy  
\label{tbl:2fgl}}
\tabletypesize{\small}
\tablewidth{0pt}
\tablehead{
   \colhead{Name} & \colhead{RA} & \colhead{Dec} & \colhead{Flux (1--100 GeV)} &\colhead{TS} 
&\colhead{Association} \\
   &h\, m\,  s  & $^0\,\,\, '\,\,\, ''$ & \colhead{(10$^{-9}$ ph cm$^{-2}$ s$^{-1}$)} & }
\startdata
2FGL J1745.6-2858&	17 45 41.6&	-28 58 43&	77.3(2.0)&	1857&	Sgr A \\
2FGL J1746.6-2851c&	17 46 40.6&	-28 51 31&	6.6(1.4)&		35	&	the Arc \\
2FGL J1747.3-2825c&	17 47 23.9&	-28 25 53&	14.2(1.4)&	112 & Sgr B	\\
1FGL J1744.0-2931c&     17 44 01.0&     -29 31 57&      10.0(1.4)&          79& Sgr C    \\
2FGL J1747.1-3000&	17 47 09.2&	-30 00 50&	25.0(1.1)&	729&	PSR J1747-2958 \\
2FGL J1745.5-3028c&	17 45 32.4&	-30 28 56&	4.3(0.9)&		26	&	H.E.S.S. J1745-303\\
2FGL J1743.9-3039c&	17 43 57.3&	-30 39 13&	3.7(0.8)&		25	&	H.E.S.S. J1745-303\\
2FGL J1748.6-2913&	17 48 39.2&	-29 13 53&	12.0(1.0) &	169&	\\
2FGL J1738.9-2908&	17 38 56.7&	-29 08 25&	6.8(0.8)&		234&	\\
2FGL J1754.1-2930&	17 54 08.9&	-29 30 33&	3.6(0.5)&		83	\\	
bkgA				 &	17 40 01.2&	-28 31 59&	3.6(0.7)&		45 &	
\enddata
\end{deluxetable}

\begin{deluxetable}{lrrrr}
\tablecaption{Comparison of Spatial Template Fits to Fermi LAT Data $\ge$1 GeV \label{tbl:spatialmaps}
}
\tablewidth{0pt}
\tablehead{
 \colhead{Model} & 
 \colhead{2 log($\mathcal{L}_1/\mathcal{L}_0$)}  &
 \colhead{$d.o.f.$} &
}
\startdata
2FGL 		 		& 0		& 12  	\\	
2FGL refit    	 		& 51		& 12		\\	
X-ray Fe K$\alpha$ 		& 68		& 8	\\	
H.E.S.S. residual 		& 101	& 8	\\	
20cm Radio big		& 113	& 8	\\	
CS  gas      			&--103	& 8	\\	
\enddata
\end{deluxetable}

\subsection {X-Ray Emission from the Galactic Center}

\subsubsection{Chandra Data}

The results of large-scale Chandra observations of the Galactic center 
focusing on the distribution of 
FeI K$\alpha$ line  emission were  described in Yusef-Zadeh et al. (2007b). 
Since the publication of these results in 2007, 
additional Chandra observations of this region 
have been  carried out. Here, we use 15 additional pointings with exposure time of 40 ks each. 
These new observations  are combined with additional  archived data 
sets described in detail by Muno et al. (2009) who presented 
a catalog of X-ray sources in the inner 2\deg$\times0.8$\deg\ of the Galactic center.  
We  reproduce below the description of data reductions that were given for earlier 
analysis of 6.4 keV line emission  (Yusef-Zadeh et al.  2007b). 

Images of the equivalent widths  of the low-ionization 6.4 keV line of FeI K$\alpha$ were constructed using the techniques 
described 
by Park et al. (2002, 2004). Adaptively-smoothed images of the diffuse line emission were generated in the same manner as 
the 
continuum image, using the 6.25--6.50 keV band for FeI K$\alpha$. The continuum under each line was computed based on 
adaptively-smoothed images of the flux in the 5.0--6.1 keV and 7.15--7.30 keV energy bands. We assumed that the flux in 
each continuum band ($F_{\rm band}$) could be described as a power-law, so that the normalization ($N$) and slope 
($\Gamma$) of the power-law could be computed from \begin{equation} F_{\rm band} = {{NE_{\rm low}^{-\Gamma+1} - NE_{\rm 
high}^{-\Gamma+1}}\over{\Gamma - 1}}. \end{equation} 
Using the fluxes in both continuum bands, the above equation was 
solved for $N$ and $\Gamma$ using Newton's algorithm and the parameters were used to estimate the continuum contribution 
to the line emission images. To derive the equivalent width  (EW) images we subtracted the estimated total continuum flux from 
the line image, and then divided the line image by the continuum flux density at the centroid of the line (6.4 keV). We 
caution that we have neglected the cosmic-ray background in generating these maps, which could account for as much as $\sim$40\% 
of the events in the 6--7 keV band and consequently biases any estimate of the EW. The assumption of a 
power-law spectrum, instead of multiple  plasma temperatures especially corresponding to the He-like 
K$\alpha$ line at 6.7 keV 
also introduces a small systematic bias in these maps. 
From Figure 5 in 
Muno et al. (2004),  the nonthermal X-ray  
flux is  estimated to be about one-third of the continuum flux 
at the FeI K$\alpha$ line band. 
We have not attempted to correct these effects 
because they are only used to search for regions of enhanced iron emission.
In order to confirm the properties suggested by Chandra images of the diffuse line and continuum emission, 
we compared the Chandra EW  map  with that constructed from 
Suzaku measurements.

\subsubsection{Suzaku Data}

To check the accuracy of the EW map measured from Chandra observations, we derived the distribution of Fe I K$\alpha$ line
emission and EW map using the Suzaku data. The details of the Suzaku observations are shown in table 1 of Uchiyama et al.
(2011), which partially covers the region between $-3^{\circ}<l<2^{\circ}$ and $-1^{\circ}<b<1^{\circ}$.

We made X-ray images in the energy bands of 5--6 and 7--8 keV for continuum emissions and 6.3--6.5 keV for Fe I K$\alpha$. We sorted 
non X-ray background (NXB) data by the cut-off rigidity with {\tt xisnxbegen} (Tawa et al. 2008) and made NXB images in the 
foregoing energy ranges for the respective observations. The NXB images are subtracted from the X-ray images. Thus, in the 
case of the Suzaku image, the effects of the NXB are removed with an uncertainty of less than $\sim$4\% (Tawa et al. 2008). 
After the NXB subtraction, the vignetting effects of the X-ray images are corrected with {\tt xissim} (Ishisaki et al. 2007). 
Both of {\tt xisnxbegen} and {\tt xissim} are include in the HEASoft package\footnote{http://heasarc.nasa.gov/lheasoft/}. 
 We 
calculated the continuum flux in the 6.3--6.5 keV band from the 5--6 and 7--8 keV band images, following 
a similar technique that was applied to  Chandra data.  
We subtracted the calculated continuum image from the 6.3--6.5 keV band image and obtained the Fe I K$\alpha$ line 
emission map. The Fe I K$\alpha$ line emission map 
was  divided by the continuum image before the EW map was 
constructed. 
Bright point sources, 2E 1743.1-2842, 2E 1742.9-2929 and 2E 1740.7-2943 are masked by circles with the radius of 
3\ddeg5.
We ignored the cosmic X-ray background (CXB) when we made the EW map.  It is because that the interstellar absorption of the
CXB is difficult to estimate.  Assuming the CXB flux of Kushino et al. (2002), the systematic errors of
the EW in Figure 9
are estimated to be less  than 20\%.



\section{Evidence for Cosmic Ray Interactions}

To examine the 
bremsstrahlung  
 model of cosmic ray electrons  interacting  with  molecular gas,  we 
studied several consequences of such an interaction to provide 
a  self-consistent  check on the applicability of this model. 
In the following sections
we show  the feasibility of 
cosmic ray electrons interacting with molecular gas to explain the 
unique  characteristics of molecular gas in the Galactic center region as well 
as production of  $\gamma$-ray and X-ray emission. 


\subsection{Cosmic Ray Ionization Rate and the Magnetic Field Strength}

Cosmic rays play an important role in star formation processes as they are a primary source of ionization of dense molecular 
clouds: driving ion-neutral chemistry, heating the gas and determining the coupling to magnetic fields.  Recent measurements 
indicate a vast amount of diffuse H$_3^+$ and H$_3$O$^+$ distributed in the Galactic center. This suggests a cosmic ray 
ionization rate $\zeta\sim 10^{-15}$\, s$^{-1}$\, H$^{-1}$, one to two orders of magnitude higher in the Galactic center 
region than in the Galactic disk (Oka et al. 2005; van der Tak et al. 2006).

Here we consider whether this high ionization rate may be produced by the low-energy tail of the population of relativistic cosmic 
ray electrons that are responsible for the observed synchrotron emission in the Galactic center.  We use the observed synchrotron 
flux to estimate their contribution to the cosmic ray ionization rate, independent of H$_3^+$ measurements.

Unfortunately the synchrotron intensity also depends not just on the cosmic-ray electron population but also on the magnetic 
field strength, which is quite uncertain.  Early estimates inferred a large-scale milligauss magnetic field permeating 
throughout the Galactic center based on the apparent resistance of nonthermal filaments to distortion by molecular clouds 
(Yusef-Zadeh, Morris and Chance 1984; Morris and Serabyn 1996; Morris 2007).  More recent estimates are somewhat lower: 
6$\mu$G was inferred from radio emission distributed over the inner 6\degree$\times$2\degree\ (LaRosa et al. 2005), whereas 
Crocker et al. (2010) inferred a minimum value B$\sim50\mu$G based on the nonthermal radio emission from the inner 
3\degree$\times$2\degree of the Galaxy.  In the face of these uncertainties we adopt a fiducial value of 100\,$\mu$G and 
consider the scaling with magnetic field in our results.

The cosmic ray ionization rate per hydrogen nucleus, $\zeta$, is directly proportional to the number density of cosmic-ray 
electrons (and also depends on their energy spectrum).  The value of $\zeta$ is dominated by the number of low-energy 
electrons, and so is sensitive to the choice of lower energy electron cutoff.  We adopt an $E^{-p}$ electron spectrum running 
between $E_\mathrm{min}=0.1$\,MeV and $E_\mathrm{max} = 10$\,GeV, a magnetic field $B$, and a line-of-sight depth $L$ of the 
source region enables us to calculate both the synchrotron intensity $I_\nu$ at frequency $\nu$ and the ionization rate 
$\zeta$ in the source region.  This yields the relationship 
\begin{equation} 
\zeta \approx \frac{3.1\times10^{-14}}{p-1}\, 
\frac{I_\nu}{\u Jy \ut arcmin -2 }\, \left(\frac{\nu}{\u GHz }\right)^{\!\alpha} \, \left( \frac{L}{30\u pc 
}\right)^{-1}\, \left(\frac{B}{100\,\mu\mathrm{G} }\right)^{\!\!-(1+\alpha)}\, \ut s -1 \ut H -1 
\end{equation} 
where $\alpha = 
(p-1)/2$ is the synchrotron spectral index and $I_{\nu}\propto\nu^{-\alpha}$ (so the product $I_\nu \nu^\alpha$ in this 
expression is constant).  We estimate a total nonthermal flux at $\nu=325$\,MHz of $2.45\times10^3$ Jy arising from the inner 
2\degree$\times$0.85\degree ($L\sim288$ pc) of the Galaxy.  This emission is patchy, and so is assumed to fill 10\% of the 
volume of an oblate spheroid with principal axes $2^{\circ}\times2^{\circ}\times0.85^{\circ}$.

Figure 5a and b illustrate the dependence of the inferred ionization rate on the spectral index, magnetic field strength, and 
the lower energy cutoff of the electron spectrum.  Figure 5a shows the ionization rate as a function of magnetic field 
strength for $\alpha$ running between 0.2 and 0.6.  The ionization rate decreases with increasing field strength because fewer 
electrons are required to yield the same synchrotron emissivity.  The black dot on each curve gives the equipartition magnetic 
field; this should be a lower limit on $B$, and so indicates the maximum ionization rate expected for each value of $\alpha$.  
The ionization rate increases for steeper spectra and a fixed emissivity at the observed frequency, as there is a successively 
larger population of lower-energy electrons radiating at lower frequencies.  Figure 5b illustrates the dependence of the 
ionization rate on these lower-energy electrons by varying the electron spectral index $p$ and the lower energy cutoff 
$E_\mathrm{min}$ assuming an equipartition field.  The dependence is weak for flat spectra ($p\la1$) but becomes marked for 
steep spectra.

For the observed range of spectral indices, $\alpha\sim 0.2-0.3$, and the equipartition magnetic field is $\sim 20\mu$G.  The implied ionization rate is $\sim10^{-15}\,- 10^{-14} \, \rm s^{-1}\,  \rm  H^{-1}$.  This is consistent with H$_3^+$ measurements toward several clouds in the Galactic center (Oka et al. 2005; Goto et 
al. 2008) which imply values of $\zeta\sim2-7\times10^{-15}\,  \rm s^{-1}\,  \rm  H^{-1}$ and  a large reservoir of warm and diffuse ionized molecular gas (H$_3^+$) in the central region of the Galaxy.
A strong mG field permeating the central regions of the Galaxy would imply $\zeta \sim 10^{-18}\,  \rm s^{-1}$ H$^{-1}$, much lower than that inferred from H$_3^+$ measurements.  Our estimate of the magnetic field is also consistent with that   of LaRosa et al.  (2005) who inferred a field strength of 
6$\mu$G over the inner  6\degree$\times$2\degree of the Galaxy.

Our estimated equipartition magnetic field strengths neglect the possible contribution of comic-ray protons to the particle energy density.  The presence of a significant proton component increases the equipartition field strength beyond the values indicated by black dots in Fig.\ 5a, by a factor of 2-4 for proton energy density is 10-100 times that for the electrons, decreasing the cosmic-ray electron density needed to explain the observed synchrotron emission and hence decreasing the ionization rate by similar factors.   The field strength would then be closer to the estimate made by Crocker et al. (2010).

\subsection {Molecular Gas Heating Rate}

Another consequence of the interaction of cosmic ray electrons with molecular clouds  is heating of molecular gas. 
This heating 
process is particularly relevant to the  population of molecular clouds showing two different 
gas temperatures 
(G\"usten, Walmsley and Pauls 1981; Mauersberger \& Henkel 1993; see the review by  Ferriere 2007 and references therein). 
A high spatial resolution study observed 36 
clouds distributed 
between $l$=$-1$\deg\,  and 3$^{\circ}$ using six transitions of NH$_3$ 
(H\"uttemeister et al.  1993).  
These observations found a 
two-temperature distribution of molecular clouds, with the warm ($T_{kin}\sim$200K) gas at low H$_2$ density of n$\sim10^3$ 
cm$^{-3}$ and cool ($T_{kin}\sim$25K) gas to dense cores with n$\sim$10$^5$ cm$^{-3}$.  In another study, ISO observations of 
rotational transitions of H$_2$ found predominantly warm molecular gas with $T \sim 150$K toward 16 Galactic center molecular 
clouds with H$_2$ column densities of $\sim$1$-$2$\times10^{22}$ cm$^{-2}$ 
(Rodriguez-Fern\'andez et al.  2001).  
The high  gas temperature of the Galactic center molecular clouds is elevated from 30-50K up to 
$T\sim200$K (Lis et 
al.  2001) and is discrepant 
with respect to the dust temperature 18-22\,K measured from observations of  the inner 2$^{\circ}\times1^{\circ}$ of the 
Galaxy (Pierce-Price et al. 2000; Molinari  et al. 2011).  
The cause of high temperature molecular gas may be due to shocks generated as a result 
of cloud-cloud collisions (Rodriguez-Fern\'andez et al.  2001). 
However, this mechanism is expected to produce warm gas only at the surface of 
the clouds where  clouds collide with each other. 
Comic rays, on the other hand, have the advantage that 
they can increase the temperature throughout the cloud, 
even the  heavily shielded  dense regions.  

Although, 
on average one ion-electron pair is produced  for every 40.1\, eV lost by a cosmic ray electron
 (Dalgarno, Yan \& Weihong 1999),  
the heating associated with cosmic ray electron  ionization
occurs because 11\%\ of the 40.1 eV cosmic ray 
energy loss associated with 
each ionization of a hydrogen molecule 
is consumed as heat 
(e.g.\ Dalgarno, Yan \& Weihong  1999).
 Another 8\,eV    
appears  as heat when H$_3^+$ recombines (e.g.\ Maloney, Hollenbach \& Tielens
1996). Thus,  each ionization of a
hydrogen molecule by a cosmic-ray electron
 is associated with the deposition of 12.4\,eV of heat
into the gas. 
Thus, the heating rate
per hydrogen nucleus $\Gamma_{\rm H}/n_{\rm H}$ is $\approx$ 25\,eV$\times \zeta_\mathrm{H}$, or
about 60\% of the energy loss for a single ionization of a hydrogen molecule,  
 
\begin{equation}
   \frac{\Gamma_\mathrm{H}}{n_\mathrm{H}}  = 2.0\ee -26 \left(\frac{\zeta_\mathrm{H}}{10^{-15}\ut s -1
    \ut H -1 }\right) \u erg \ut s -1 \ut H -1 .
    \label{eq:Gamma}
\end{equation}

Previous estimates  (Yusef-Zadeh, Wardle and Roy 2007a) have shown that 
for $n(\mathrm{H}_2) = 5000\ut cm -3 $ the
equilibrium temperatures are approximately 60, 130 and 280\,K for 
$\zeta_H = 10^{-15}$, $10^{-14}$ and $10^{-13}\ut s -1 \ut H -1 $, 
respectively ($\zeta_H$ is the same $\zeta$ as defined in previous section). 
However, what is emerging from studies of molecular clouds in the Galactic center is 
that the diffuse component with low density and warm gas is 
significant, implying that the volume filling factor 
of dense gas is at a 1\% level (Oka et al. 2005; Dahmen et al. 1997; Sawada et al. 2001; Magnani et al. 2006). 
In this picture,  a dense cool gas with $T< 50$K is surrounded by 
a  warm diffuse  gas with $T> 70$K. H$_3^+$ measurements indicate that the 
warm diffuse gas must have a density of $<100$ cm$^{-3}$  (Goto et al. 2011). 


To estimate the temperature of diffuse molecular gas subject to ionization by cosmic-ray electrons,  we
compute the cooling rate based on the calculations for $T\geq 100$\,K by Neufeld \& Kaufman (1993) and
$T\leq 100$\,K by Neufeld, Lepp \& Melnick (1995).  At the densities $\sim$100 cm$^{-3}$ of the molecular
gas under consideration here, cooling is dominated by rotational transitions of CO below 100\,K (e.g.
Neufeld, Lepp \& Melnick 1995; Goldsmith and Langer 1978) and by rotational transitions of H$_2$ at
higher temperatures (e.g. Neufeld \& Kaufman 1993).  The CO cooling is in the low-density, optically
thin limit, with each collisional excitation radiated away by the subsequent radiative transitions down
the rotational ladder back to the ground state; the cooling rate per CO molecule is therefore
proportional to the H$_2$ density. Collisional de-excitation of H$_2$, on the other hand, is important
even at these low densities  because of the much smaller Einstein $A (s^{-1}$) coefficient 
values for rotational transitions. Consequently, 
this results in  cooling by H$_2$ which is partly saturated and does not increase as strongly with density.
We used spline interpolation of the CO (optically thin, low density limit)  and H$_2$ rotational cooling
rates tabulated by Neufeld \& Kaufman (1993) and Neufeld, Lepp \& Melnick (1995) to construct the
cooling function plotted in Figure 5c.  We added 0.08 dex to the CO cooling function  tabulated by
Neufeld \& Kaufman (1993) above 100\,K to eliminate a discontinuity at the 100\,K boundary with the
function listed by Neufeld, Lepp \& Melnick (1995).   

The total cooling rate per H  is plotted in Figure 5c  for representative H$_2$ densities 
of 50,
100, and 200\,cm$^{-3}$.  As noted above, CO rotational cooling dominates below 100\,K and scales
linearly with H$_2$ density, whereas H$_2$ rotational cooling dominates above 100\,K and does not scale
as strongly with increasing density.   Note that the CO cooling rate is also directly proportional to
the assumed CO abundance, which here is assumed to be that appropriate for gas of twice solar
metallicity, i.e. $n(\mathrm{CO})/n(\mathrm{H}_2) = 2.8\times 10^{-4}$.
The cooling rate can then be mapped to the cosmic ray ionization rate that would supply heat at the
same rate, and this is indicated by the right-hand axis.  We see that ionization rates of $10^{-15}$
and  $10^{-14}$\,s$^{-1}$ would yield gas temperatures of $\sim 100$ and $\sim 200$\,K respectively.

The mass of the molecular nuclear disk is estimated to be $2 - 6
\times10^7 \msol$ (Oka et al.  1998; Pierce-Price et al.  2000). 
The diffuse molecular gas is estimated to have 
a density of $\sim$100 cm$^{-3}$, as H$_3^+$ absorption studies  indicate. 
Assuming  the  gas temperature $\sim 150$\,K and  
the cosmic ray ionization  rate $\zeta\sim5\times10^{-15}$ s$^{-1}$, Figure 5c gives a
cooling rate 
$10^{-25}$ ergs s$^{-1}$ H$^{-1}_2$. The total energy that needs to 
be resupplied to keep the gas warm is  estimated to be 
$6 \times 10^{38}\u erg \ut s -1 $.  
Assuming  that $\sim 10^{50}\u ergs $ corresponding to 10\% of the energy of a typical supernova goes into
particles and the magnetic field (Duric et al.  1995), this is equivalent 
to one supernova per $7\times10^3$ years.


\subsection{Ionization Fraction}

One of the consequences of higher  cosmic ray ionization rate in 
the Galactic center environment is an increase in  
the ionization fraction 
$x_e={n_e}/{n_H}$ 
in molecular clouds. 
\begin{equation}
x_e = \frac{ne}{n_H}  \sim  10^{-7}{\frac{(\zeta/10^{-17} \rm s^{-1})^{0.5}} {(n_H/10^4 \rm cm^{-3})^{0.5}}}
\end{equation}
For typical values of $\zeta\sim10^{-15}\, \rm to\, 10^{-14}$ s$^{-1}$ H$^{-1}$, 
the ratio $x_e$ is (1--3.2)$\times10^{-5}$, respectively,  for a gas density of 100 cm$^{-3}$.
Assuming an order of magnitude lower value of $\zeta$ for dense gas, the 
inferred value of $x_e$  is 1 to $3.2\times10^{-6}$ assuming that the gas density is 
10$^4$ cm$^{-3}$.  
Such a high ionization fraction 
couples  the magnetic field and the gas, and consequently  slows down star formation 
due to an  increase in  the time scale for 
ambipolar diffusion (e.g., Yusef-Zadeh, Wardle and Roy 2007a).
Another consequence of this interaction that can be 
studied in the future is  the
abundance of  ionized molecular species such as H$_3^+$ in the Galactic center molecular clouds.

\subsection {Neutral FeI 6.4 keV Line Emission}

Although the X-ray irradiation resulting from a variable fluorescent echo of the X-ray flash is considered more 
``favorable" to explain FeI K$\alpha$ line emission (e.g., Ponti et al. 2010), we show here that the 
cosmic ray picture can also explain the 6.4 keV K$\alpha$ line, especially, from 
molecular clouds that are considered to be interacting with nonthermal radio sources (e.g.,  G0.11-0.11).  
In the cosmic ray irradiation picture,  the expected 
low EWs of the K$\alpha$ line (Yusef-Zadeh et al. 2007b) place a strong constraint on the 
applicability of this model for all observed clouds which show a range of low (few hundred eV) and high 
($\ge10^3$ eV) K$\alpha$ EW values. The cosmic ray electron picture has difficulty in explaining the origin of clouds with high 
K$\alpha$ EW values, unless FeI abundance is increased and/or the energy index of electrons is harder in a thick target 
(Tatischeff, Decourchelle \& Maurin  2012). 
Low energy cosmic ray ions, interacting with molecular gas can also explain the origin of the
steady FeI 6.4 keV line emission with  
high EW values and no need to increase the abundance of iron
(Tatischeff, Decourchelle \& Maurin  2012). 

The remarkable distribution of 
diffuse fluorescent FeI K$\alpha$ line emission, as measured by Suzaku X-ray Observatory (Uchiyama et al. 2011), is 
continuous and is arising from both dense and diffuse clouds in the central molecular zone (CMZ). It is difficult to 
date the transient events that are responsible for production of K$\alpha$ line emission from diffuse clouds. Given 
present observations, the cosmic ray picture is a natural explanation. 
Future studies should be able to determine the relative contributions of X-ray 
and cosmic ray models. Our aim here is to study the cosmic ray electron picture 
which is mainly motivated by a large reservoir of cosmic ray electrons radiating from the Galactic center at radio 
frequencies.


We turn to the  scenario in which a high flux of relativistic
particles from diffuse and localized nonthermal sources in the Galactic center inject
electrons into neutral clouds. Here, we  focus only on the cosmic ray electron model 
which was applied to individual 
Galactic center clouds (Yusef-Zadeh et al. 2003, 2007b) 
and expand the model to the inner 2\deg$\times$1\deg\ of
the Galactic center. 
The motivations for investigating the interacting cosmic ray  picture
are observationally driven since there is a  large reservoir of radio emitting 
relativistic particles
as well as a high concentration of warm molecular gas distributed in this region. 
Cosmic rays interact with molecular gas to produce 
FeI K$\alpha$ line emission, and should consequently be accompanied by 
an enhanced cosmic ray ionization rate and  elevated molecular gas temperature.

\subsubsection{The Low Energy Cosmic Ray  Model}

The role of low-energy cosmic ray electrons in producing X-rays has been discussed in the context of impulsive solar flares  
(Zarro, Dennis \& Slate 1992) and the background X-ray emission from the Galaxy (Valinia et al. 2000). A low energy cosmic ray electron model was subsequently applied to the 6.4 keV X-ray emission from the Galactic center molecular cloud G0.11--0.11 
(Yusef-Zadeh, Law, \& Wardle 2002). The emission was explained in terms of the impact of low-energy cosmic ray electrons with neutral gas associated with the G0.11-0.11 molecular cloud. 


To investigate the applicability of the cosmic ray interaction, we made a systematic study of molecular clouds showing 
fluorescent 6.4 keV line emission (Yusef-Zadeh et al. 2007b). We present below a strong spatial correlation between nonthermal 
radio  sources and molecular clouds.  We account for the distribution of FeI K$\alpha$ line emission arising from 
relativistic particles diffusing from both nonthermal filaments and compact nonthermal sources and impacting on neutral 
gas, producing both nonthermal X-ray bremsstrahlung and 6.4 keV line emission. The energy density of 
low-energy cosmic rays was estimated to be in the range 20 to  $\sim$10$^3$ eV cm$^{-3}$ for all Galactic center 
clouds (Yusef-Zadeh et al. 2007b).

The production rate of FeI K$\alpha$ photons depends on the electron spectral index and iron abundance.  For a given particle 
spectral index $p$, the efficiency of K$\alpha$ production $q$ increases with column density, eventually flattening when the 
column is sufficient to stop the bulk of the injected electrons within the cloud.  Figure 6a shows the dependence of Fe 
K$\alpha$ production per erg of electron energy injected into a cloud with column density $N_\mathrm{H}$ for different values 
of the electron energy spectral index $p$. The curves indicate that hard particle spectra ($p=1-1.5$) show high efficiency of 
producing K$\alpha$ line emission corresponding to high column densities whereas the soft particle spectra ($p=2.5-3$) show 
high efficiency at low column densities. Steeper spectra have particle energies increasingly concentrated towards low energies 
and so the flattening of the K$\alpha$ production rate occurs at successively lower column densities.  However, for the range 
of column densities between 10$^{23}$ and 7$\times10^{24}$, electrons with the power-law energy index $p=2$ are most efficient 
in producing K$\alpha$ line emission.  This spectrum, has equal energy per decade between 10\,keV and 1\,GeV, or energy 
density is the same in each frequency decade. We conclude that typically $q\sim 50 \, z/z_\odot$\,ph\,erg$^{-1}$, where z is 
the metallicity and z$_\odot$ the solar metallicity, for typical spectral indices and cloud column densities, increasing to 
$\sim 100\, z/z_\odot$ for hard electron spectra and high column densities, as observed in the Galactic center. 


In order to determine the EW of K$\alpha$ line emission, the bremsstrahlung emission at 6.4 keV was also estimated, as plotted in Figure 6b.  For $z = z_\odot$, the EW varies between 250 and 450  eV for particle spectral index $p$ varying between 3 and 1. There is a  dependence of EW on column density in excess of $10^{24}$ cm$^{-2}$, where we note an increase in EW for hard particle spectra. A similar study also concluded that the EW of FeI K$\alpha$ line emission at 6.4 keV produced by electrons having a hard spectrum $p=1.5$ ranges between 450 to 500 eV for high column density of 10$^{25}$ to $10^{24}$\,cm$^{-2}$, respectively (see figure 4 of Tatischeff, Decourchelle \& Maurin 2012).  These parameters apply to Galactic center clouds since the spectrum of electrons at low energies is hard and the column density of neutral gas is generally high in this region. 

We apply the LECR model to the inner 2$^{\circ}\times0.85^{\circ}$ by estimating the  K$\alpha$ line and bremsstrahlung emission from the interaction of the extrapolated low energy tail of the nonthermal electrons responsible for synchrotron radio emission. 
We assume that the nonthermal radio emission is produced by a power-law spectrum $n(E)\propto E^{-p}$ of electron energies between 10\,keV and 1\,GeV, and that the depth of the emitting region is of the order of its diameter 288\,pc and calculate the energy density of cosmic-ray electrons assuming that they are in equipartition with the magnetic field.

First, we note that the interaction of these electrons simultaneously heats and ionizes the cloud, as well as giving rise to fluorescent emission in the K$\alpha$ line due to the ejection of inner K-shell electrons from FeI.   Noting that each ionization is associated with deposition of energy $W\approx 40.1$\,eV in the cloud and that $q$ K$\alpha$ photons are produced for each erg deposited, the K$\alpha$ photon production rate per hydrogen nucleus is $qW\zeta$ \,ph\,s$^{-1}$\,H$^{-1}$ where $W$ is expressed in erg.  The intensity of K$\alpha$ photons received from a hydrogen column $N_\mathrm{H}$ subject to a cosmic-ray electron induced ionization rate per hydrogen nucleus $\zeta$ is then
\begin{equation}
I_{\mathrm{K}\alpha} =  \frac{qW\zeta\,  N_\mathrm{H}}{4\pi} \approx 8.7\ee -8 \left(\frac{\zeta}{10^{-14}\ut s -1 }\right)
\left(\frac{N_\mathrm{H}}{10^{23}\ut cm -2 }\right) {\rm \ ph \rm \ s^{-1} \ cm^{-2} \ arcmin^{-2}}\,,
\end{equation}
where we have assumed $q=200\u ph \ut erg -1 $, corresponding to a metallicity twice solar.   The unabsorbed FeI K$\alpha$ photon flux determined by Suzaku observations is about $3.8\times10^{-7}$ ph s$^{-1}$ cm$^{-2}$ arcmin$^{-2}$ (Uchiyama et al. 2011).  Typical Galactic center clouds such as Sgr C have column densities in the range 3--8$\times10^{23}$ cm$^{-2}$ (e.g.\ 
Yusef-Zadeh et al. 2007b).  Adopting a typical value N$_H=4\times10^{23}\,  \rm cm^{-2}$, we infer that the required ionization rate $\zeta\sim10^{-14}$ s$^{-1}$ H$^{-1}$.

This cosmic ray ionization rate can be compared with the minimum value of the product of the cosmic ray ionization rate and the path length L in diffuse molecular gas determined by H$_3^+$ absorption measurements over 8 sight lines toward the Galactic center (see Table 4 of Goto et al. 2008).  The average minimum value of $\zeta$L is $\sim1.2\times10^5$ cm s$^{-1}$ H$^{-1}$. Multiplying $\zeta$L by the density of neutral gas $n(\mathrm{H}_2) \sim 100$\,cm$^{-3}$ inferred from the H$_3^+$ measurements and comparing this to $\zeta \rm N_H$, we estimate that the diffuse molecular gas component responsible for the $H_3^+$ absorption contributes about 10\% of the  FeI K${\alpha}$ line emission at 6.4 keV. The remaining 90\% of the 6.4 keV emission can be explained by the interaction of electrons with $\sim10^3$ cm$^{-3}$ molecular gas. Thus, the cosmic ray ionized rates inferred from H$_3^+$ and FeI K$\alpha$ line emission probe diffuse and dense phases of the molecular ISM, respectively.  This is consistent with ammonia measurements which also infer a two-phase structure to the molecular gas
based on a bimodal temperature distribution within Galactic center molecular clouds (H\"uttemeister et al.\ 1993).  

The contribution of protons has been ignored here, but can also be significant (see Tatischeff, Decourchelle \& Maurin  2012). 
In a thick target, the efficiency of producing the 6.4 keV line emission is similar for both protons and electrons with energies $\sim$20-200 MeV and $\sim$10-100 keV, respectively. For proton-to-electron ratios of 40, the production of the 6.4 keV line emission could be as  important as electrons as long as the spectrum of the particles is hard (see Figs 6 and 13 of 
(Tatischeff, Decourchelle \& Maurin  2012).

Another important issue is the energetics required by this picture.  The total energy of cosmic ray electrons contained within a volume of 
2$^{\circ}\times2^{\circ}\times0.85^{\circ}$ is 2.7$\times10^{50}$ ergs,  assuming that the observed radio flux comes from an oblate spheroid with the emission having a volume filling factor of 0.1 and that the electrons are in equipartition with a 20\,$\mu$G field.  The K$\alpha$ flux emitted over the inner 
2\deg$\times1$\deg\, 
is estimated to be 2.7$\times10^{-3}$ ph\, cm$^{-2}$ s$^{-1}$ (Uchiyama et al. 2011), corresponding to a 6.4 keV luminosity  $\sim 2\times 10^{35}$ erg s$^{-1}$.  Given that the efficiency for producing the FeI K$\alpha$ line by cosmic ray electron impacts is $\sim2\times10^{-6}$ 
(Tatischeff, Decourchelle \& Maurin  2012), 
the total kinetic power lost by the LECR electron population is $\sim10^{41}$ erg s$^{-1}$. This implies that 
the electrons depositing their energies onto molecular clouds from the inner 2\deg$\times1$\deg\, must be resupplied 
on a 100-year timescale.  A longer timescale can be achieved if the electrons that lose their energies are resupplied 
by the leakage of cosmic rays. In this picture, a higher fraction of the electrons will interact with molecular gas 
and the electrons need to be resupplied on a longer time scale of $\sim10^3$ years.  The time scale estimate could be 
a lower limit as the average K${\alpha}$ line at 6.4 keV surface brightness could be lower than that estimated here due to 
contamination by unresolved point sources. In addition, transient X-ray sources such as Sgr A* could have contributed 
in producing a fraction of the FeI K${\alpha}$ line emission.

\subsubsection {The Correlation of FeI K$\alpha$ Line and Radio Continuum Emission}

The production of FeI K$\alpha$ line emission in the 
LECR scenario implies a correlation  between the distribution of 
nonthermal radio continuum and FeI K$\alpha$ line emission.  Combining equations (3) and (6) yields 
\begin{equation}
I_{\mathrm{K}\alpha} \approx  
\frac{2.7\ee -7 }{p-1}\,
\frac{I_\nu}{\u Jy \ut arcmin -2 }\,
\left(\frac{\nu}{\u GHz }\right)^{\!\alpha}
\left(\frac{B}{100\,\mu\mathrm{G} }\right)^{\!\!-(1+\alpha)}
\left(\frac{N_\mathrm{H}}{10^{23}\ut cm -2 }\right)
\left(\frac{L}{30 pc}\right)^{-1}\,
\\
{\rm \ ph \rm \ s^{-1} \ cm^{-2} \ arcmin^{-2}}\,,
\end{equation}
where we have again adopted $q=200$\,ph\,erg$^{-1}$ for the efficiency of K$\alpha$ production.

To test this correlation, we used the intensity of K$\alpha$ line emission listed in Table 2 of Uchiyama et al. (2011),  and compared with radio emission at 325 MHz based on combined GBT and VLA data.  We convolved the 325 GHz map by a 6$'\times6'$ Gaussian before  it was compared with Suzaku data. Figure 7(a-i) show nine cross cut plots  of X-ray and radio data  made in the direction parallel (Fig. 7a) and perpendicular (Fig. 7b-i) to the Galactic plane. Given that there is contamination by thermal radio continuum sources at 325 MHz, these plots suggest that the flux of nonthermal continuum emission does indeed track the  6.4\, keV line flux, as predicted by equation 7.  
For a typical flux of 50 Jy at 325 MHz and $p=1.4$, the predicted K$\alpha$ line intensity is $\sim 2\times10^{-7}$ ph s$^{-1}$ cm$^{-2}$
arcmin$^{-2}$ which is similar to what has been observed in most plots shown in Figure 7. 
The radio flux in Figure 7a shows the highest  value of $\sim250$ Jy but K$\alpha$ intensity is 
similar to the values shown in a typical slice. 
The peak radio flux mainly arises from the strong  radio emission associated with 
Sgr A East  where  K$\alpha$ line intensity is weak. 
The well-known Sgr A East supernova remnant is known to be interacting with 
the 50 \kms\ molecular cloud. 
The clouds that  show a lack of  6.4 keV emission 
in the context of the cosmic ray irradiation picture 
have either a low column density for production of 
X-ray photons or 
a high extinction inside a dense cloud that 
substantially reduces the flux of 
K$\alpha$ emission. The 50 \kms\,  molecular is one example in which 
the column density is sufficiently large for extinction to reduce  the emission. 
Column densities ranging between $\sim5-10 \ee 23 \ut cm -2 $ reduce the X-ray flux 
at 6.4 keV by a factor of 2-5 and 5.6-31 for one and two times solar metallicity, respectively
(Morrison and McCammon 1983). The assumption that all of 
synchrotron emitting particles interact with molecular gas can also contribute to explaining this 
discrepancy. If only a fraction of the 
electrons interact with molecular gas, the radio flux in equation 6 will be smaller, and thus 
the flux of the K$\alpha$ line will be reduced.  Lastly, the time scale 
for the variability of 6.4 keV line emission is relatively short 
in the case where electrons propagate through a dense molecular cloud. 
Thus, the discrepancy can  be explained by a combination of effects 
that can potentially suppress   the 6.4 keV line emission from dense clouds with high 
column densities. However, if the metallicity is twice solar in the 
50 \kms\, cloud, the extinction could be the dominant effect in suppressing the 6.4 keV emission. 


\subsubsection {EW Distribution  of FeI K$\alpha$  Line Emission}

One of the key parameters that can distinguish LECR and X-ray irradiation 
scenarios of FeI K$\alpha$ line production is the strength of  the EW of 
FeI K$\alpha$ line.  The 
large EW$\sim 10^3$ eV is more consistent with the X-ray irradiation picture of 
producing FeI K$\alpha$ emission (Sunyaev \& Churazov 1998) whereas the 
low EW of about few hundred  is more consistent with that of the cosmic ray 
irradiation picture. Given that both   emission mechanisms depend on the abundance of FeI, 
the predicted EWs  should be higher in the Galactic center due to 
an increase in the  metallicity gradient toward the Galactic center (Giveon et al. 2002; Rudolph et al. 2006). 
 Recently, 
spectral analysis of  emission lines from several neutral atoms 
based on  Suzaku observations
showed  the  EW of K$\alpha$ line emission 
at 6.4 keV to be   $\sim1.1$ keV 
(Nobukawa et al. 2010). 
Figure 8a
shows a 
grayscale distribution of the 6.4 K$\alpha$ line emission based on Chandra 
observations with a resolution of $2''$.  Figure 8b
shows contours of the 
EW superimposed on a 20 cm continuum image of the Galactic center. A range 
of EWs are noted between $\sim$100 and $\sim2\times10^3$ eV associated with Galactic 
center molecular clouds. These images shows a number of compact 
sources with high value of EWs (e.g., Sgr B2, the Arches cluster) as well 
as large diffuse structures associated with extended molecular clouds. 
The 
low EW distribution of the FeI K$\alpha$ line emission suggests that diffuse 
envelopes  are consistent with typical EW values 
expected in the context of bombardment of cosmic 
ray and X-ray photon irradiation.
Recent work shows that LECR electrons can produce a significant 
 FeI  K$\alpha$  line emission from 
diffuse  molecular clouds with N$_H < 10^{22}$   cm$^{-2}$ , especially in 
clouds  with  strong particle diffusion (see fig. 4b of 
Tatischeff, Decourchelle \& Maurin  2012). 
An observation of a 6.4 keV line emission 
from a cloud with 
N$_H\sim 10^{21}$ cm$^{-2}$ is  potentially  a promising signature of 
LECR electrons, as the efficiency of  production of this line by hard 
X-ray irradiation of the cloud is  low (Yaqoob et al. 2010).
For higher values of the EW, the Fe abundance in 
the gas phase or in dust has to increase in order to be consistent with 
the LECR electron model.


\subsubsection {Comparison of EW: Chandra vs. Suzaku}

We also measured the EW distribution using 
Suzaku with its  moderate resolution from the same region shown  in Figure 8. 
Suzaku data is sensitive to diffuse 6.4 keV  line emission whereas Chandra is 
clearly able to identify compact features in the EW distribution. The contribution 
of point sources have been removed in the Chandra image before the EW map is constructed. 
To compare these two different measurements, we show several cross cuts parallel to the Galactic 
ridge at  constant latitudes. 
Figure 9 shows plots of these cross cuts in blue and red corresponding to 
Suzaku and Chandra  measurements.  
We also averaged the EW over longitude intervals of 
$3'$ and made a
one-dimensional distribution of EW based on Suzaku and Chandra.
Both Chandra and Suzaku data 
indicate that the  EW values vary substantially in the Galactic center from a few hundred eV to 
close to 1-2 keV. 
The EW profiles were binned by averaging over 4$'$
and the variation from the mean for both  Chandra and Suzaku 
data are consistent with each other. 
However, there is an offset  EW$\sim$100-150 eV in the mean 
value    the EW 
that is present especially at  
higher galactic latitudes.  Chandra data is averaging over low intensity 
pixels that  have no available data in Suzaku measurements. 
This is because the  spatial coverage  of Suzaku observations 
measurements was not uniform.  
Additional cause of this discrepancy could be 
due to the uncertainty of properly subtracting the hot 
background 
continuum emission (Koyama et al. 2009) 
in construction of the 6.4 keV EW map based on 
Chandra observations, as described in $\S2.5.1$. 
In a recent study of the Galactic center, 
the anti-correlation between the EW values of the 6.7 and 6.4 keV line 
emission based on Suzaku measurements suggest that the EW values of 6.4 keV line emission 
when averaged over the region -2\degree $< l < 1\degree$  is about 700 eV which is 
more consistent with Suzaku measurements shown in Figure 9 (Uchiyama 2011).
These measurements 
suggest  that the low  value of the EWs   can 
be explained in the context of  the cosmic ray  model or a mixture 
of cosmic ray and X-ray echo models. 
As pointed out earlier, our radio continuum observations of the inner 
2\degree$\times$1\degree\, suggest that 
the energy spectrum of 
cosmic ray electrons is $\sim1.5$ and that the the column density is 
$\ge10^{23}$ cm$^{-2}$. Under these conditions, the typical EW is 
predicted to be about 450 eV for solar abundance of FeI K$\alpha$ line emission.   
In this case, 
the spectrum of injected particles diffusing 
into a cloud 
with a higher metallicity can account for high and low  values of the  EW.  






\subsubsection {Radio Arc and G0.11-0.11}

One of the most interesting Galactic center molecular clouds showing strong 
6.4 keV emission  is  
G0.11-0.11 which is positioned at the edge of the nonthermal filaments 
of the Arc.  Thus, it is an important cloud as it can be used to test 
the cosmic ray irradiation picture. 
Figure 10a shows contours of the EW (eV) based on Chandra observations 
and are superimposed on a grayscale continuum image at 20 cm. This figure  
illustrates  the long nonthermal radio filaments that run perpendicular to the Galactic 
plane.  Figure 10b shows the same region of the sky as that of Figure 10a except that 
contours of  K$\alpha$ line emission are superimposed on a grayscale K$\alpha$ line image. 
The morphology the eastern  edge of the G0.11-0.1 molecular cloud
led Tsuboi, Ukita \& Handa  (1997) to argue that radio filaments 
and G0.11-0.11 are dynamically interacting with each other. 
The EW ranges between $\sim$110 to $\sim$1100 eV peaking on diffuse and compact sources within the 
G0.11-0.11 cloud.  The low value of the  EW of K$\alpha$ line emission is consistent with the interaction picture. 
The high value of the EW of FeI 6.4 keV line toward G0.11-0.11 requires to have a high abundance of 
iron.  It is interesting that spectroscopic studies  of gas toward the radio Arc bubble immediately adjacent 
the G0.11-0.011 cloud 
shows a 6.5 fold increase in the abundance of iron in the gas phase  compared to the surrounding ISM 
(Simpson et al. 2007). 
Using the observed value of the EW  toward G0.11-0.11 (Nobukawa et al. 2010), 
we find that the Fe abundance has to be  $\geq$2.2 in the context of the LECR  picture.

Another isolated cloud at the edge of the nonthermal filaments in Figure 10a,b 
shows a cloud with a low value of K${\alpha}$ line emission. This cloud G0.16-0.22 
coincides with recent 
Suzaku observations which first  reported the discovery of K$\alpha$ line emission 
(Fukuoka  et al. 2009), lying  at the southern end of the nonthermal filaments. 
The low value of the EW led them to suggest that cosmic ray irradiation by electrons 
is responsible for 6.4 keV emission. Our Chandra measurements are 
consistent with the low value of EW which peaks at $\sim$130 eV.

Another interesting 6.4 keV cloud revealed in Figure 10 is the presence 
of K$\alpha$ line emission peaking at the location where the longest nonthermal radio filament 
(Fig.\  10a) 
crosses  the  well-known   Sickle HII region  G0.18-0.04 (Yusef-Zadeh and Morris 1987) 
at $l$=10.6$'$, b=-3$'$. 
This  peak has EW$\sim$ 200 eV and  lies 
at the interface of nonthermal  filaments and the 40 
km s$^{-1}$\, molecular cloud  which is associated with the Sickle. 
The kinematics of ionized gas exactly at the interface between nonthermal filaments and
the sickle at G0.18-0.04 shows a forbidden velocity of  $-30$ \kms (Yusef-Zadeh, Roberts and Wardle 1997).  
This anomalous motion of ionized gas in the Sickle can be interpreted in the context of the interaction of the
magnetized filaments and the 40 \kms cloud. 
A more detailed 
account of this result which  is 
consistent with the  cosmic ray interaction picture will be given elsewhere.






\subsubsection {Time Variability of FeI K$\alpha$  Line Emission}

One of the key issues related to  the cosmic ray interaction picture 
in the Galactic center  is the observed 
time variability of FeI K$\alpha$ emission 
from several clouds (Inui et al. 2009; Terrier et al. 2010; Ponti et al. 2011). 
Although the echo X-ray picture can explain the variable emission well, 
it turns out that the cosmic ray electrons picture can also explain this 
remarkable property of Galactic center molecular clouds.  
In the  cosmic ray picture, the flux of electrons 
diffuse from their acceleration site to the cloud edge, and then
freely stream into the cloud because ion-neutral damping suppresses   
the magnetic fluctuations responsible for the diffusion. 
As the electrons penetrate molecular clouds, the time scale for 
energy  loss of electrons at low energies is quite  rapid. 
Consequently,   
the  energy spectrum  of particles become harder.

As a demonstration of this idea, 
Figure 11a  shows spectra of LECR electrons 
propagating in a medium of density n$_H=10^4$   cm$^{-3}$, 
over a time interval that is very short compared to the 
lifetime against energy losses. 
The energy spectrum of electrons is assumed to have 
a power-law energy spectral index  p=2.4 
above a minimum energy E$_{min}$=100 keV  and E$_{max}$=1 GeV. 
The electrons of energies less than  E$_{min}$
are expected not  to escape their acceleration region and do not penetrate into molecular 
clouds. The 
normalized  electron injection spectrum corresponds to  the total electron energy content of  10$^{48}$ ergs. 
Since the electrons losses are proportional to the density, 
if we choose  a number density n$_H=100$  cm$^{-3}$, then the 
energy loss time scale increases by a factor of 100. 
The energy loss time scale also depends on the energy spectrum of 
electrons. For a harder spectrum of electrons p=1,  
the energy loss time scale does not vary as fast as shown in Figure 11a.   
So for n$_H=100$  cm$^{-3}$, 
the typical timescale of evolution is 100 yr, 
any significant variation in the 6.4 keV line emission on shorter time scales would 
rule this model out.

Figure 11b shows nonthermal X-ray spectra produced by LECR electrons with the assumption 
that the metallicity of the ambient medium is twice solar. 
We have accounted for photoelectric absorption along the line of 
sight with a  hydrogen 
column density N$_H=6 \ee 22 \ut cm -2 $. 
We have ignored the FeI K$\alpha$ line  emission 
generated by the impact of secondary electrons. These electrons 
 can account for   no more than 20\% of X-ray emission. 
The observed variability measurements  toward dense clouds, 
such as G0.11-0.11 and Sgr B2, are consistent with  the prediction of 
flux variability  of  the 6.4 keV line, as presented in Figure 11b. 


Lastly, Figure 11c 
shows the time evolution of 
the EW of the neutral Fe K$\alpha$ line and
the luminosity in 
this line.  
To illustrate how the  time evolution of X-ray luminosity varies, 
the injected spectrum of electrons  are varied  between 
p=1.5 and 2.4 as well as different minimum energy 10 and 100 keV. 
In the bottom panel of Figure 11c, we note a constant luminosity of $\sim10^{42}$ photons 
$s^{-1}$
corresponding to $\sim3\times10^{-4}$ photons  cm$^{-2}$  s$^{-1}$  at 8 
kpc   followed by  a decrease in time as $\sim t^{-1.2}$. 
The top panel shows that the EW of the 6.4 keV line stays almost constant at  ~0.8  keV.
These figures show clearly that the interaction of LECR particles with ambient gas can 
produce a fast variability of the FeI K$\alpha$ line emission at 6.4 keV. 
There are numerous nonthermal radio filaments distributed in the Galactic center region, some of which will 
be interacting with  molecular gas in this region. 
At the site of the interaction, the nonthermal electrons are impulsively injected into the cloud
as they follow the distorted magnetic field lines. In this picture, the injected particles 
lose  their energies on a time scale that is similar to the observed decay time scale of 
K$\alpha$ line emission. The particles diffusing into the cloud are not resupplied. 
Another consequence of the cosmic ray interaction picture is that the 
value of $\zeta$L measured from H$_3^+$ absorption study 
should be time variable. This is 
because low energy  cosmic  ray electrons are responsible for production of this important ionized molecule. 

Another possibility that could account for the variability of K$\alpha$ line emission is 
to accelerate electrons to TeV energies at the interface of the interaction.  
Chandra observations detected a filamentary X-ray structure (G0.13-0.11) sandwiched between
nonthermal radio filaments of the Arc and the 6.4 keV line emission from the extended molecular
cloud G0.11-0.11 molecular gas (Yusef-Zadeh, Law and Wardle 2002). 
This filament has a
photon index $\Gamma$=1.4 to 1.5 (Wang, Gothelf \& Lang 2002; Johnson et al. 2009). 
The nature of this filament,  
which has neither 6.4 keV line nor radio continuum counterparts,
suggests  that the nonthermal X-ray emission is consistent with a synchrotron
emission from electrons with TeV energies.
The variability of the emission due to short synchrotron lifetimes place
strong constraints on the energy and the magnetic field of the particles at the interaction site.  
For example, for E=4 TeV, B= 1 mG, t$_{synch}\sim$ 2 years at 1 keV energies. 
The variable X-ray emission 
at the site of the interaction
could in principle  produce the variable X-ray fluorescent line emission detected from 
G0.11-0.11. This should be observed in future X-ray observations.





\section {Diffuse GeV and TeV $\gamma$-ray  Emission}

In previous sections, we have argued that the  interaction of  cosmic ray electrons 
with the gas in the Galactic center can explain  the measured  high values of cosmic ray
ionization and heating rates. We also explained 
that the interacting low energy tail of cosmic ray particles 
contribute 
nonthermal   
bremsstrahlung continuum emission as well as 
the  FeI  K$\alpha$ line at 6.4 keV. 
We now examine whether nonthermal bremsstrahlung from a population of synchrotron emitting 
electrons in the Galactic center  can explain the 
GeV and TeV $\gamma$-ray emission from the Galactic center. 

\begin{deluxetable}{lllrrrrrrrrrr}
\tablecaption{Parameters of the fit to $\gamma$-ray sources using $Fermi$ and H.E.S.S. data}
\label{tbl:fit}
\tabletypesize{\small}
\tablewidth{0pt}
\tablehead{
  \colhead{Source} & 
  \colhead{B} & 
  \colhead{n$_H$} & 
  \colhead{F$_{325MHz}$} & 
  \colhead{p1} &  
  \colhead{p2} & 
  \colhead{$\nu_{break}$} & 
  \colhead{Flux$_{}$(H.E.S.S.)} & 
  \colhead{$\Gamma$(H.E.S.S.)} \\ & 
  \colhead{($\mu$G)} &  
  \colhead{(cm$^{-3}$)} & 
  \colhead{(Jy)}& & & 
  \colhead{(GHz)} & 
  \colhead{(MeV cm$^{-2}\, s^{-1}$)}
}
\startdata
GC diffuse &  8 &  12.5 & 508 &  1.5 &  4.4 &  3.3& (2.58$\pm1.39)\times10^{-5}$ & 2.27$\pm$0.07 \\
Sgr A      & 70 &  770  & 185 &  1.4 &  3.2 &  5  & (6.37$\pm4.18)\times10^{-6}$ & 2.20$\pm$0.09 \\
Sgr B2     & 30 &  2600 &   9 &  0.4 &  4.4 &  10 & (5.62$\pm4.24)\times10^{-6}$ & 2.19$\pm$0.10 \\
Radio Arc  & 40 &  450  & 156 &  2.4 &  2.8 &  20 &  & \\
Sgr C      & 60 &  470  &  96 &  0.4 &  3.6 &  10 &  & \\  
\enddata
\end{deluxetable}

Recent observations with the H.E.S.S. have discovered large-scale diffuse TeV
emission from the inner 200 parsecs of the Galaxy (Aharonian et al.  2006).
The morphology of diffuse emission correlates well with the
distribution of CS molecular clouds, thus suggesting that the
$\gamma$-ray emission is a product of the interaction of cosmic rays  
with interstellar gas near the Galactic center.  
These authors show   
that the spectrum of TeV emission from resolved clouds toward the
Galactic center has  a photon index $\Gamma\sim 2.3$ which is harder
than that in the Galactic disk.  They note that the $\gamma$-ray
flux above 1 TeV is a factor of 3-9 times higher than in the Galactic
disk and argue for an additional population of cosmic rays in this
unique region.  They propose that the TeV emission is due to hadronic
interaction of cosmic rays with the target material. 
Given that the target material is the same, we recently argued (Yusef-Zadeh et al. 2007b) 
a spatial correlation between 
the distribution of TeV emission with those  of the 6.4 keV emission 
and submillimeter emission 
tracing  molecular gas toward the Galactic center. 
An argument 
against the importance of TeV electrons in Sgr B2 is the short lifetime of few decades for  
their energies with a magnetic field of an order of 0.5 mG  (Crocker et al. 2010).

Morphological distribution of diffuse  $\gamma$-ray emission 
detected by $Fermi$ is remarkably similar 
to  that of radio continuum emission at 1.45 GHz, as shown in Figure 4a,b. 
It is also known that  the distribution of TeV emission traces Galactic center 
molecular clouds.  
These correlations suggest that the $\gamma$-ray emission is a 
product of the interaction of cosmic rays with interstellar gas 
near the
Galactic center.   
Adopting a
power-law electron energy spectrum  E$^{-p}$ (Schlickeiser 2003)  
, the bremsstrahlung photon flux is
\begin{equation}
F_{\gamma} \approx  
\frac{3.3\ee -13 }{p-1}\,
\left(\frac{S_\nu}{Jy}\right)\,
\left(\frac{\nu}{\u GHz }\right)^{\!\alpha}
\left(\frac{B}{100\,\mu\mathrm{G} }\right)^{\!\!-(1+\alpha)}
\left(\frac{n_\mathrm{H}}{\ut cm -3 }\right)
\left(\frac{E_{\gamma}}{1\, GeV}\right)^{-p}
{\rm \ ph \ cm^{-2} \ s^{-1} \ GeV^{-1} }
\end{equation}
where E$_{\gamma}$  is the photon energy in GeV, $\nu$ 
is the synchrotron observing frequency in GHz, 
         S$_{\nu}$  is  synchrotron flux
and 
          n$_H$  = n(HI) + 2n(H2) is the number density of hydrogen nuclei in atomic or
molecular form, in cm$^{-3}$.  
We  assume here   that  all of the synchrotron-emitting electrons are interacting with the gas.

We interpret the Fermi and H.E.S.S. data 
in the context of nonthermal bremsstrahlung radiation mechanism. 
We present  the spectra 
from  several sources from  the inner 
2\deg$\times$1\deg\ of the Galactic center. 
Figure 12a shows the spectrum of the diffuse emission from 
the inner 2\deg$\times$1\deg\ by 
excluding the source that lies very near Sgr A*. 
This spectrum includes 
both Fermi and H.E.S.S. data points which are represented in red and blue, respectively.
The red  triangles are  3-sigma upper limits from Fermi data. 
The red dashed curve is the $\gamma$-ray 
bremsstrahlung, 
as predicted by our nonthermal radio 
spectrum with $\alpha$ = 0.25
below 3.3 GHz and 1.7 above it.  These  values correspond
to synchrotron spectrum of 
electrons with p=2$\alpha$+1=1.5 and below 3.3 GHz and 4.4 above it. 
  To 
convert from radio
synchrotron flux to $\gamma$-rays,  
an equipartition field of 8$\mu$G is selected,  which is 
about a factor of two higher than 16$\mu$G estimated 
from  the cosmic ray ionization rate (Fig. 5a). 
In other words, the source volume and energetic electron density 
are  similar to those  used to estimate the ionization rate. One difference is that the 
$\gamma$-ray flux from Sgr A* is not included here.   
Given that $\gamma$-ray flux depends on the product of neutral gas density 
and radio flux,  we adjust 
the assumed hydrogen  number density  in the source volume 
to normalize the curve to give a reasonable fit to the data.  
We chose  n$_H=12.5$ cm$^{-3}$ and a radio flux of 508 Jy at 325 MHz (using n$_H=2.5$ cm$^{-3}$ 
and a radio flux of 2450 Jy at 325 MHz could 
also give the same fit). 
Note that  this could correspond to a higher density in a fraction of  the source volume.
A break in the radio spectrum reflecting  the spectrum of $\gamma$-rays 
provides  compelling evidence that the interaction of radio emitting 
GeV electrons with neutral gas 
 is  also responsible for production of  GeV emission. 
The change in the spectral index can be attributed to  rapid cooling of electrons 
at  high energies. 

The blue dashed power-law fit to the H.E.S.S. photon spectrum  in Figure 12a is of the form
F$_0$(E/GeV)$^{-\Gamma}$ with E$^2 \times \rm F_0 = (2.58\pm1.39) \times10^{-5}$\,   MeV\,  cm$^{-2}$\, s$^{-1}$
and $\Gamma=2.274\pm0.072$. 
Finally, 
the black solid curve is the sum of the two bremsstrahlung contributions, giving a reasonable 
fit to the H.E.S.S. and Fermi data. 
We assume that the  emission at TeV energies  is also produced by 
bremsstrahlung mechanism 
from an additional hard electron population with an E$^{-2.274}$
 spectrum extending all the way 
from GeV to TeV.   What is interesting about model  fitted   H.E.S.S. spectrum 
is that it has  a hard spectrum ($\Gamma$=2.27), as  is in   the part of the modeled 
fitted Fermi spectrum ($\Gamma$=1.25) that is not affected by cooling losses.  
We reconcile the GeV and TeV emission with similar spectrum in terms of 
a population of electrons that is young, thus producing $\gamma$-ray emission with 
a hard spectrum. The spectrum of the GeV emission showing a 
steep spectrum arises from an older generation of electrons that  has 
experienced cooling. 

As discussed above,  the equipartition  magnetic field of $\sim15\mu$G 
gives longer lifetime for TeV particles to lose their energies. 
For example, for E=1 TeV, B= 15 $\mu$G, t$_{synch}\sim$ 800 years, during which electrons 
will diffuse in the reservoir of molecular gas distributed in the central 
few hundred parsecs of the Galaxy, in the  so-called  central molecular zone (Morris and Serabyn 1996). 
We consider that 
the flux of TeV electrons inside molecular clouds  
stream freely along the magnetic field lines close to speed of light. 
This is  because ion-neutral damping suppresses   
the magnetic fluctuations responsible for the diffusion. 

Given the short lifetime of TeV electrons, the question arises as to what 
is responsible for  resupplying  these highly relativistic particles.  
One possibility for production of TeV electrons is the site 
at which nonthermal Galactic center  radio filaments interact 
with molecular clouds. There is at least one cloud in which 
nonthermal X-ray continuum emission has been detected at the edge 
of the cloud G0.11-0.11 (Wnag, Lu and Lang 2002).  Another possibility is 
that a source like Sgr A* is  responsible for production of 
TeV electrons which then propagate through the central molecular zone of the 
Galactic center.

As a demonstration of the concept, model fitting of  Fermi and H.E.S.S. emission from Sgr A and Sgr B$_2$ 
as well Fermi emission from the radio arc near l$\sim0.2^{\circ}$  
are shown in Figures 12b-d, respectively. 
Table 4  shows the parameters of the fit to all  these sources including Sgr C as well as 
the 
diffuse emission from the central  2\deg$\times$1\deg.
Column 1 shows the source names and columns 2-7 show the fit to $Fermi$ data whereas the last two columns 
give the observed  H.E.S.S. flux and the  photon index ($\Gamma$)  required by the H.E.S.S. observations.  
The magnetic field (column 2), the density of neutral gas (column 3), radio flux at 325 MHz (column 3)
and the break frequency (column 7) 
are  adjusted in order to match the data. The indices of the broken power-law distribution of electrons p1 and p2 
are shown in columns 5 and 6. 
There is a degeneracy in the number density of 
neutral gas and the nonthermal radio flux. A more detailed measurements 
of the flux from individual sources and the possible contribution of electrons 
upscattering Galactic center background radiation will be given elsewhere. 
In this study, we use Table 4 as a demonstration of the concept that 
bremsstrahlung mechanism can explain the observed $\gamma$-ray emission. 
In the context of this picture, we predict that TeV electrons interacting with molecular gas 
are themselves the source of synchrotron emission. The synchrotron luminosity by TeV electrons 
is predicted to be 1 and 3$\times10^{38}$  erg s$^{-1}$  for diffuse emission  form the inner 
2\deg$\times$1\deg\ and Sgr B2, respectively. 
This should be observed in X-rays with the new generation of X-ray observatories like Chandra, 
Suzaku  and  XMM.  
Another prediction of this model is that  the population of young electrons with a hard spectrum 
should contribute 900 Jy to the radio flux at 330 MHz. 
These are robust predictions of the TeV electrons with a specific spectrum in X-rays  
that can be investigated in future studies.


\section{Conclusions}

We have explored different aspects of diffuse emission from the Galactic center in the context of cosmic ray 
electrons interaction with Galactic center clouds. We began by presenting $\gamma$-ray observations and data 
reductions using $Fermi$ LAT, followed studying the relativistic and the nonrelativistic components of nonthermal 
particles in the interstellar medium of the Galactic center. We presented nonthermal radio flux over the inner 
2\degree$\times1$\degree of the Galactic center at four different radio frequencies. We used the reservoir of 
relativistic and non-relativistic cosmic rays electrons in the Galactic center region as seed particles interacting 
with neutral gas to determine cosmic ray ionization rate, molecular gas heating rate, the production of K$\alpha$ 
line and diffuse X-ray and $\gamma$-ray emission. The origin of high energy X-ray and $\gamma$-ray emission was 
explained in the context of bremsstrahlung mechanism, as had previously been used to explain the origin of 6.4 keV 
line emission from neutral iron. In addition, we investigated the time variability of low energy cosmic ray flux by 
discussing  that the cosmic ray flux should vary on a short time scale because of ionization losses of electrons 
diffusing through a molecular cloud. The ionization losses of electrons are particularly dramatic at 100 keV which 
could diffuse for roughly ten years in a medium with molecular density of 100 cm$^{-3}$ before they lose most of 
their energy. Assuming that diffusion of low energy cosmic ray particles is not hindered by magnetic field 
fluctuations in molecular clouds, the fluorescent 6.4 keV line emission was predicted to vary on such a time scale 
unless there is constant acceleration of particles at these energies. Future studies will determine the importance of 
cosmic ray irradiation model for individual Galactic center molecular clouds when compared to that of the X-ray 
irradiation model. Another characteristic that can place constraints on the applicability of these two models is to 
study the chemistry of the cosmic ray dominated region of the central molecular zone.

In summary, we explained the origin of $\gamma$-ray emission based on $Fermi$ and the H.E.S.S. observations. The mechanism 
for production of $\gamma$-ray emission is similar to that invoked to explain the production of K$\alpha$ line 
emission except that high energy particles are involved for production of bremsstrahlung $\gamma$-ray radiation. 
Another byproduct of the impact of cosmic rays with gas clouds is ionization losses suffered by interacting electrons 
with gas particles. The estimate of the ionization rate was compared with that measured from H$_3^+$ absorption 
lines. Lastly, cosmic rays heat molecular gas, increasing  the temperature and ionization fraction of molecular gas. The 
required cosmic ray heating rate was estimated to explain an increasing molecular gas temperature in the Galactic 
center region. These physical processes placed constraints on the strength of the magnetic field, the cosmic ray 
ionization rate and cosmic ray heating rate of molecular gas in the inner region of the Galaxy. Observed synchrotron 
emission from the Galactic center at radio wavelengths indicated a magnetic field of $\sim15\mu$G and a large 
population of relativistic GeV electrons. The interaction of these electrons with neutral gas explained the GeV 
emission observed with $Fermi$. We were also able to explain the origin of the TeV emission and the FeI K$\alpha$ 
line emission at 6.4 keV which required high cosmic ray ionization rate with some uncertainties related to the 
extrapolation of the electron spectrum to 10 keV and TeV energies.

\acknowledgements

J.W.H. is supported by an appointment to the NASA Postdoctoral Program 
at the Goddard Space Flight Center, administered by Oak Ridge Associated 
Universities through a contract with NASA. We thank Mike Muno for his 
contribution in the early phase of this project. We are thank 
Johann Cohen-Tanugi, Andrew W. Stong and Luigi Tibaldo for helpful 
comments that improved the text. This research is supported in part by 
a grant from the $Fermi$ Guest Investigator Program and 
the National Science Foundation.

\begin{figure}
\center
\includegraphics[scale=0.55,angle=0]{f1a_20cm_gbt.ps}\\
\center
\includegraphics[scale=0.45,angle=-90]{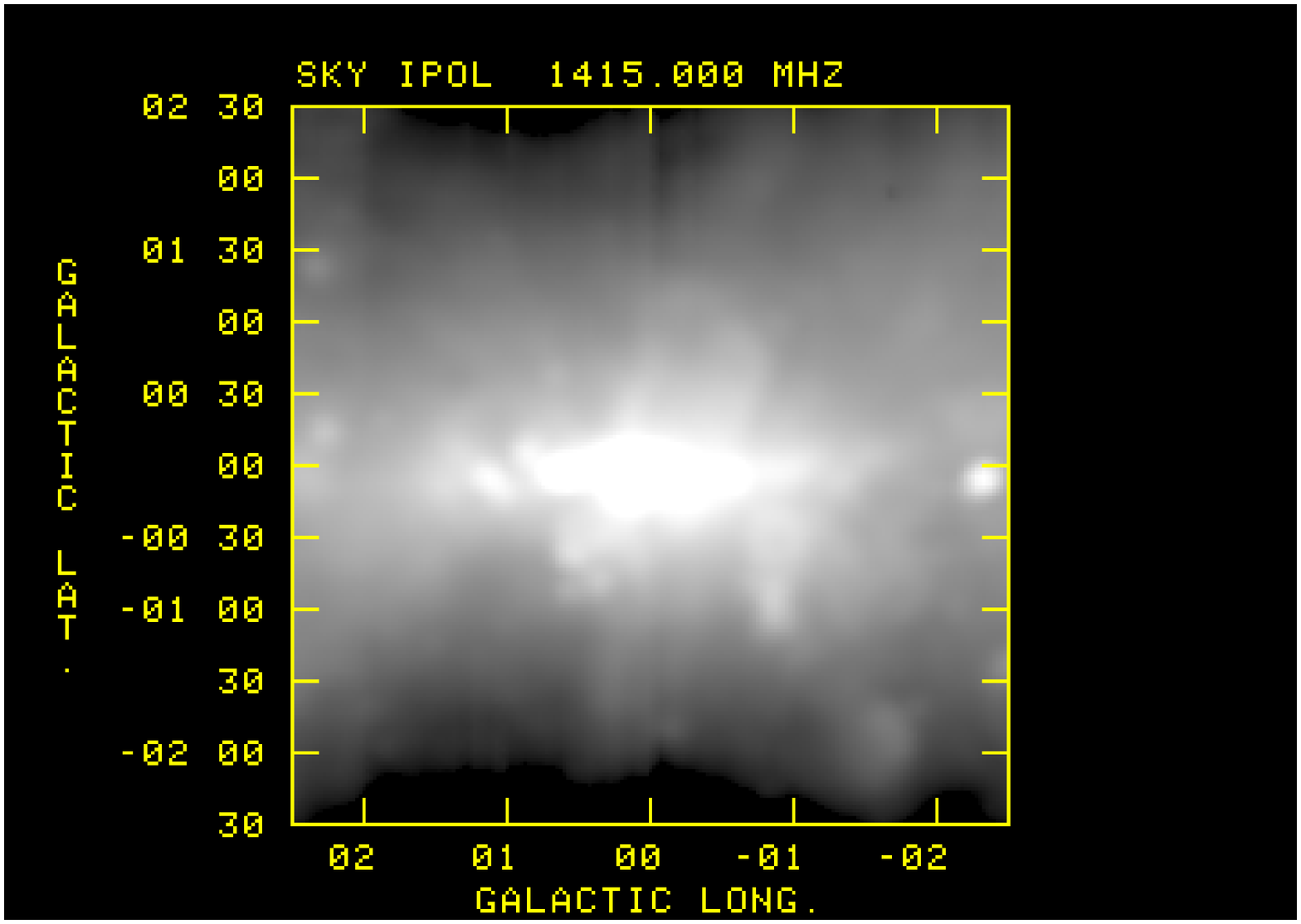}
\caption{
{\it (a - Top)}
Contours of background-subtracted continuum emission 
at 2, 4, 6, 8, 10, 12.5, 15, 20, 30, 40, 50, 75, 100, 
150, 200 
Jy beam$^{-1}$ 
are superimposed on a grayscale image at 1.415 GHz with a spatial resolution of 
539$''$.
 The greyscale ranges between 1 and 7 Jy beam$^{-1}$. 
{\it (b - Bottom)} 
The same as (a) except that only the greyscale image is shown. 
The prominent vertical features toward negative latitudes b$\sim-2$\deg, 
are best displayed in this image. 
}\end{figure}

\begin{figure}
\center
\includegraphics[width=3in,angle=0]{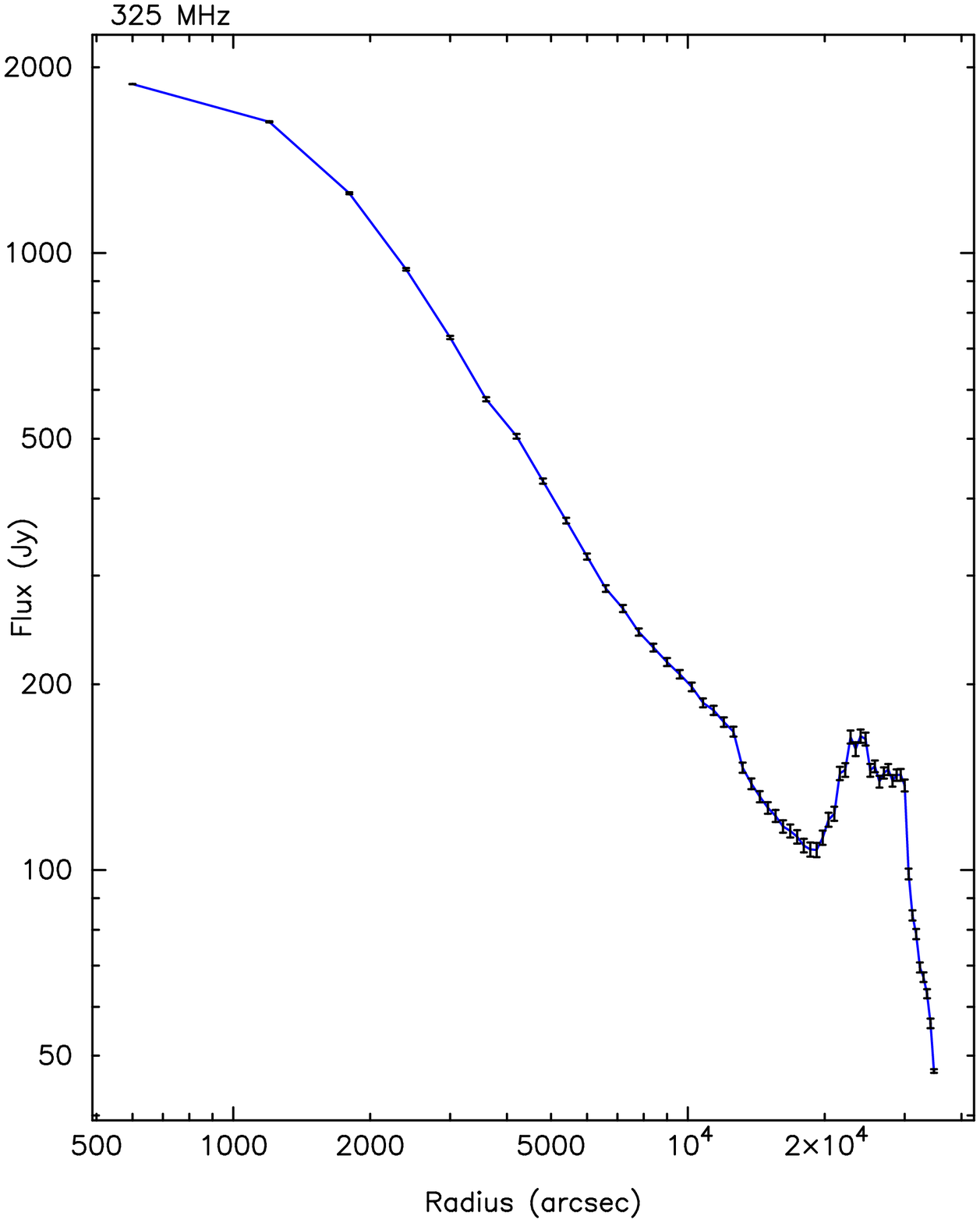}
\includegraphics[width=3in,angle=0]{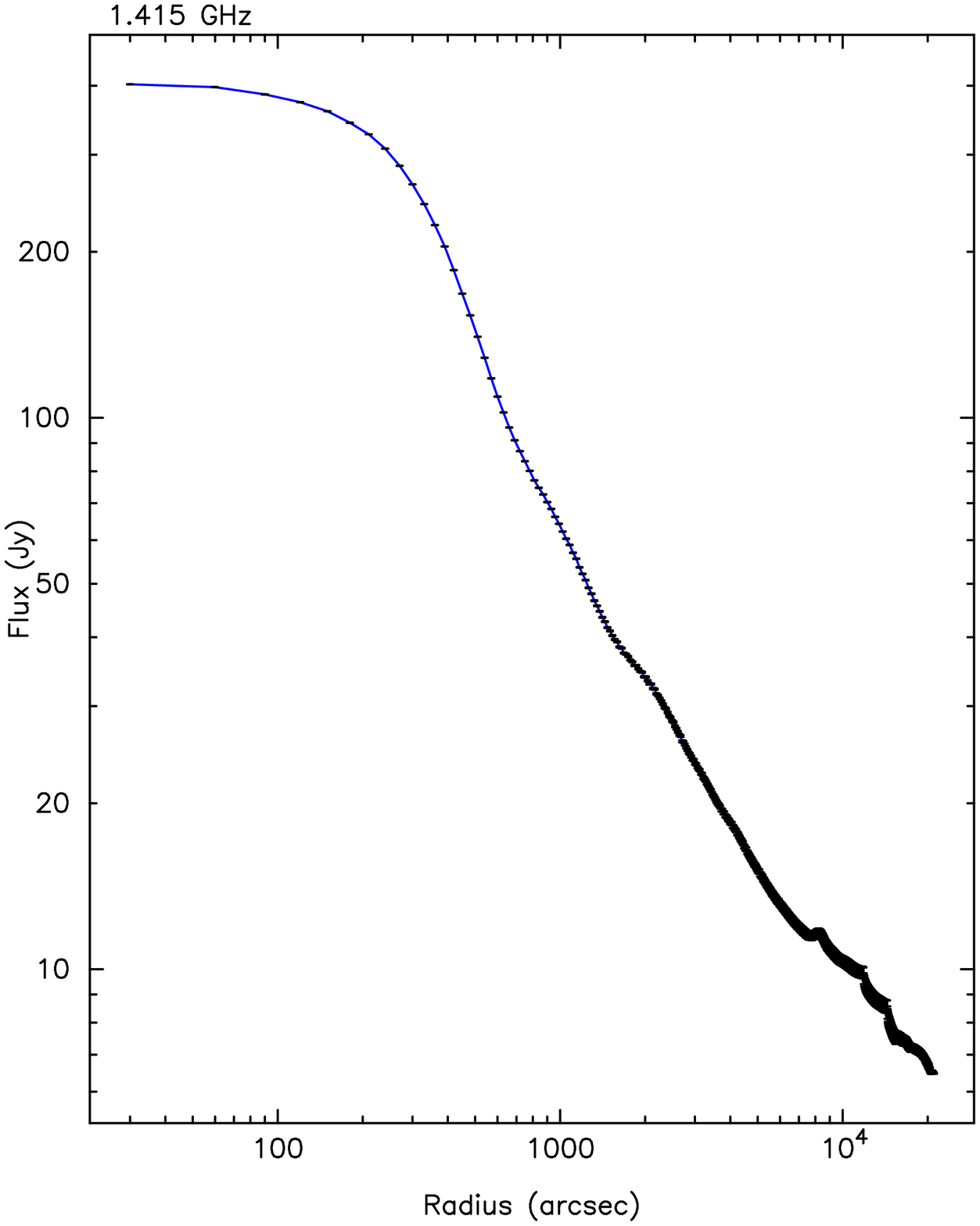}
\includegraphics[width=3in,angle=0]{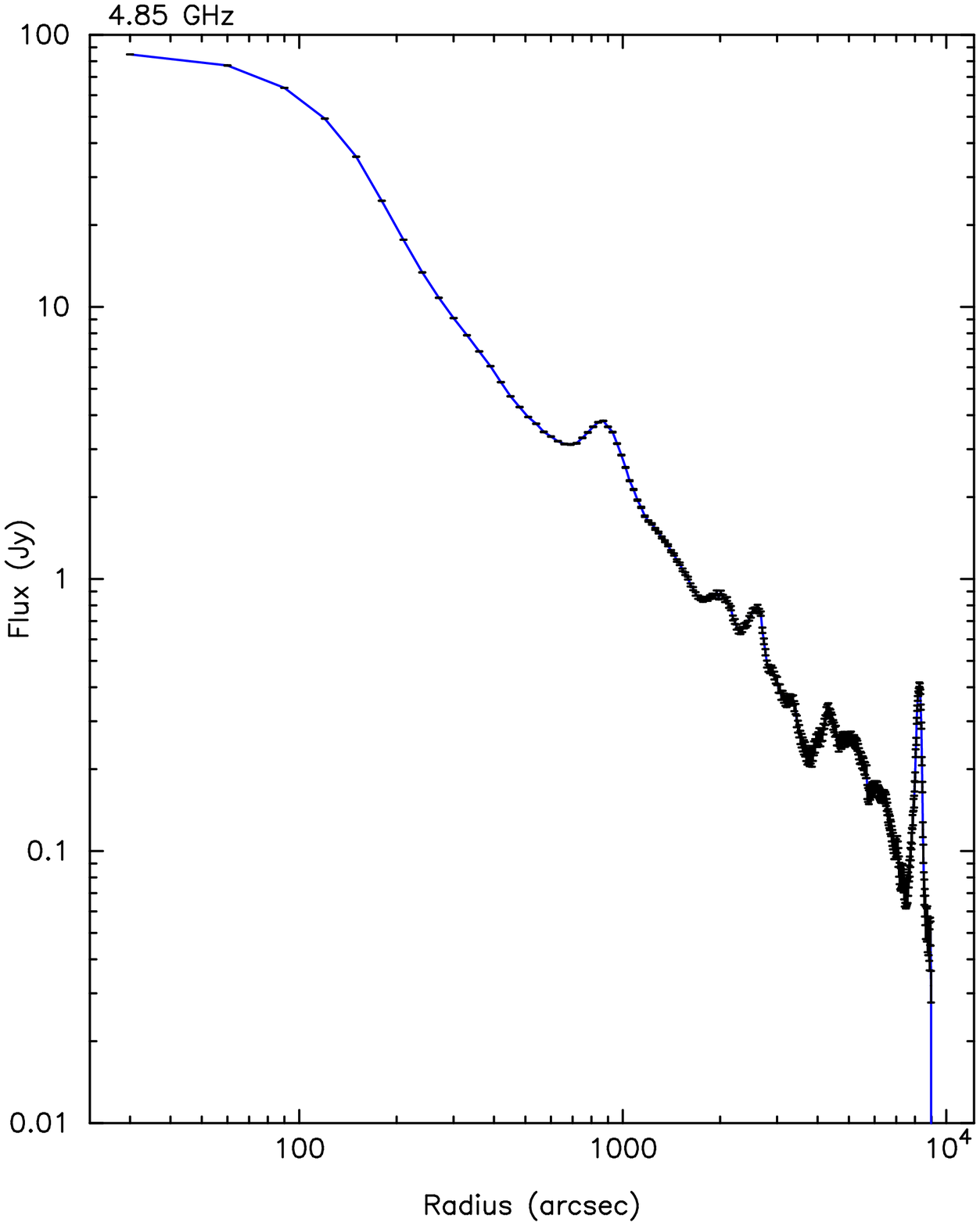}
\includegraphics[width=3in,angle=0]{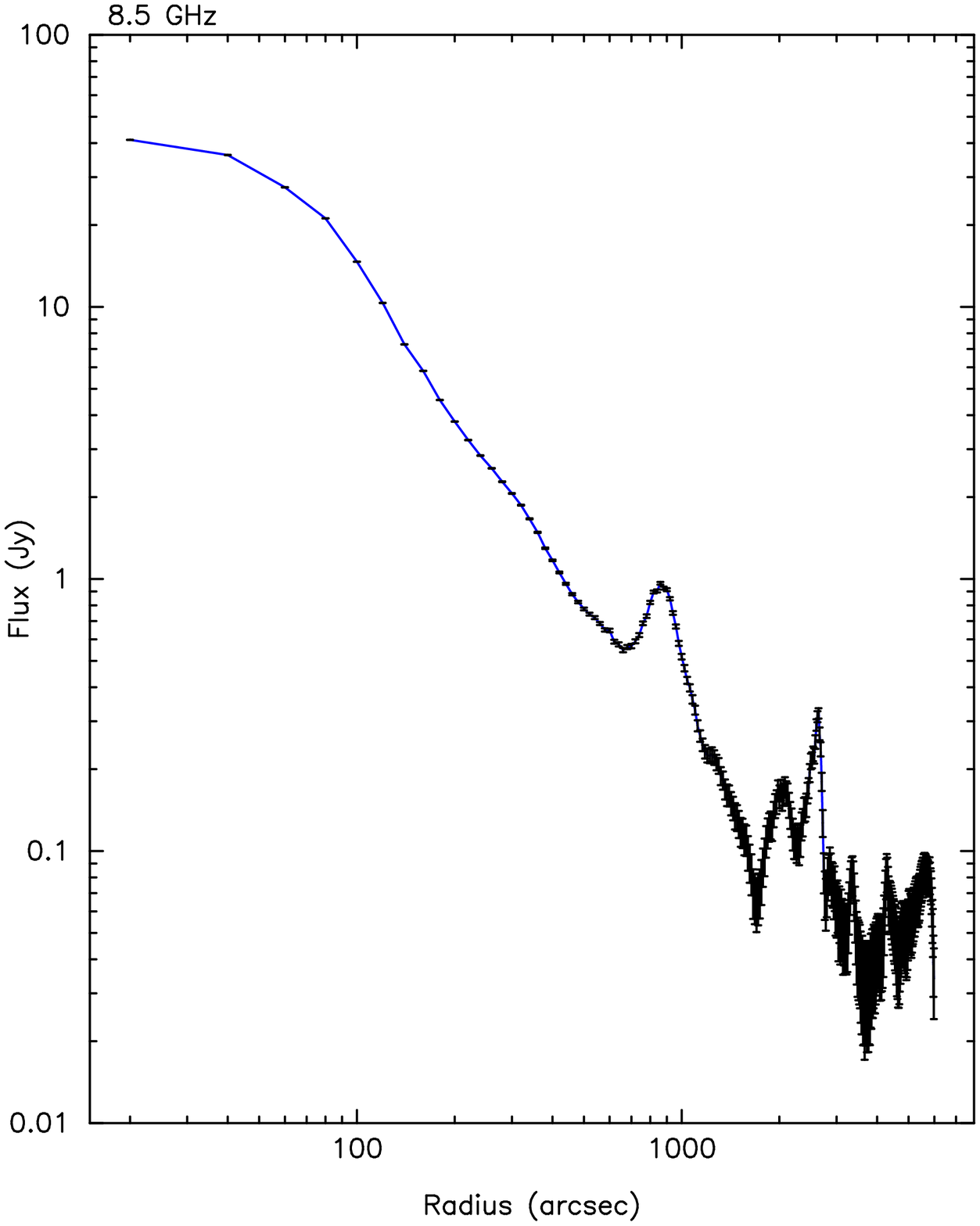}
\caption{
The distribution of flux as a function of radius   from the Galactic center is 
made by 
azimuthally averaging 
radial profiles of radio emission at 325 MHz, 1.415 GHz, 4.85 GHz and 8.5 GHz 
with spatial resolutions of 
$2328'', 539'', 
153''$ and $88''$ and are shown 
in (a), (b), (c) and (d), respectively. 
The  error bars of azimuthally averaged flux  are also superimposed. 
The plots and the errors are  logarithmic.
}\end{figure}

\begin{figure}
\center
\includegraphics[width=5in,angle=0]{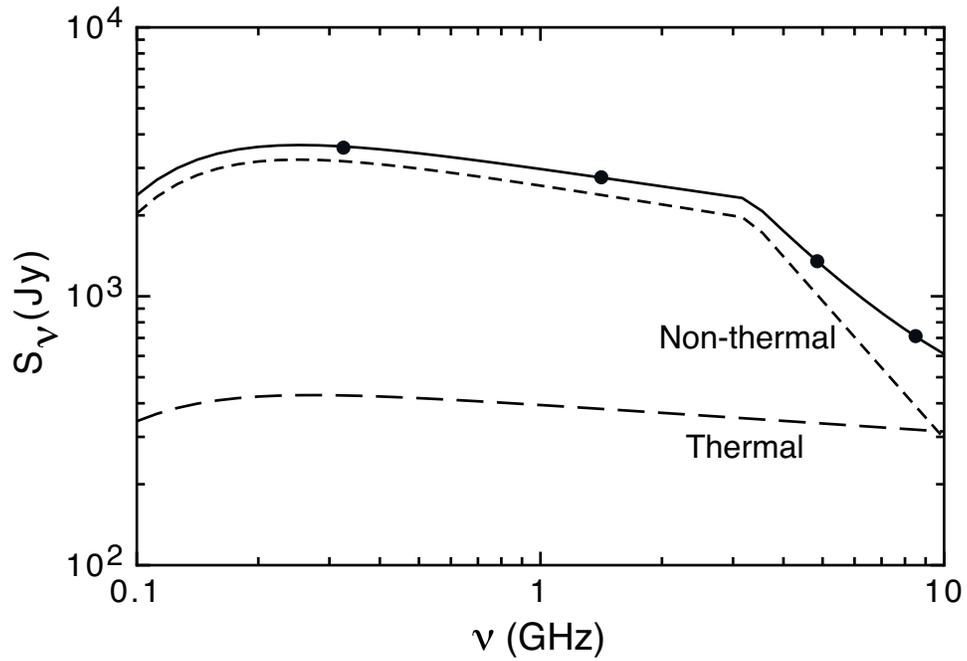}
\caption{
A plot of the decomposition of the diffuse radio flux from the Galactic center into thermal and 
nonthermal components.  
We fixed the thermal contribution at 4.85 GHz to be 25\% of the total flux 
and assumed a kinetic  temperature of 5000K.
We used a broken power-law for the  unabsorbed nonthermal emission which is considered 
to  lie behind the thermal screen.  The black dots represent the observed flux at a given frequency. 
}\end{figure}

\vfill\eject

\begin{figure}
\center
\includegraphics[scale=0.6,angle=0]{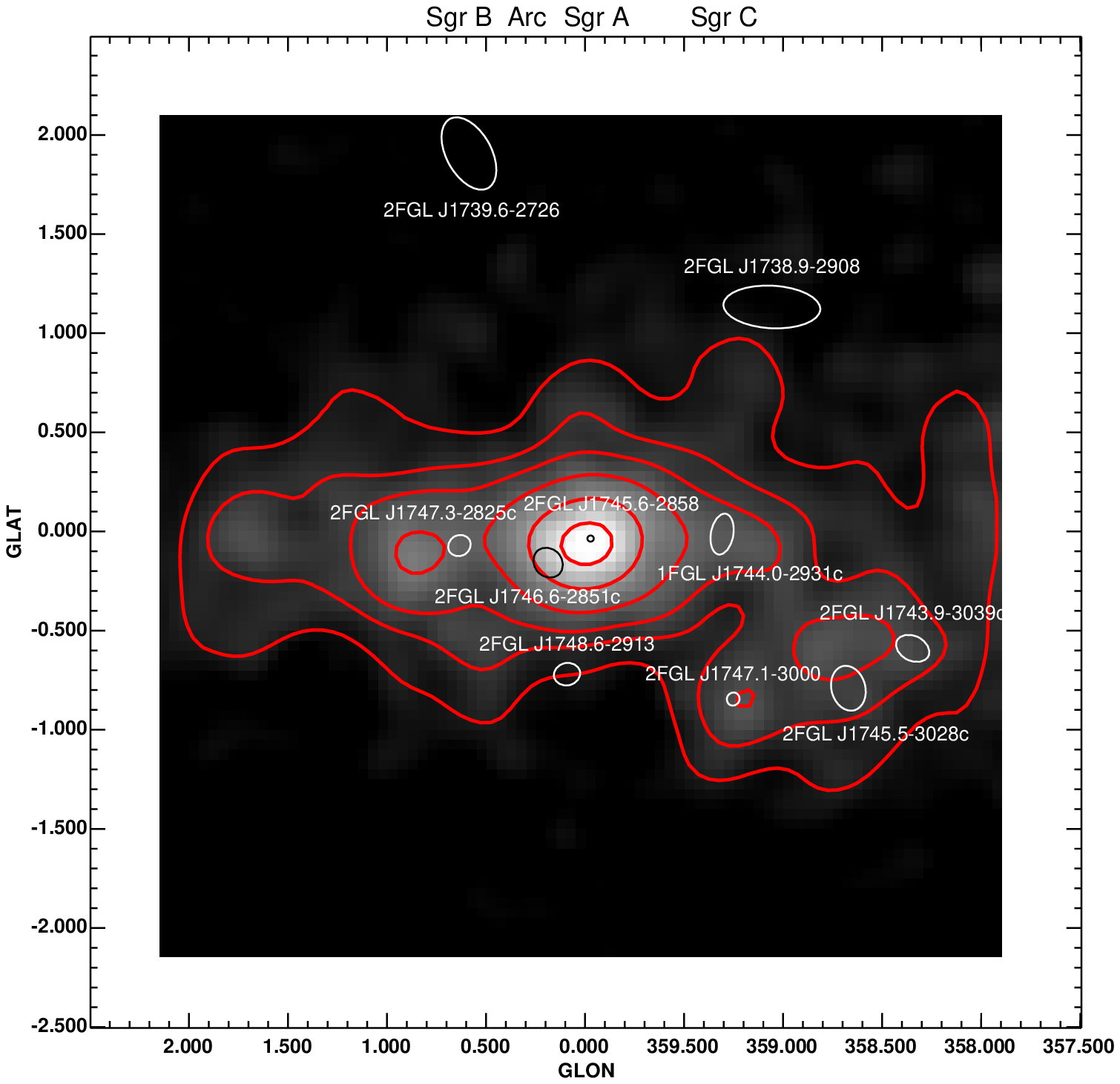}
\includegraphics[scale=0.6,angle=0]{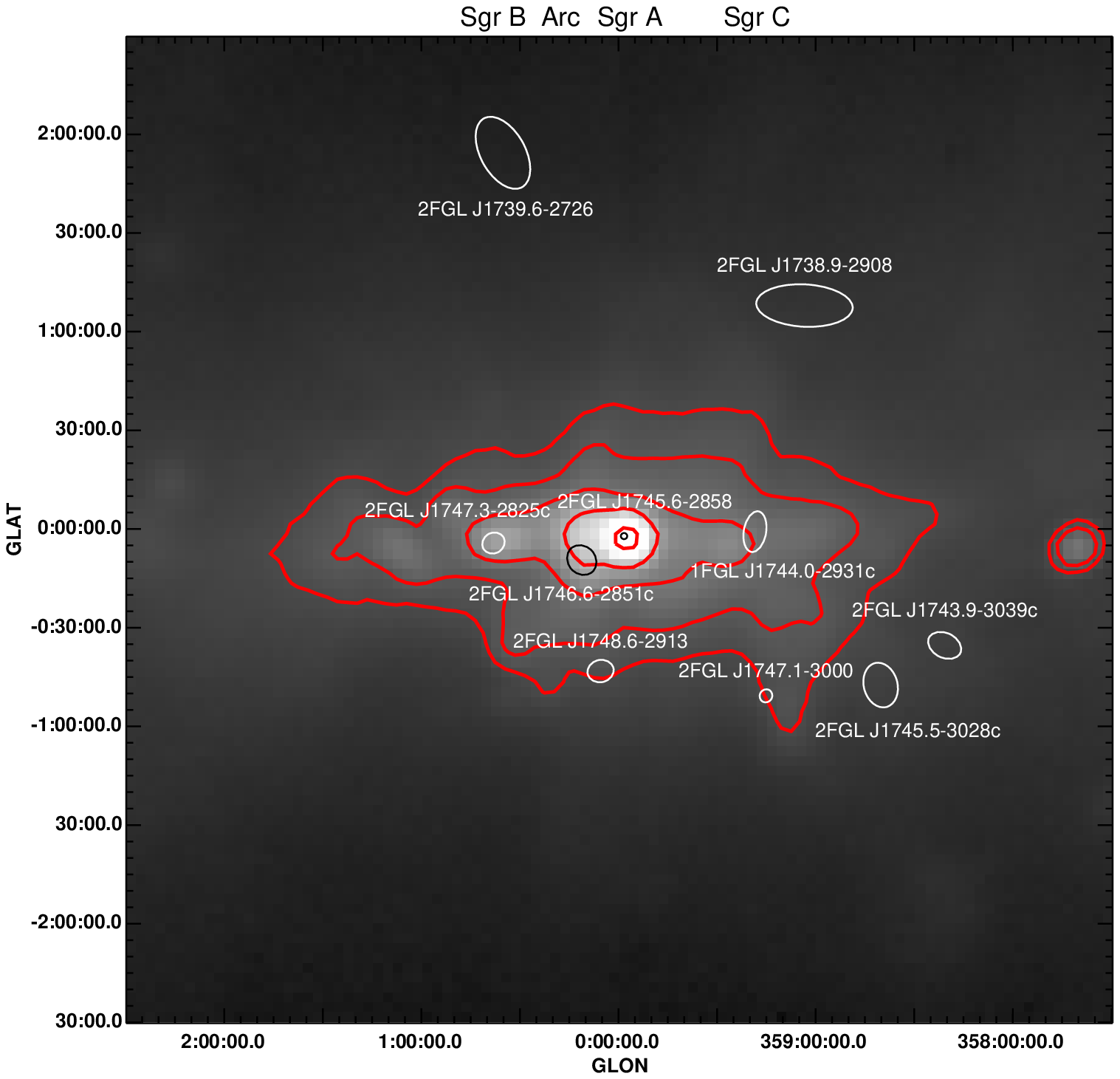}
\label{fig:fermi}
\caption{{\it (a - Top)} 
Contours of $Fermi$ LAT 
$\gamma$-rays between 1-300 GeV are 
superimposed on grayscale background subtracted image of
$Fermi$ LAT $\gamma$-rays between 1-300 GeV. 
The sources are
identified by the centroid 95
Contour levels are set at 2000, 3600, 4800, 6400, 9000, 13000 counts deg$^{-2}$.
{\it (b - Bottom)} 
Similar to (a) except that
contours of radio emission at 1.4 GHz based on GBT observations
are superimposed on a grayscale image at 1.4 GHz. 
The longitudes  of Sgr A, Sgr B, Sgr C and the radio Arc are labeled. 
Contour levels are set at 13, 20, 40, 100, 300 Jy beam$^{-1}$.
}\end{figure}


\begin{figure}
\center
\includegraphics[scale=0.55,angle=0]{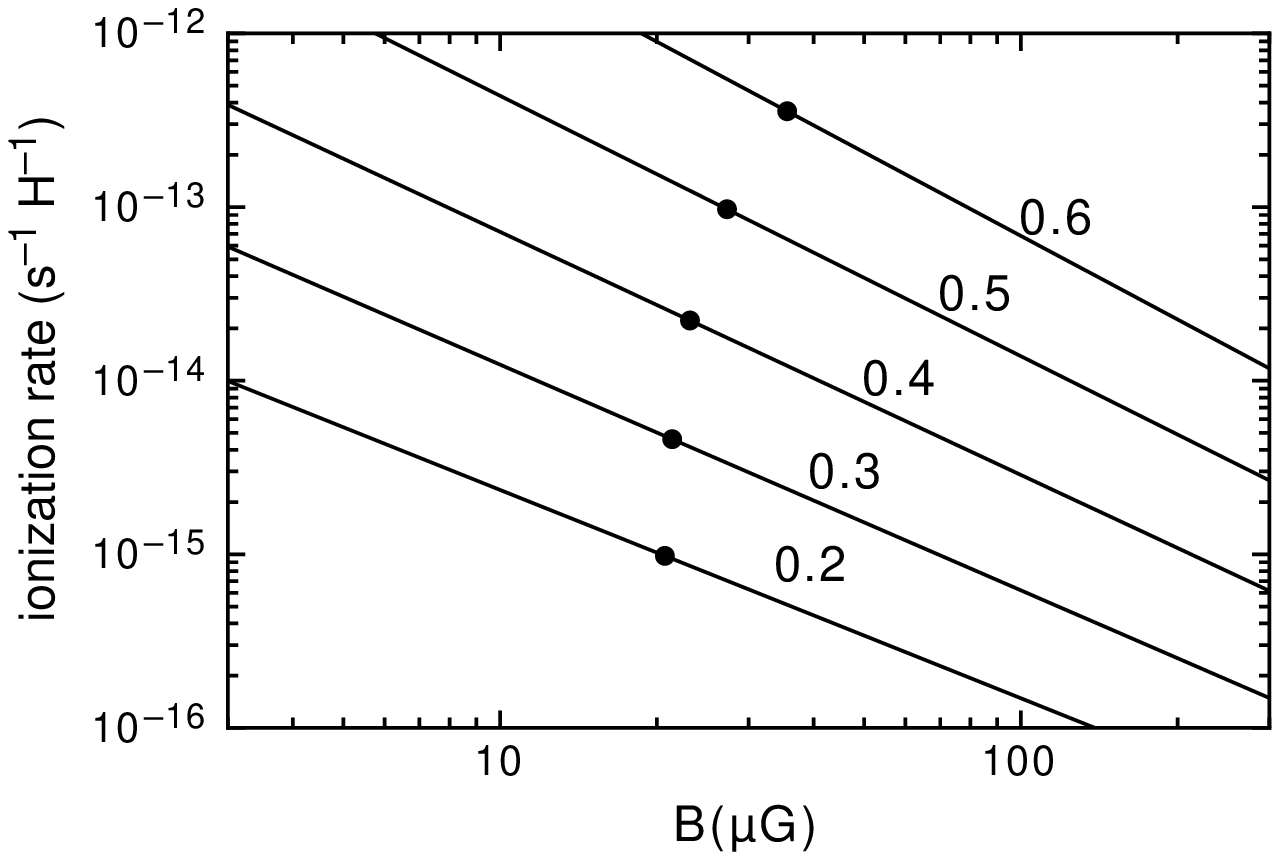}
\includegraphics[scale=0.55,angle=0]{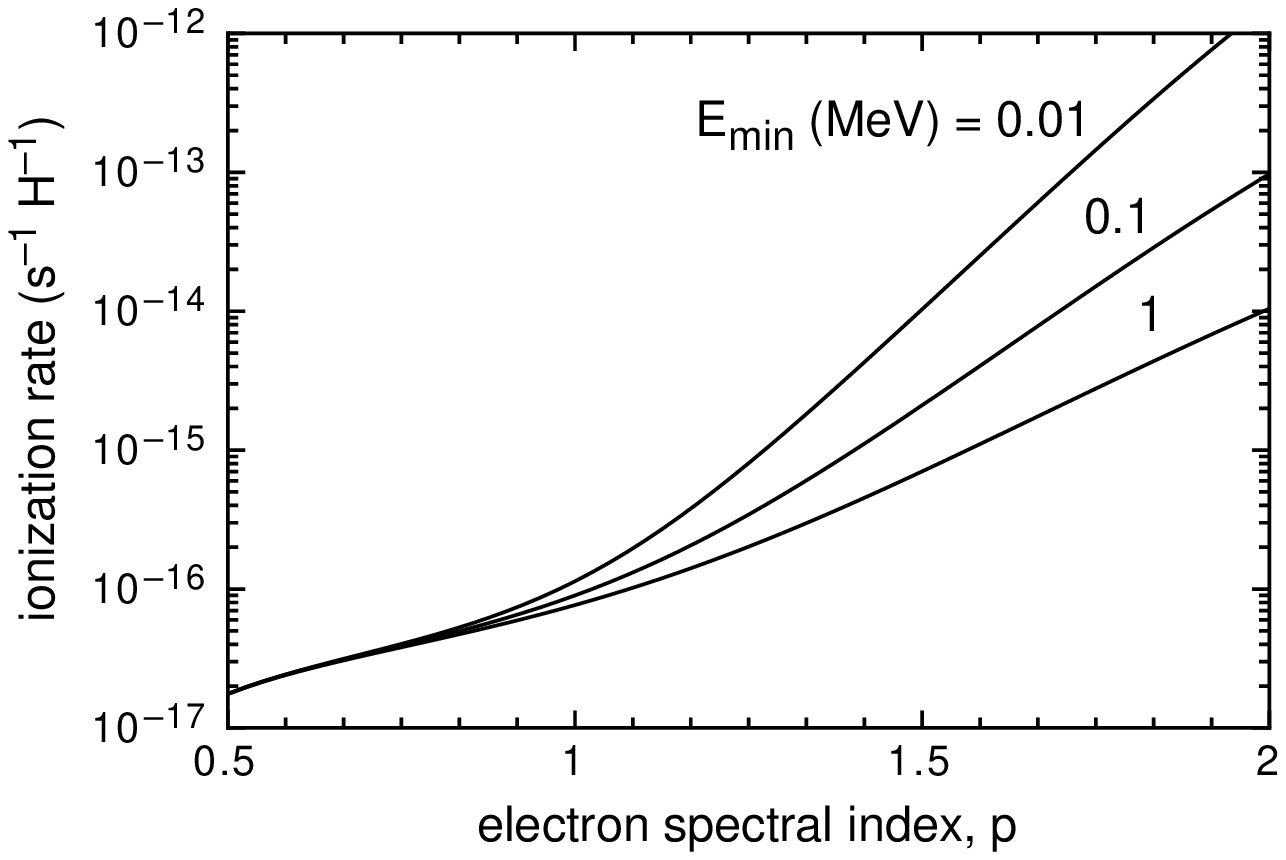}\\
\label{fig:ionization}
\includegraphics[scale=0.9,angle=0]{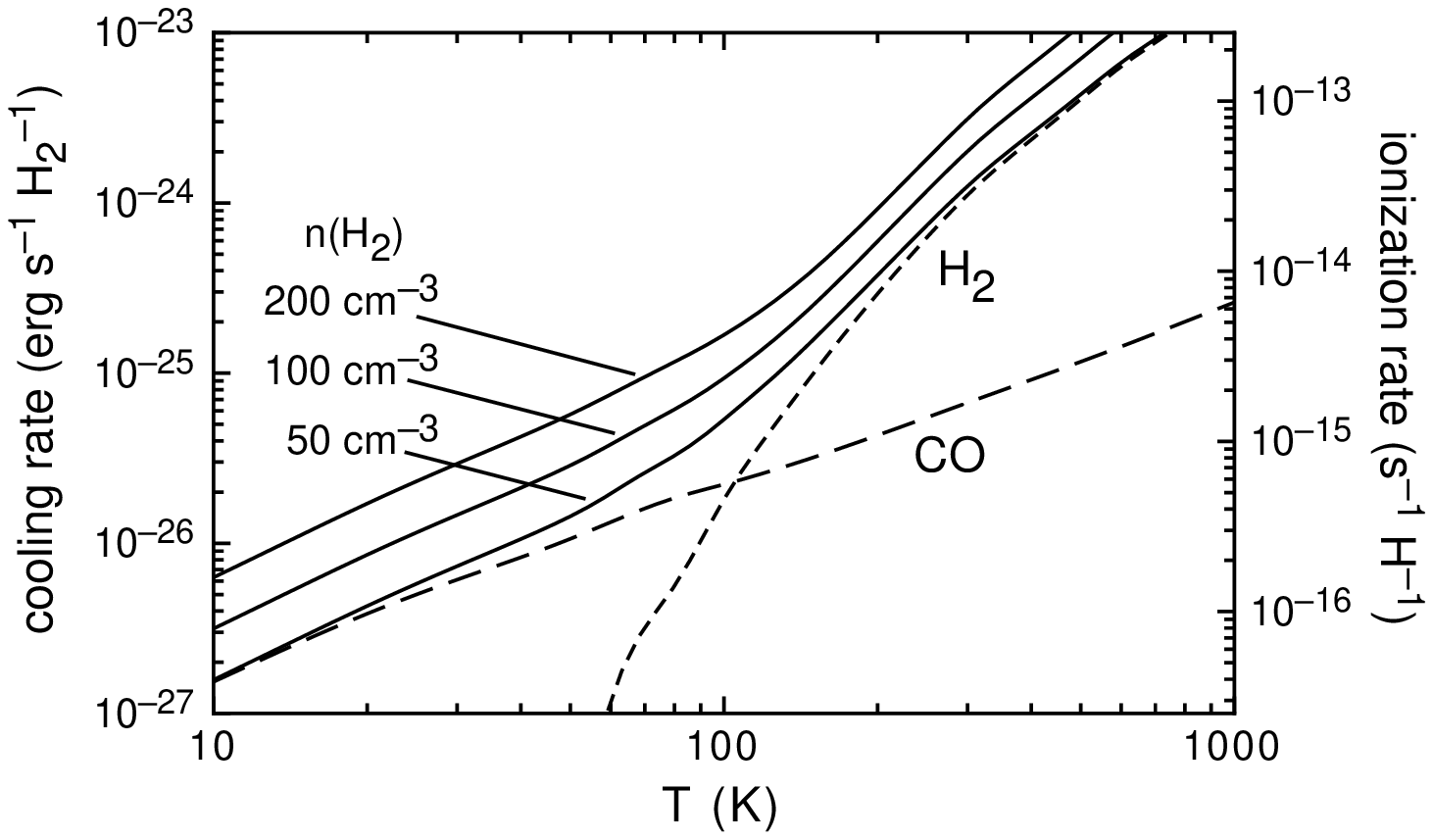}
\caption{
{\it (a - Top Left)}
The variation of cosmic ray ionization rate as a function of the magnetic field 
for different values of the spectral  index of the 
radiation  $\alpha$. 
The black dot on each curve gives the value at which the magnetic field 
and particle energies  are in equipartition. 
{\it (b - Top Right)} 
This plot shows how the ionization rate inferred from the nonthermal diffuse Galactic center flux at 325 
MHz depends on electron energy spectral index p and the lower energy cutoff E$_{min}$.
{\it (c - Bottom)} 
Solid curves show the total cooling rate for diffuse molecular gas for H$_2$ densities
of 50, 100, and 200\,cm$^{-3}$.  Only the dominant cooling, by rotational transitions of H$_2$ and
CO, has been included; these contributions for $n(\mathrm{H}_2)=50\,\mathrm{cm}^{-3}$ are
indicated by the short-dashed and long-dashed curves, respectively.  A gas-phase abundance of
CO/H$_2 = 2.8\times 10^{-4}$ has been adopted,  appropriate for a metallicity twice that of the
sun.  The right hand axis shows the ionization rate by cosmic-ray electrons needed to supply the
corresponding heating rate, assuming that each ionization is associated with the deposition of
12.4\,eV of heat (see text).
}\end{figure}

\begin{figure}
\center
\includegraphics[scale=0.8,angle=0]{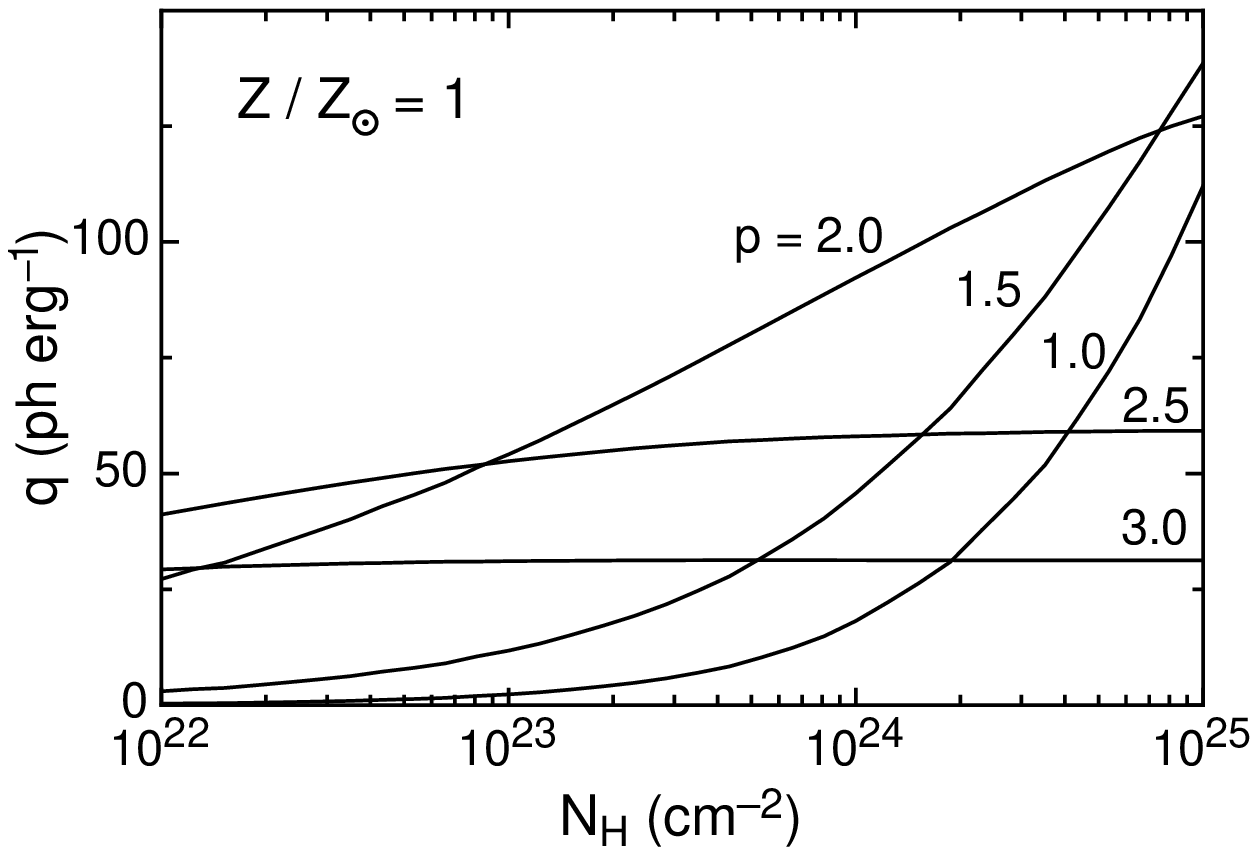}
\label{fig:q-NH}
\center
\includegraphics[scale=0.8,angle=0]{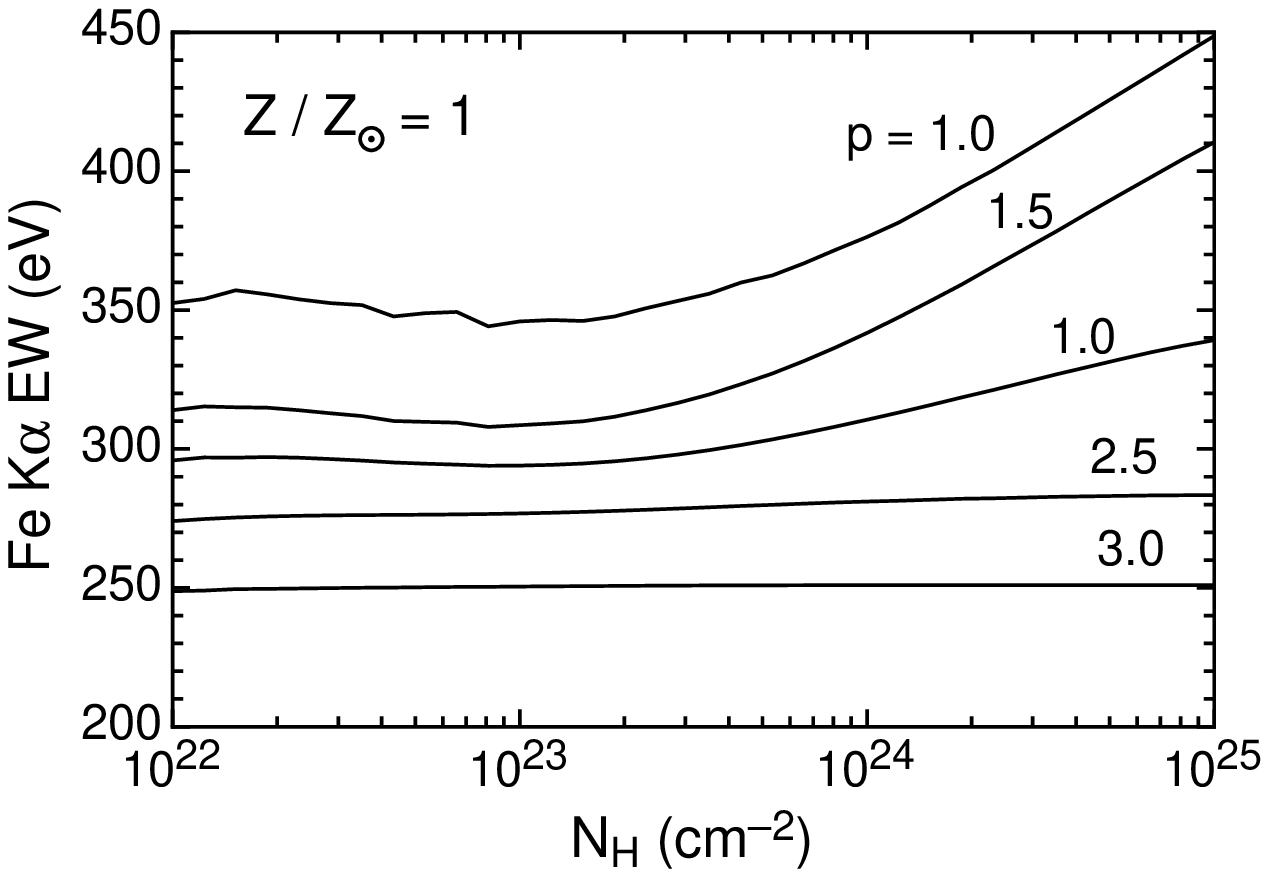}
\label{fig:EW-NH}
\caption{
{\it (a - Top )}
Fe K$\alpha$ production per erg of electron energy injected
into a cloud of given column density $N_\mathrm{H}$ and a solar iron
abundance (Fe/H = $2.8\ee -5 $).  The curves are
labeled by the power-law index $p$ of the electron energy spectrum
($\propto E^{-p}$), which is assumed to run from 10\,keV to 1\,GeV.
{\it (b - Bottom )}
Similar to (a) except that the
EW of Fe K$\alpha$ is shown as a function
of column density for different values of the energy  spectral index.
}\end{figure}

\begin{figure}
\center
\includegraphics[scale=0.25,angle=0]{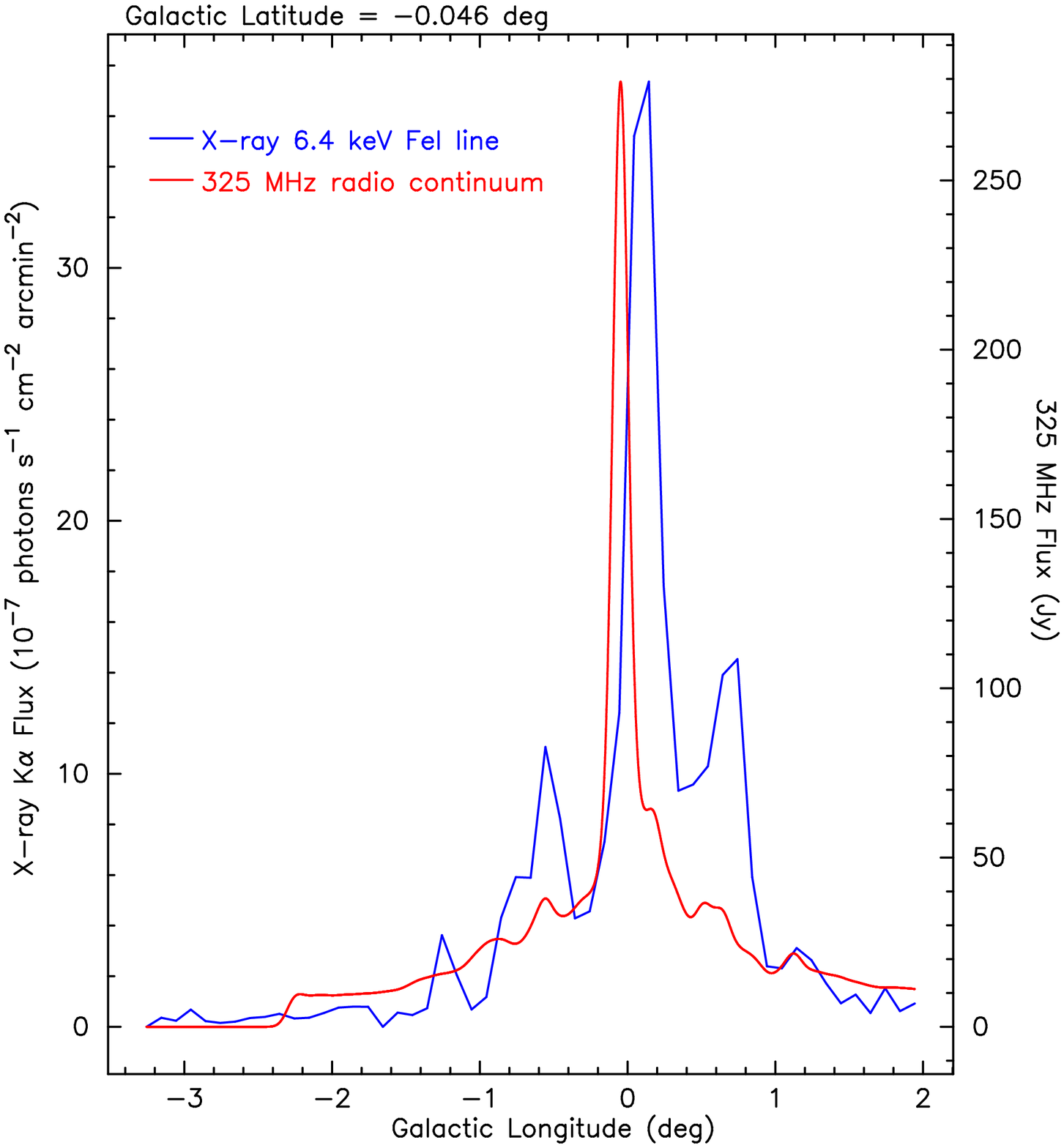}
\includegraphics[scale=0.25,angle=0]{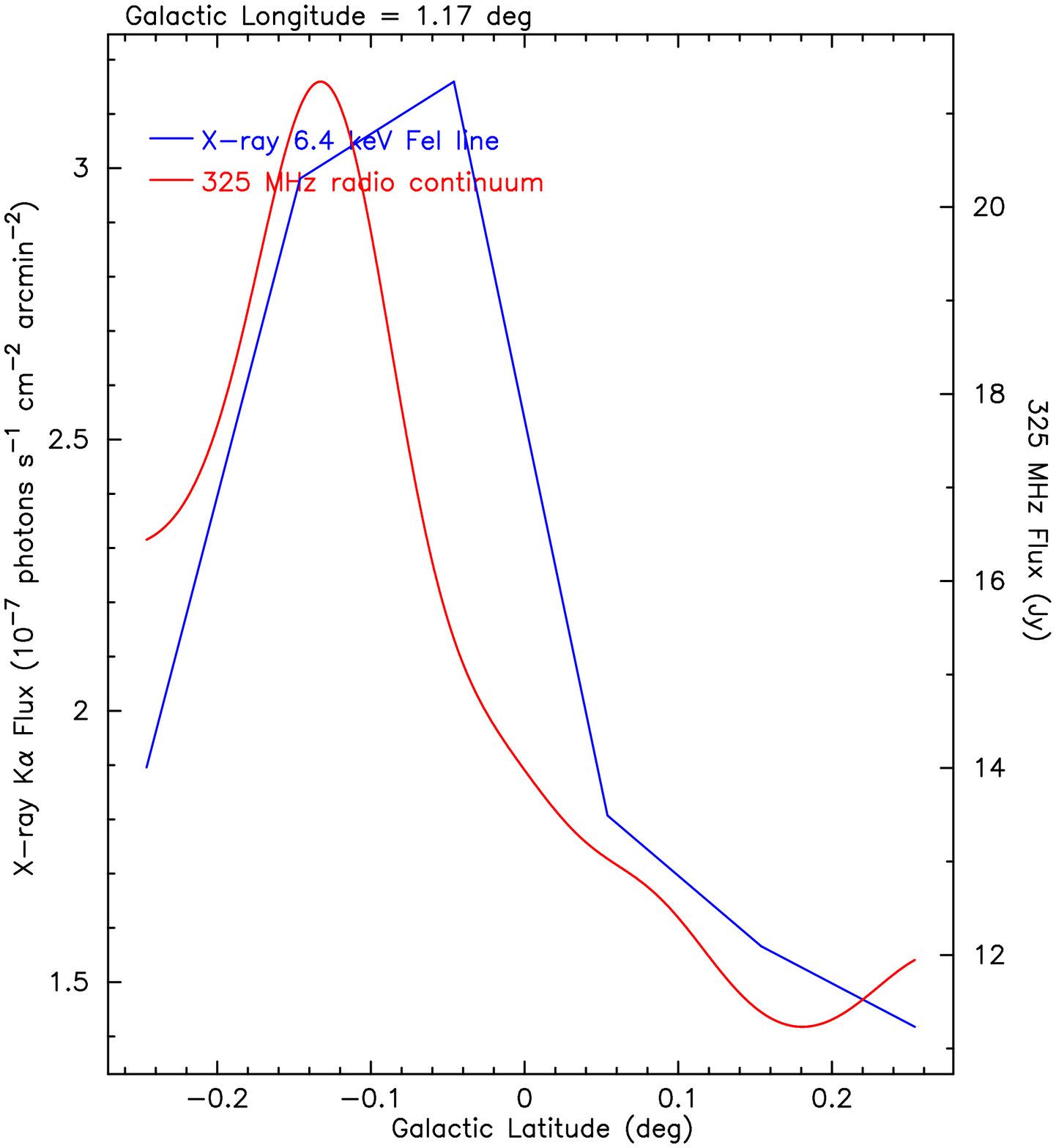}
\includegraphics[scale=0.25,angle=0]{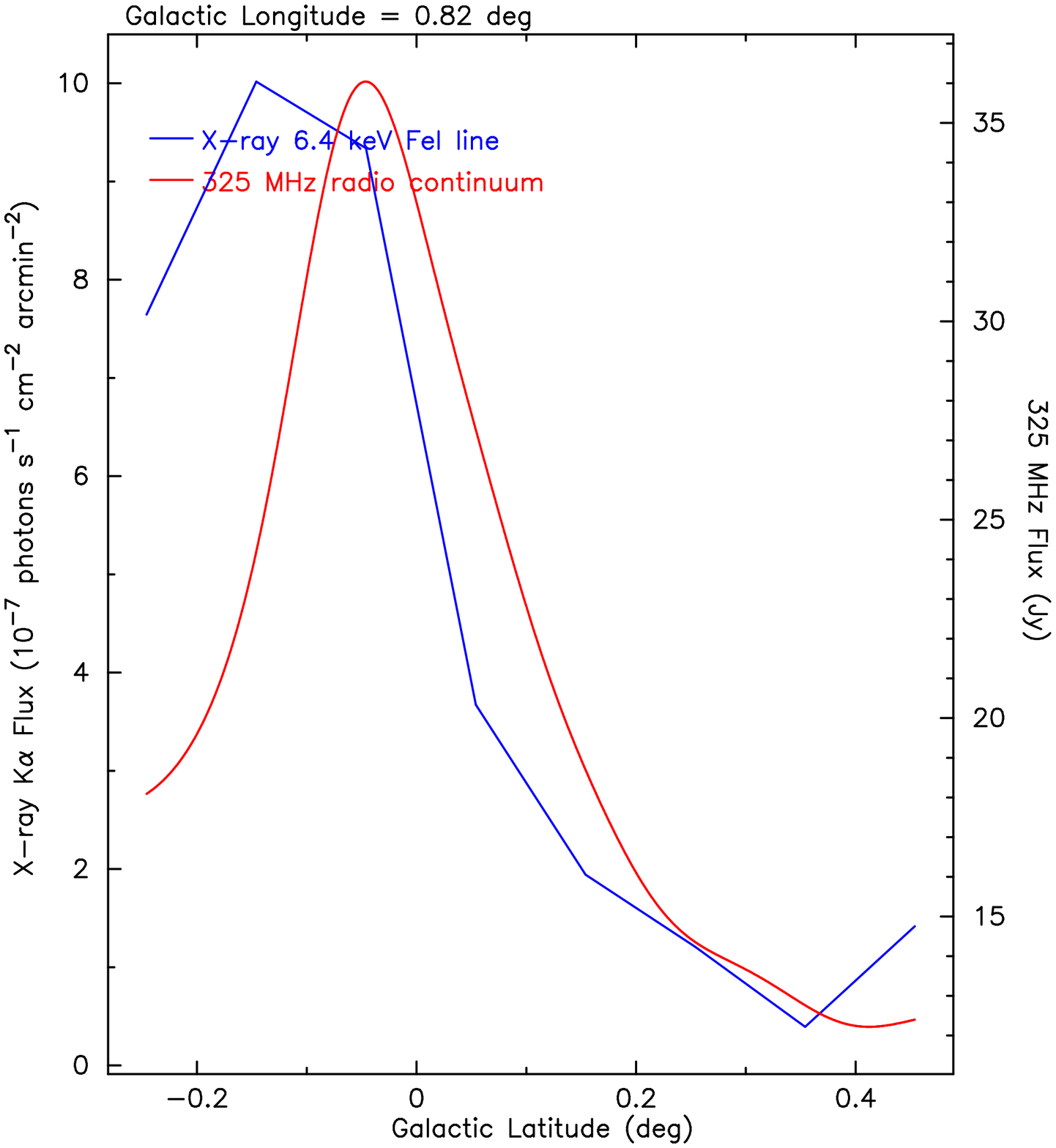}\\
\includegraphics[scale=0.25,angle=0]{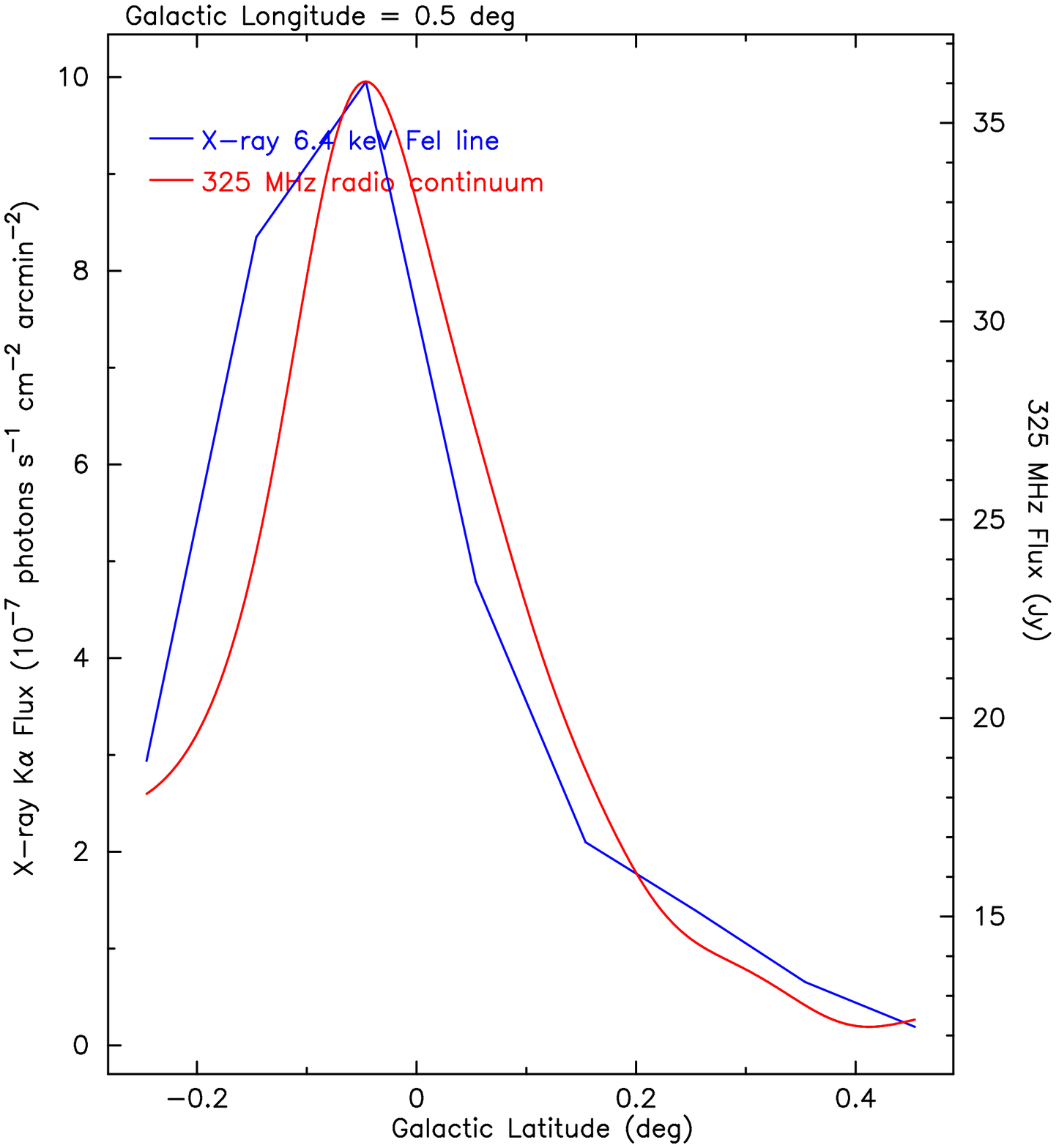}
\includegraphics[scale=0.25,angle=0]{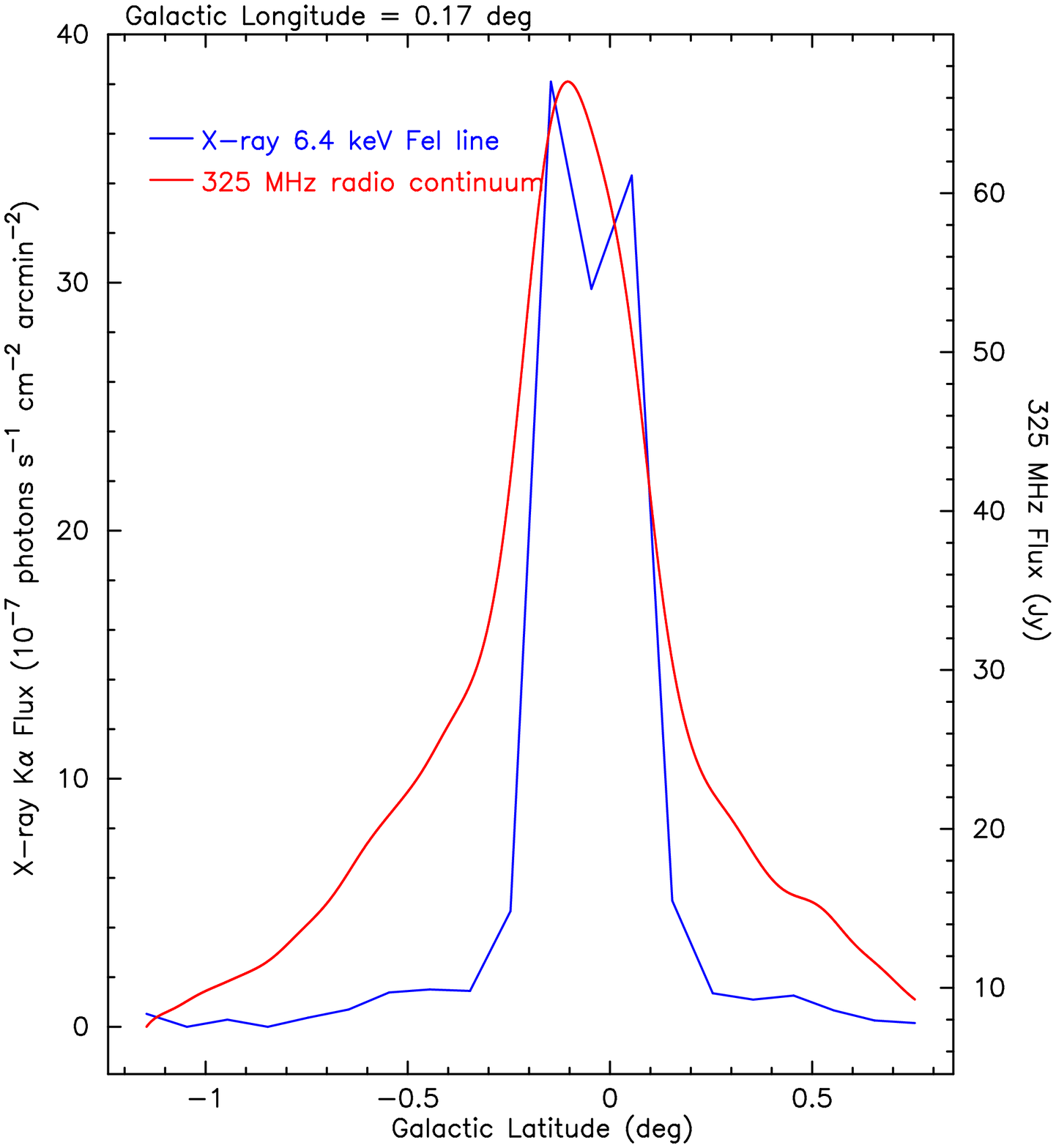}
\includegraphics[scale=0.25,angle=0]{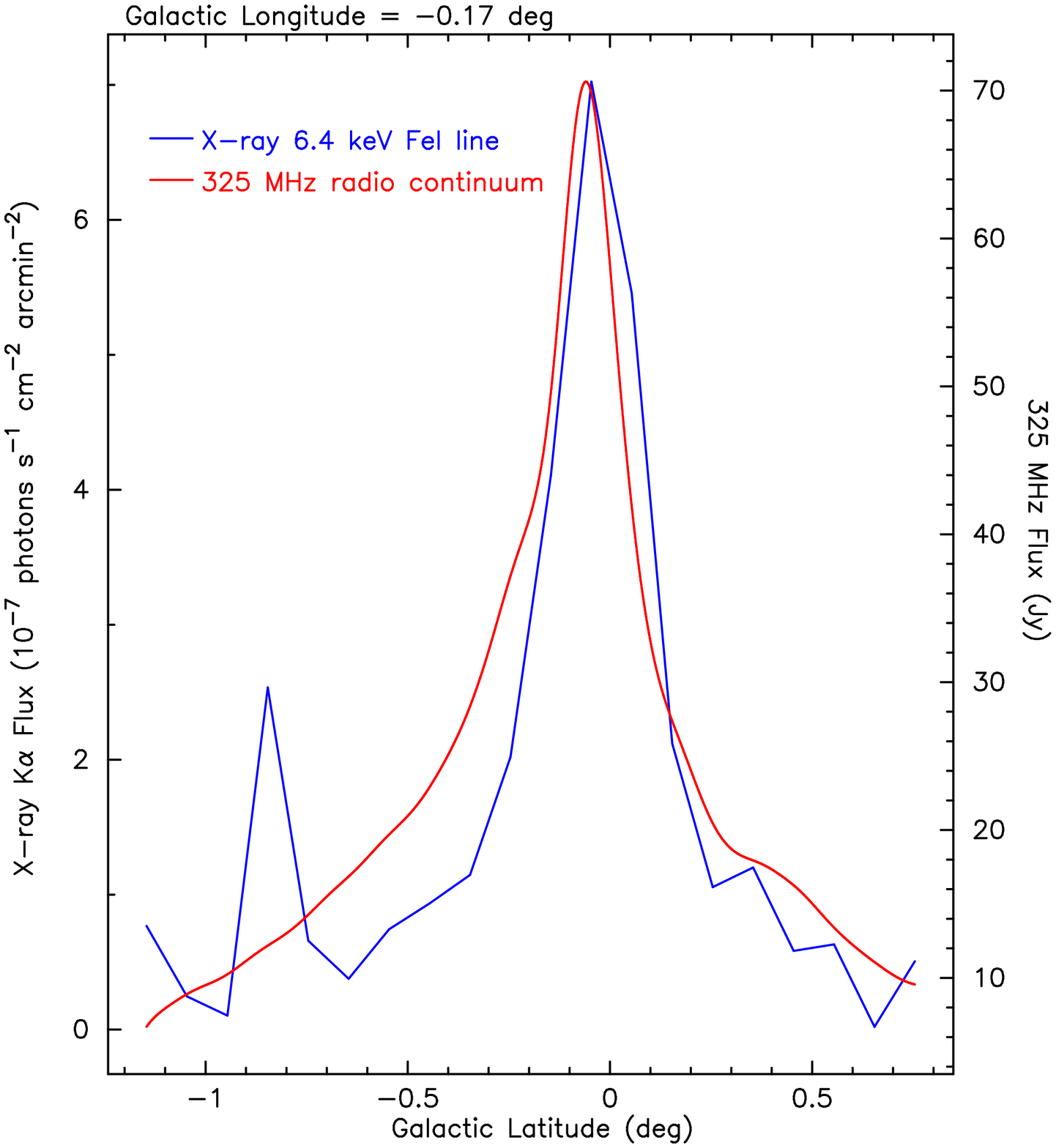}\\
\includegraphics[scale=0.25,angle=0]{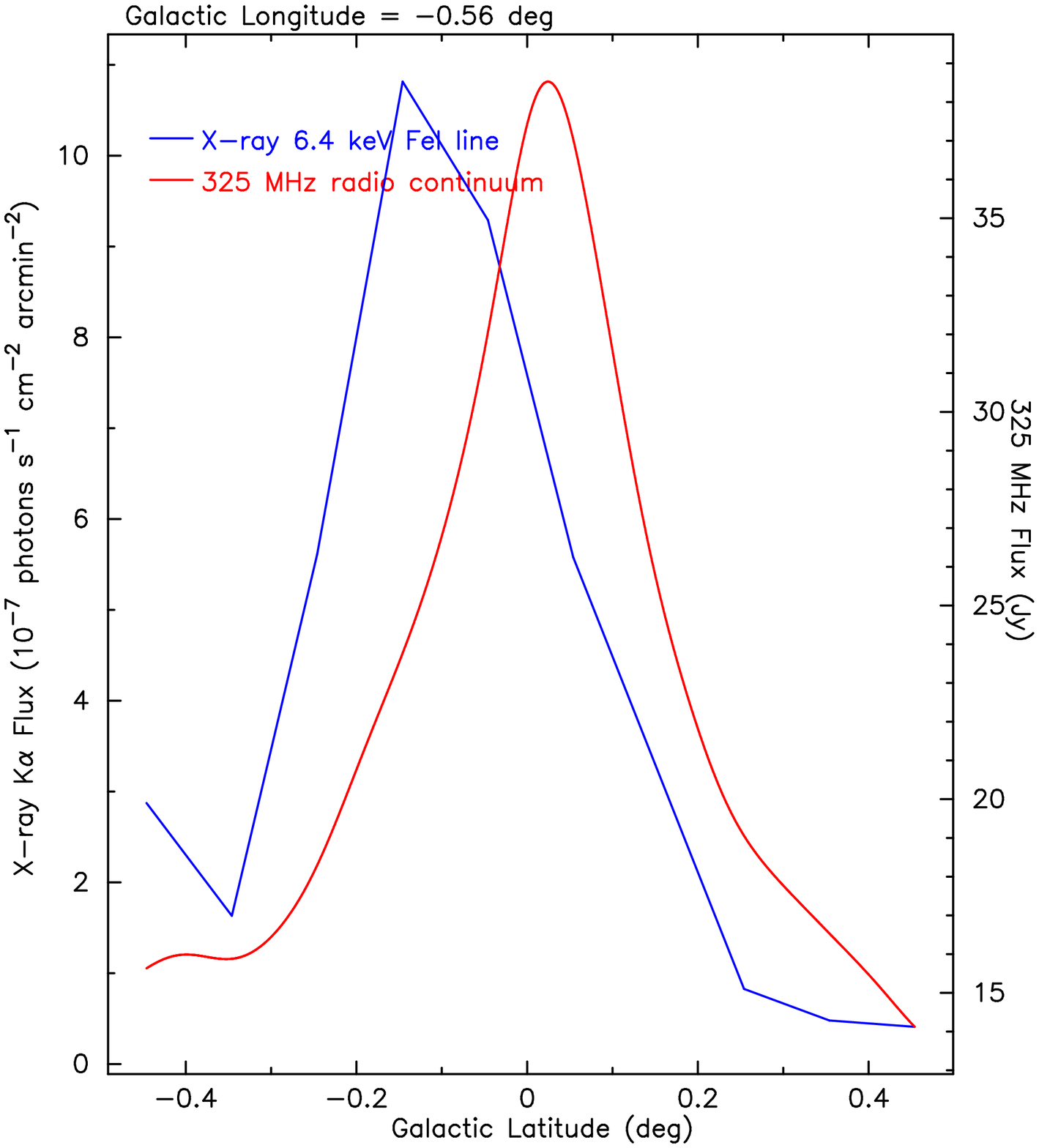}
\includegraphics[scale=0.25,angle=0]{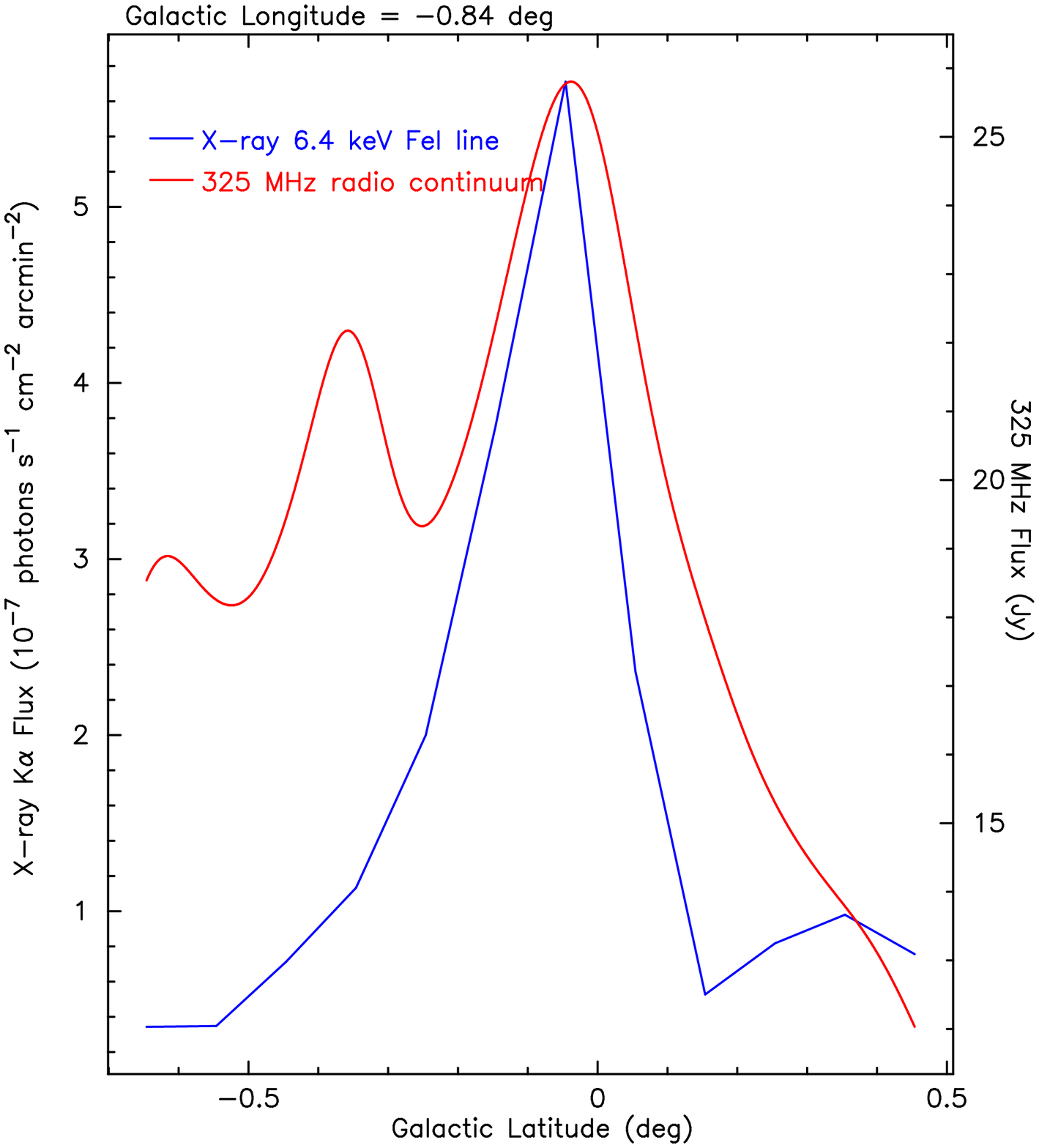}
\includegraphics[scale=0.25,angle=0]{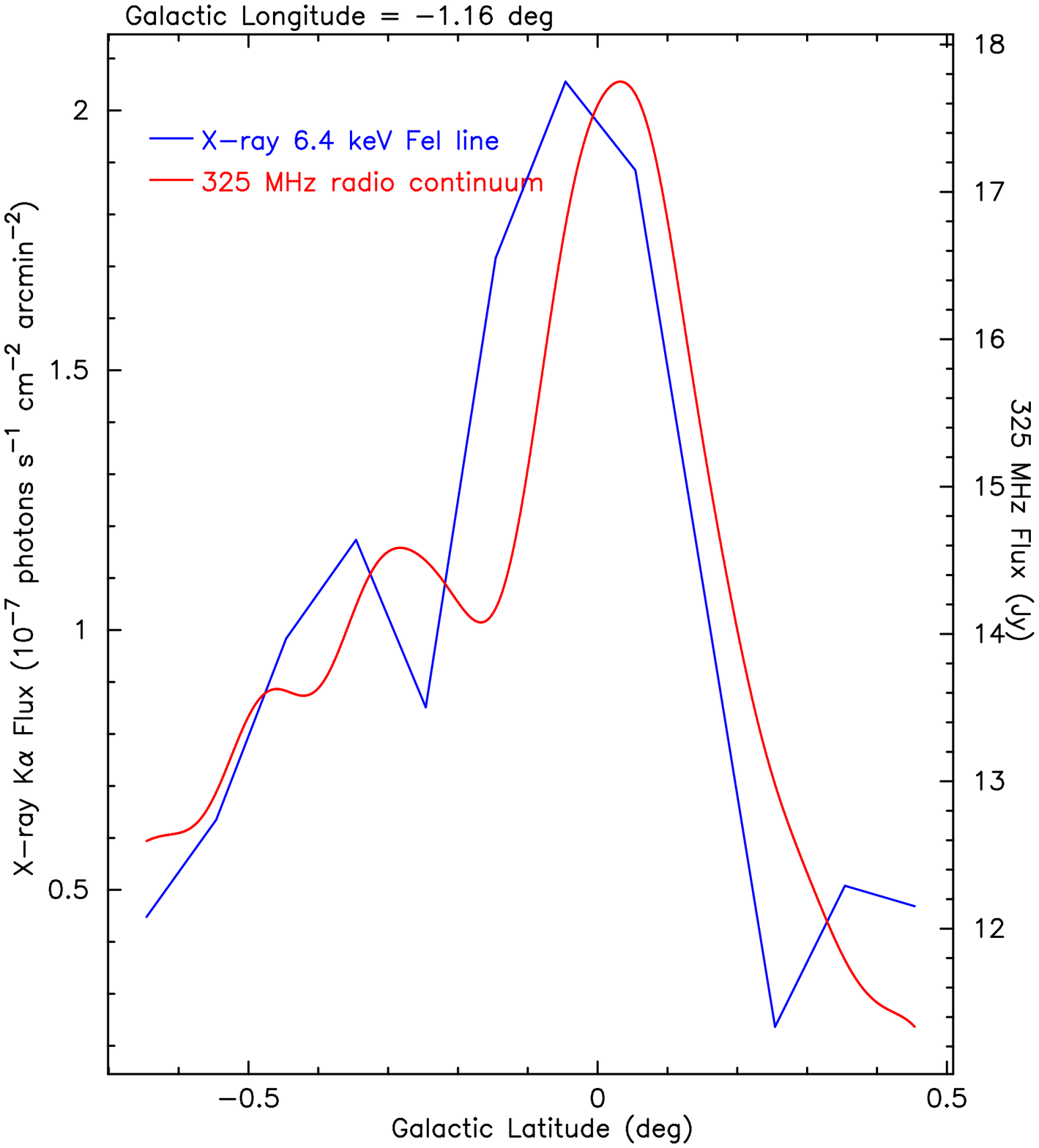}
\label{fig:plot_xray_cuts}
\caption{
{\it (a - i)}
Cross cuts made on the 
distribution of K$\alpha$ line (blue)  and 325 MHz (red) 
with different position angles. 
The 325 MHz data is based on GBT and VLA observations 
that have been combined before the map was convolved with 
a 6$'$ Gaussian in order to match the distribution of 
X-ray flux measured with Suzaku. 
There is  one cross cut at constant latitude b=$-0.046$\deg\ shown in (a) 
whereas eight other   cross cuts are shown in (b--i) corresponding 
to  constant longitudes. 
}\end{figure}

\begin{figure}
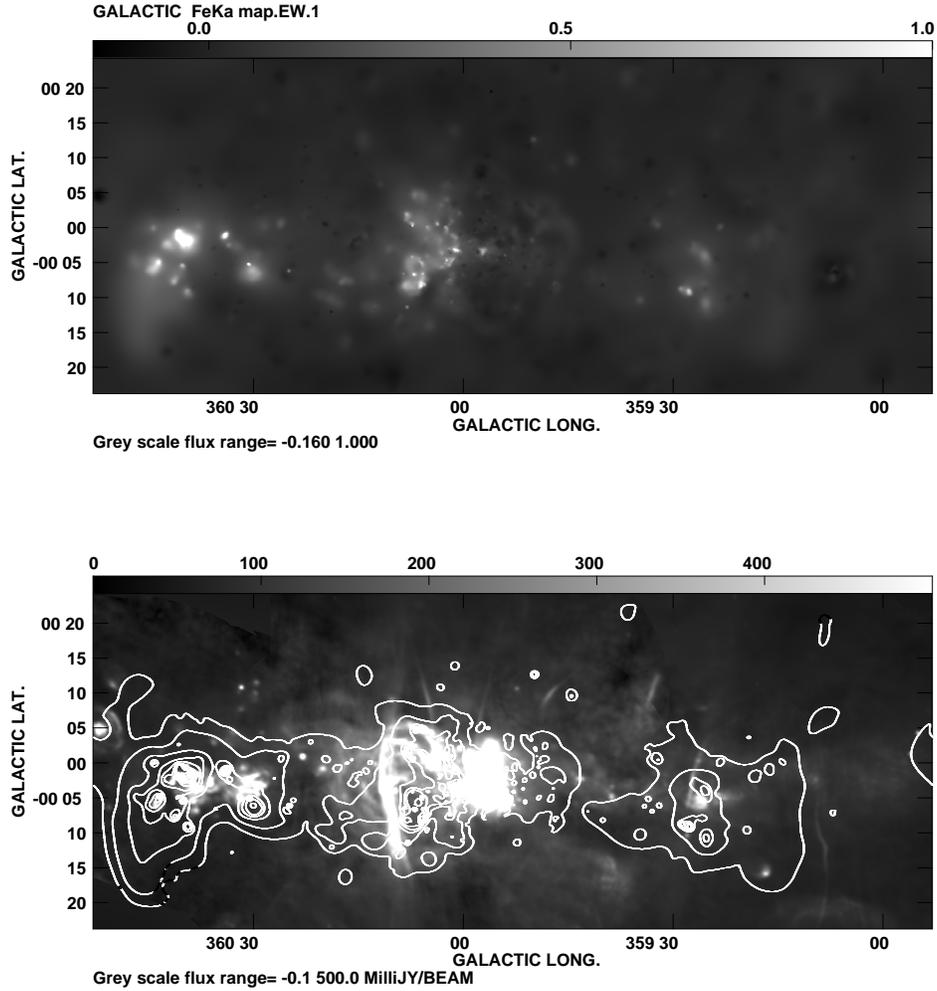

\center
\includegraphics[scale=0.5,angle=-90]{f8a_fekaew_ew.ps}
\center
\includegraphics[scale=0.5,angle=-90]{f8b_20cm_fekaew_cont.ps}
\caption{
{\it (a - Top)}
An adaptively smoothed  greyscale image of the EW of FeI K$\alpha$ line 
emission 
from the inner -0.8\deg$<l<0.7$\deg. The range is between 50 to 2$\times10^3$  eV. 
{\it (b - Bottom)} 
Contours of the EW of FeI K$\alpha$ line emission with values 
set at
$100\times$ (0.5, 1, 2, 3, 4, 6, 8, 10, 14, 18 and 20) eVs
are  superimposed 
on a greyscale continuum image at 1.4 GHz with a resolution of 30$''$. 
}\end{figure}

\begin{figure}
\center
\includegraphics[scale=0.4,angle=0]{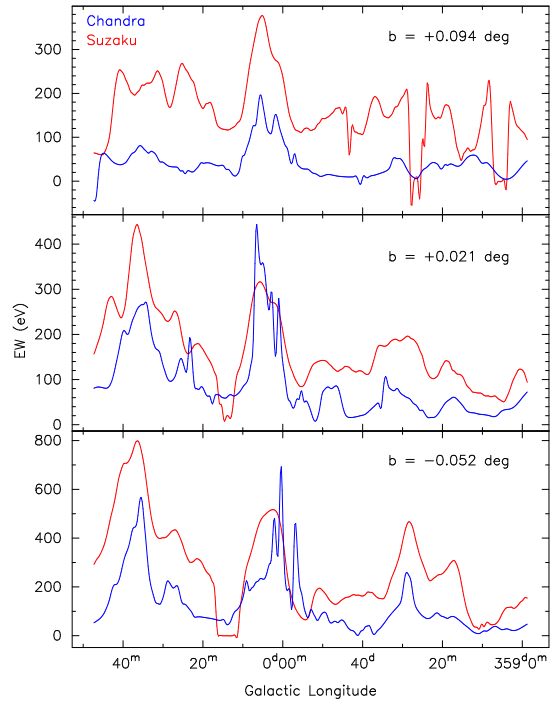}
\caption{
Plots of cross cuts of EW as a function of Galactic longitudes 
at three different latitudes. 
in blue and red corresponding to 
Chandra and Suzaku measurements.  
Each slice is  an 
average of all the data in 4$'$ of galactic latitude 
and a range of galactic longitude divided in  19 steps  between $l$=53$'$ to $l$=--53$'$. 
A part of the Suzaku data are missing  because regions of bright point
sources, 2E 1743.1-2842, 2E 1742.9-2929 and 2E 1740.7-2943 are masked.
}\end{figure}


\begin{figure}
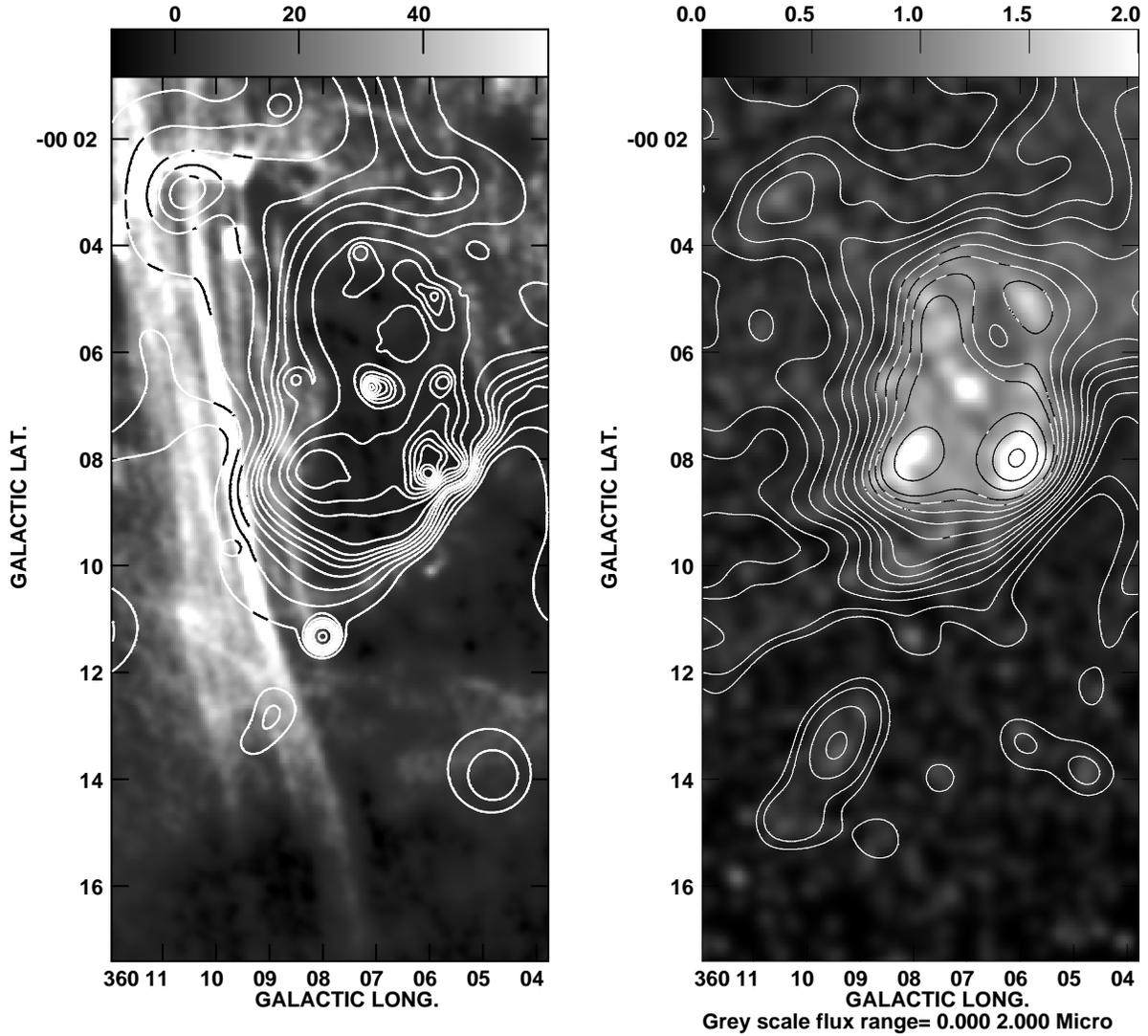

\center
\includegraphics[scale=0.65,angle=0]{f10a_feka_ew_new.ps}
\includegraphics[scale=0.65,angle=0]{f10b_arc_fekaflux.ps}
\caption{
{\it (a - Left)}
Contours of EW based on Chandra observations 
with levels set at 
100$\times$ (1.1, 1.2, 1.5, 1.75, 20, 22.5, 25, 30, 40, 50, 60, 70, 80, 90, 100, 110) eV
are superimposed on a grayscale continuum image at 1.415 GHz 
based on VLA observations. 
{\it (b - Right)} 
Similar to (a) except that contours of K$\alpha$ line emission are
superimposed  on its grayscale.
}\end{figure}

\begin{figure}
\center
\includegraphics[scale=0.5,angle=0]{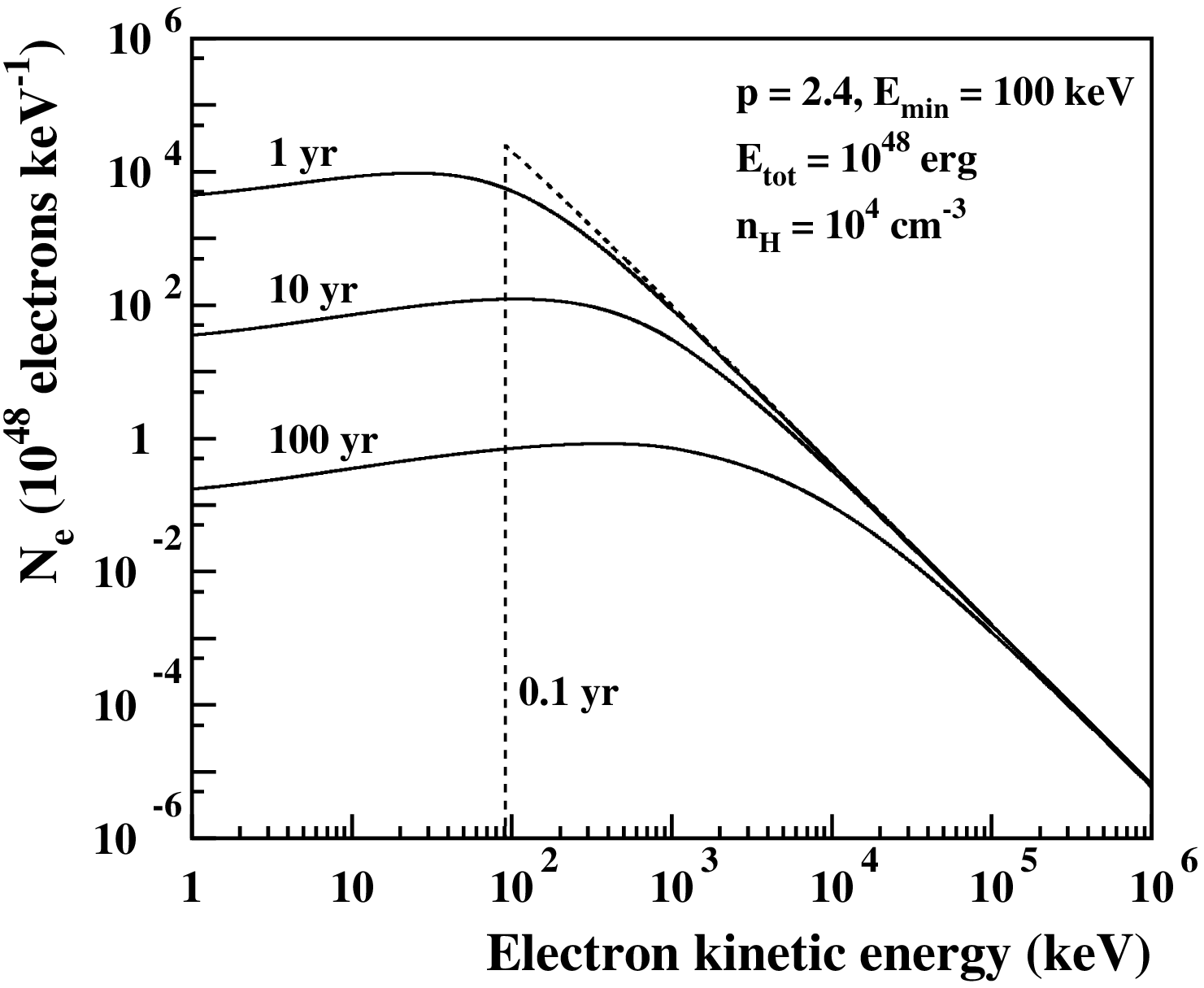}
\label{fig:spece3}
\includegraphics[scale=0.5,angle=0]{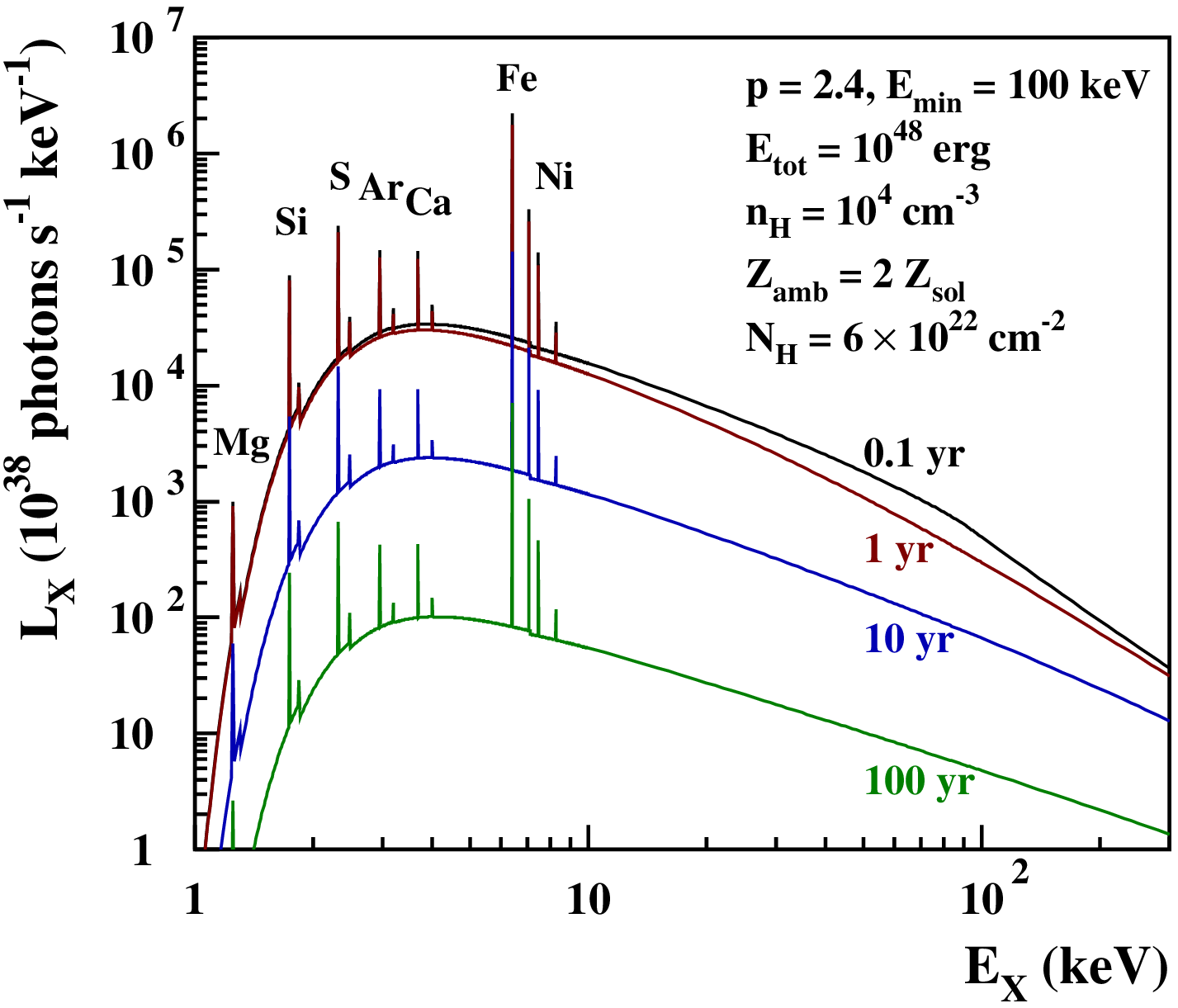}
\center
\includegraphics[scale=0.5,angle=0]{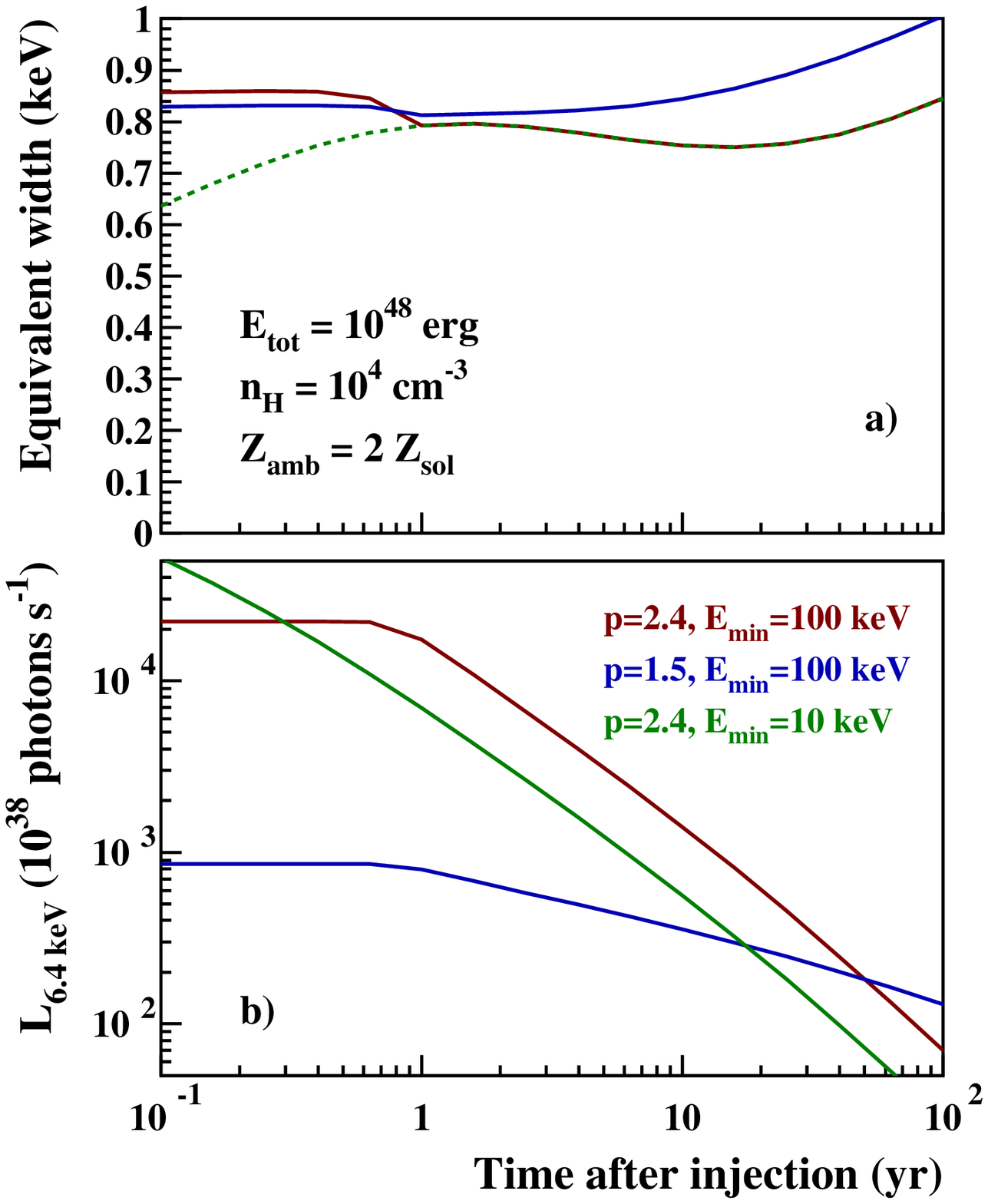}
\caption{
{\it (a - Top Left)}
The energy loss of electrons in a medium with ambient density 
of $10^4$ cm$^{-3}$ are shown 
after 1, 10 and 100 years. The initial spectrum with a power-law 
distribution is shown as a dotted line between low and high energy cutoffs 
of 100 keV and 1 GeV, respectively. 
{\it (b - Top Right)} 
Nonthermal X-ray spectra produced by LECR electrons with the assumption
that the metallicity of the ambient medium is twice solar.
Photoelectric absorption along the line of sight with a
column density N$_H=6\times10^{22}$ cm$^{-2}$ have been assumed.
{\it (c - Bottom)} 
The top  panel shows the time evolution of  the EW of the neutral FeI  K$\alpha$  line 
whereas the bottom panel shows the luminosity in the 6.4 keV line.
}\end{figure}


\begin{figure}
\center
\includegraphics[scale=0.6,angle=0]{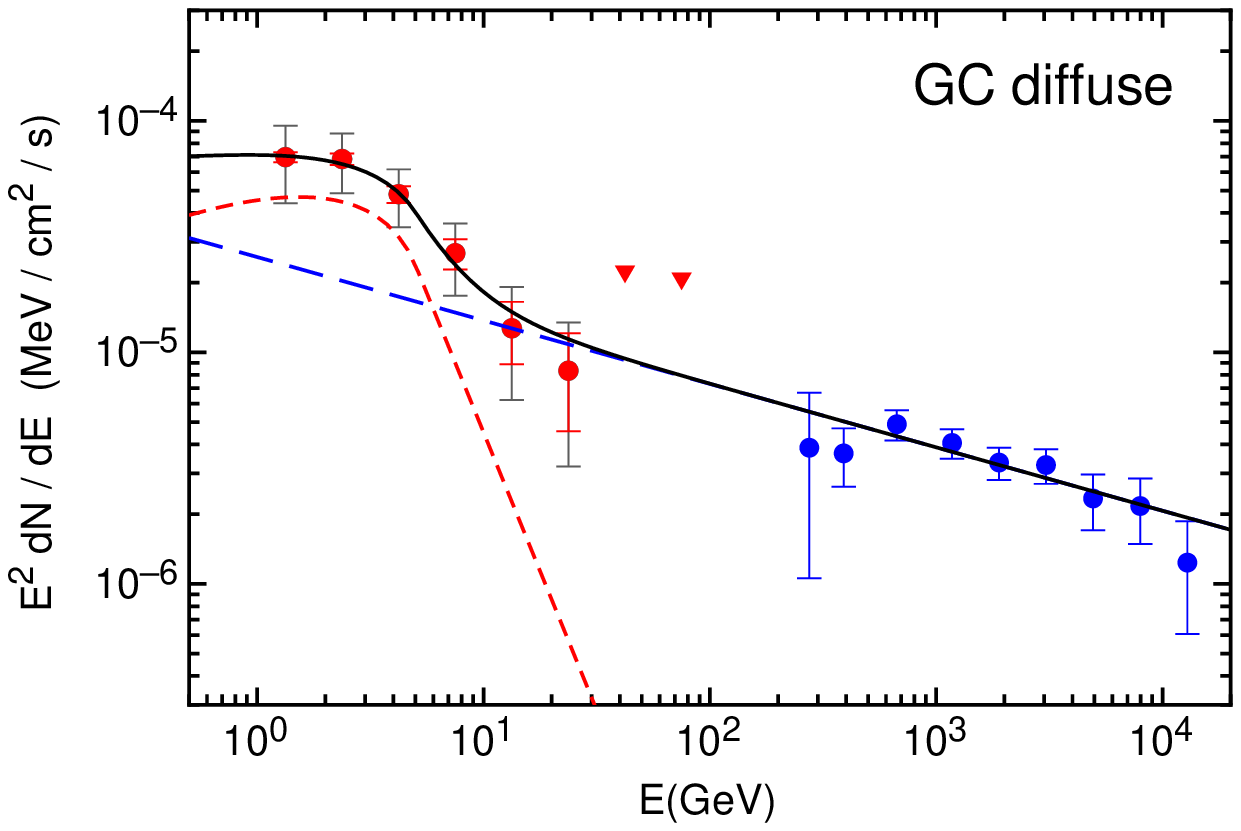}
\includegraphics[scale=0.6,angle=0]{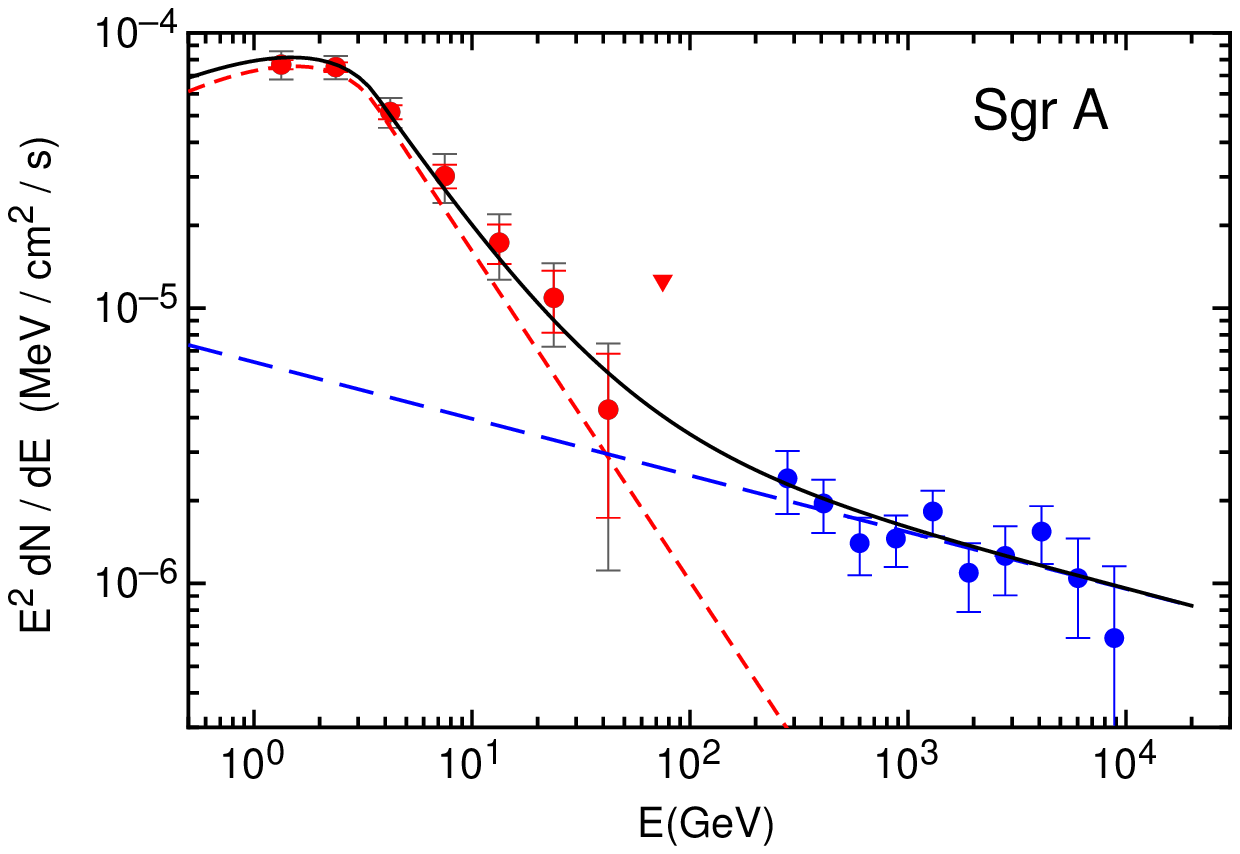}\\
\includegraphics[scale=0.6,angle=0]{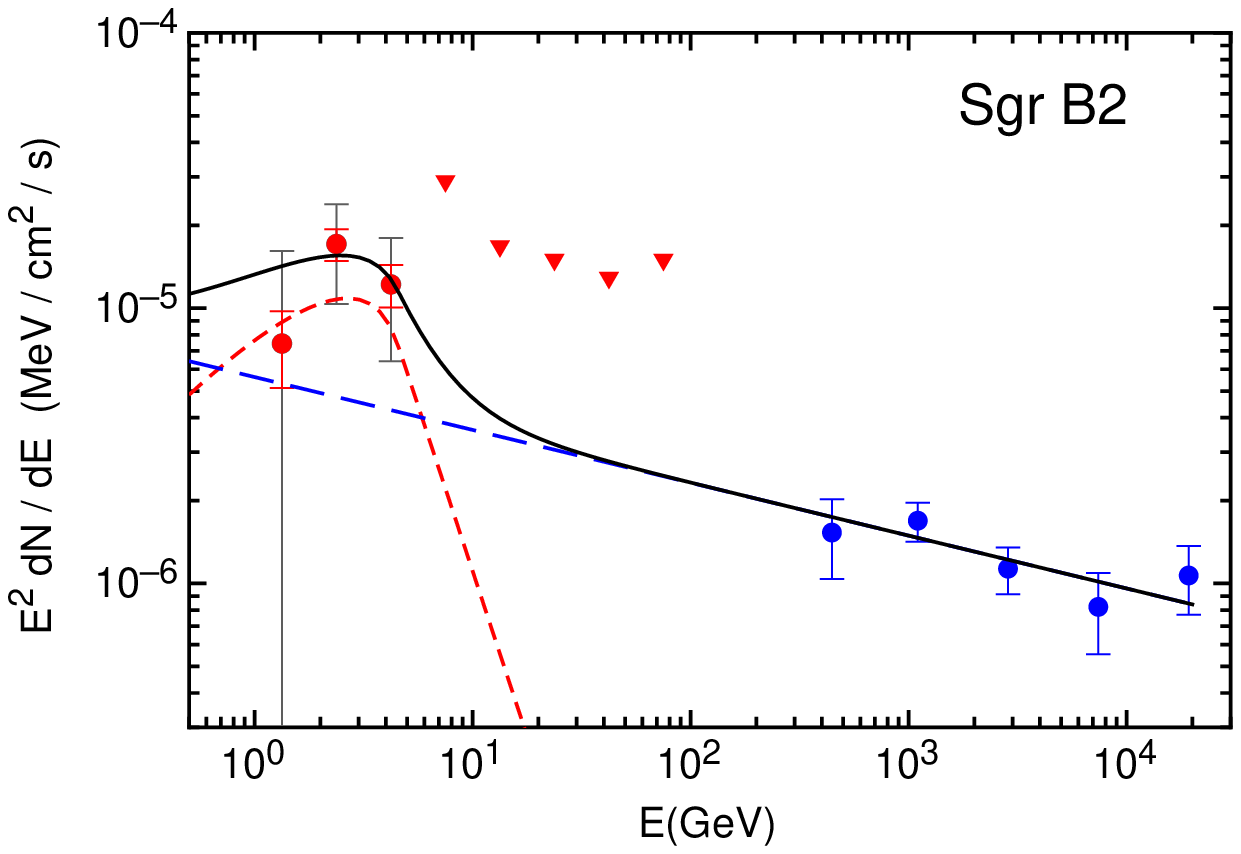}
\includegraphics[scale=0.6,angle=0]{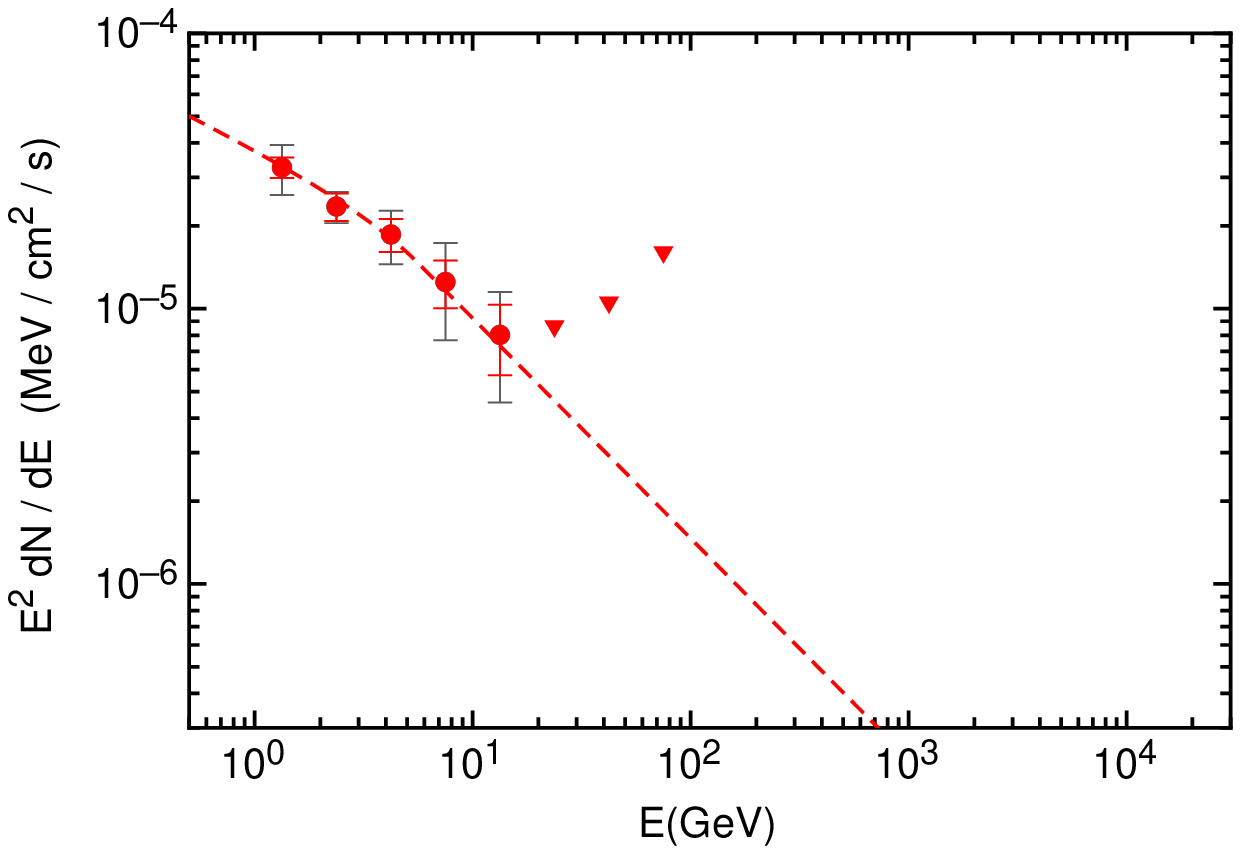}\\
\label{fig:spec_model}
\caption{
{\it (a - Top Left)}
The plot for Galactic center  diffuse emission 
shows $Fermi$ and H.E.S.S. data points 
in red and blue, respectively.
Red triangles show  3-$\sigma$  upper limits of $Fermi$ data points. 
The red dashed curve is the $\gamma$-ray bremsstrahlung predicted from
nonthermal radio spectrum using two different spectral index values. 
The blue dashed power-law fit to the H.E.S.S. 
data. 
The black solid curve is the sum of the two bremsstrahlung
contributions. 
{\it (b- Top Left)}
Similar to (a) except that  the plot shows the spectrum of $\gamma$-ray emission from 
Sgr A. 
{\it (c- Middle  Left)}
Similar to (a) except that  the plot shows the spectrum of $\gamma$-ray emission from 
Sgr B2. 
{\it (d- Middle  Right)}
Similar to (a) except that  the plot shows the spectrum of $\gamma$-ray emission from 
the radio Arc. There is no H.E.S.S. data for this source. 
The parameters of the fit to all the plots are given  in Table 4. 
}
\end{figure}

\end{document}